\documentclass[letterpaper,usenglish, 11pt]{article}
\usepackage[letterpaper, margin = 1 in]{geometry}
\usepackage[utf8]{inputenc}
\usepackage{amsmath,amssymb,amsthm}
\usepackage{placeins}
\usepackage{caption,subcaption}
\usepackage{tikz}
\usepackage{bbold} %mathbb{1}
\usepackage{hyperref,cleveref}
\usepackage{thm-restate}
\usepackage{todonotes} %while work in progress
\usetikzlibrary{calc}
\usetikzlibrary{patterns}
\usetikzlibrary{decorations.pathreplacing}
\usetikzlibrary{fadings}
\usepackage[linesnumbered,ruled,vlined]{algorithm2e}
\title{Passing the Limits of Pure Local Search for Weighted $k$-Set Packing} %TODO Please add
\author{Meike Neuwohner}%TODO mandatory, please use full name; only 1 author per \author macro; first two parameters are mandatory, other parameters can be empty. Please provide at least the name of the affiliation and the country. The full address is optional
\date{}
\newtheorem{theorem}{Theorem}
\newtheorem{lemma}[theorem]{Lemma}
\newtheorem{proposition}[theorem]{Proposition}
\newtheorem{corollary}[theorem]{Corollary}

\newtheorem{claim}[theorem]{Claim}
\theoremstyle{definition}
\newtheorem{definition}[theorem]{Definition}

        \newcommand*{\chrg}[2]{\mathrm{charge}(#1,#2)}
        
        \newcommand*{\contr}[2]{\mathrm{contr}(#1,#2)}

\begin{document}

\maketitle

\begin{abstract}
We study the weighted $k$-Set Packing problem, which is defined as follows: Given a collection $\mathcal{S}$ of sets, each of cardinality at most $k$, together with a positive weight function $w:\mathcal{S}\rightarrow\mathbb{Q}_{>0}$, the task is to compute a sub-collection $A\subseteq \mathcal{S}$ of maximum total weight such that the sets in $A$ are pairwise disjoint. For $k\leq 2$, the weighted $k$-Set Packing problem reduces to the Maximum Weight Matching problem, and can thus be solved in polynomial time~\cite{edmonds1965maximum}. However, for $k\geq 3$, already the special case of unit weights, the unweighted $k$-Set Packing problem, becomes $NP$-hard as it generalizes the 3D-matching problem~\cite{karp1972reducibility}. The state-of-the-art algorithms for both the unweighted and the weighted $k$-Set Packing problem rely on local search. In the unweighted setting, the best known approximation guarantee is $\frac{k+1}{3}+\epsilon$~\cite{FurerYu}. For general weights, Berman's algorithm SquareImp, which yields a $\frac{k+1}{2}+\epsilon$-approximation, has remained unchallenged for twenty years~\cite{Berman}. Only recently, Neuwohner managed to improve on this by obtaining approximation guarantees of $\frac{k+\epsilon_k}{2}$ with $\lim_{k\rightarrow\infty}\epsilon_k=0$~\cite{neuwohner2021limits}. She further showed her result to be asymptotically best possible in that no algorithm considering local improvements of logarithmically bounded size with respect to some fixed power of the weight function can yield an approximation guarantee better than $\frac{k}{2}$~\cite{neuwohner2021limits}. 

In this paper, we finally show how to beat the threshold of $\frac{k}{2}$ for the weighted $k$-Set Packing problem by $\Omega(k)$. We achieve this by combining local search with the application of a black box algorithm for the unweighted $k$-Set Packing problem to carefully chosen sub-instances. In doing so, we do not only manage to link the approximation ratio for general weights to the one achievable in the unweighted case. In contrast to previous works, which yield an improvement over Berman's long-standing result of $\frac{k+1}{2}+\epsilon$ either only for large values of $k\geq 2\cdot 10^5$~\cite{neuwohner2021limits}, or by less than $6\cdot 10^{-7}$~\cite{NeuwohnerLipics}, we achieve guarantees of at most $\frac{k+1}{2}-2\cdot 10^{-4}$ for all $k\geq 4$. 
\end{abstract}
\newpage
 \section{Introduction}
 For a positive integer $k$, the weighted $k$-Set Packing problem is defined as follows: As input, we are given a collection $\mathcal{S}$ of sets that contain at most $k$ elements each, and are equipped with a strictly positive weight function $w:\mathcal{S}\rightarrow\mathbb{Q}_{>0}$. The task is to find a sub-collection $A\subseteq \mathcal{S}$ such that the sets in $A$ are pairwise disjoint and the total weight of $A$ is maximum. We refer to the special case where $w\equiv 1$ as the \emph{unweighted $k$-Set Packing problem}. 
 
 For $k\leq 2$, the weighted $k$-Set Packing problem can be solved in polynomial time since adding dummy elements to all sets of size less than two yields a reduction to the Maximum Weight Matching problem~\cite{edmonds1965maximum}. On the other hand, for $k\geq 3$, already the unweighted $k$-Set Packing problem is $NP$-hard as it is a generalization of the 3D-matching problem~\cite{karp1972reducibility}. Even more, unless $P=NP$, there is no $o\left(\frac{k}{\log(k)}\right)$-approximation algorithm for the unweighted $k$-Set Packing problem~\cite{LowerBoundKSetPacking}.
 
 The technique that has proven most successful in devising approximation algorithms for both the unweighted and the general case is \emph{local search}. Given an instance $(\mathcal{S},w)$ of the weighted $k$-Set Packing problem and a collection $A\subseteq \mathcal{S}$ consisting of pairwise disjoint sets, we call $X\subseteq \mathcal{S}\setminus A$ a \emph{local improvement of $A$ of size $|X|$} if the sets in $X$ are pairwise disjoint as well, and \[w(X) > w(\{a\in A:\exists x\in X:a\cap x\neq \emptyset\}),\]
 that is, the total weight of $X$ exceeds the total weight of the sets in $A$ that intersect sets in $X$. Note that these are the sets we need to remove from $A$ in order to add the sets in $X$.
 
  A generic local search algorithm starts with an arbitrary solution, e.g.\ the empty one, and iteratively applies local improvements until no more exist. While whenever $A^*$ is an optimum solution, and $A$ is not, $A^*\setminus A$ defines a local improvement of $A$, when aiming at a polynomial running time, it is certainly infeasible to check every subset of $\mathcal{S}$. Instead, a common strategy is to either bound the maximum size of the local improvement by a constant, or to impose a certain structure that can be searched for efficiently. 
 
 In the unweighted case, the state-of-the-art $\frac{k+1}{3}+\epsilon$-approximation algorithm by F\"urer and Yu~\cite{FurerYu} searches for local improvements of constant size $\mathcal{O}(k)$, as well as a certain type of local improvement of logarithmically bounded size. A polynomial running time is achieved by means of color coding. We will employ this result as a black box in our improved local search algorithm for the weighted $k$-Set Packing problem.
 
 In the weighted setting, Berman's algorithm \emph{SquareImp}~\cite{Berman} has been unchallenged for the last twenty years. It considers local improvements with respect to the \emph{squared weight function} that are of bounded size $\leq k$ and bear a special structure that we will describe in section~\ref{SecSquareImp}. In doing so, SquareImp yields an approximation guarantee of $\frac{k+1}{2}+\epsilon$, where the $\epsilon$-term arises from scaling and truncating the weight function in order to ensure a polynomial number of iterations (see \cite{ChandraHalldorsson}, \cite{Berman}). By extending the algorithm to check for arbitrary local improvements (again with respect to $w^2$) of size at most $k^2+k$, Neuwohner~\cite{NeuwohnerLipics} could slightly improve this to an approximation guarantee of $\frac{k+1}{2}-\frac{1}{63,700,993}$. Moreover, she observed that instances on which the analysis of Berman's algorithm SquareImp is almost tight are close to being unweighted in a certain sense. By adapting the way local improvements of logarithmic size are constructed in the unweighted case, she managed to obtain approximation guarantees of $\frac{k+\epsilon_k}{2}$ with $\lim_{k\rightarrow \infty} \epsilon_k =0$~\cite{neuwohner2021limits}. In addition, she proved her result to be asymptotically best possible in that no local improvement algorithm that searches for improvements of logarithmically bounded size with respect to some fixed power of the weight function can provide an approximation guarantee better than $\frac{k}{2}$~\cite{neuwohner2021limits}.
 
  On the one hand, this result seems to conclude the story of local search algorithms for the weighted $k$-Set Packing problem. On the other hand, it appears quite striking that the ``difficult'' instances actually seem to be those that are close to the unweighted case, in which the best known approximation guarantee of $\frac{k+1}{3}+\epsilon$ is by a factor of almost $\frac{2}{3}$ better than what is known for general weights. 
 
 In this paper, we leverage this observation towards improvements in the approximation ratio both for small values of $k$ and asymptotically. More precisely, our main result is given by Theorem~\ref{TheoSimpleMainTheorem}.
 \begin{restatable}{theorem}{TheoSimpleMainTheorem}
 	For $k\geq 4$, there exists a polynomial time algorithm for the weighted $k$-Set Packing problem that obtains an approximation ratio of $\min\{0.5\cdot (k+1)-0.0002, 0.4986\cdot (k+1) + 0.0208\}$.\label{TheoSimpleMainTheorem}
 \end{restatable}
  The high-level approach is as follows: We start with the empty solution. In each iteration of our new algorithm, building on the works of Berman and Neuwohner, we first search for certain types of local improvements with respect to the squared weight function directly. If none of these exist, we consider certain sub-collections $\mathcal{S'}$ of our given set family $\mathcal{S}$ that locally resemble unweighted instances in a certain sense. For these sub-instances, we apply a black box algorithm for the unweighted $k$-Set Packing problem by simply dropping the weights. If the output of one of these computations (after some post-processing) results in a local improvement with respect to the squared weight function, we apply it. Otherwise, our algorithm terminates.
 
  As the sum of the squared weights of the sets in our solution strictly increases within every iteration of our algorithm except for the last one, it is guaranteed to terminate. A polynomial number of iterations can be ensured by scaling and truncating the weight function as in \cite{ChandraHalldorsson}, \cite{Berman}, losing only a factor of $1+\delta$ with $\delta>0$ arbitrarily small in the approximation guarantee.
  
   In order to bound the approximation ratio of our algorithm, the broad idea is that we can apply Berman's and Neuwohner's analyses. If they are close to being tight, i.e.\ yield a worse approximation guarantee than what we are aiming at, then we show that the instance must be ``locally unweighted'' and in particular, in one of the sub-instances $\mathcal{S'}$ we consider, the number $|\mathcal{S'}\cap A^*|$ of sets in $\mathcal{S'}$ from a fixed optimum solution $A^*$ is by a factor of roughly $\frac{k+1}{2}$ larger than the number $|\mathcal{S'}\cap A|$ of sets in $\mathcal{S'}$ that come from our current solution $A$. Hence, the $\frac{k+1}{3}+\epsilon$-black box approximation for the unweighted case will return a collection $X$ of sets with the property that $|X|$ is by a factor of almost $\frac{3}{2}$ larger than $|\mathcal{S'}\cap A|$. By exploiting the fact that weights between intersecting sets from $A$ and $\mathcal{S'}$ differ by less than, say $\sqrt{\frac{3}{2}}$, we can show that we can obtain a local improvement w.r.t.\ $w^2$ from $X$. Hence, we know that when our algorithm terminates, we must have reached the desired approximation guarantee stated in Theorem~\ref{TheoSimpleMainTheorem}.
 
 \section{Preliminaries}
 \subsection{Conflict graphs and the Maximum Weight Independent Set problem}
 In order to phrase the state-of-the-art algorithms for weighted $k$-Set Packing, it is convenient to introduce the notion of the conflict graph $G_{\mathcal{S}}$ representing an instance $(\mathcal{S},w)$ of the weighted $k$-Set Packing problem. Its vertices correspond to the sets in $\mathcal{S}$ and are equipped with the respective weights. The edges in $G_{\mathcal{S}}$ model non-empty set intersections. It is easy to see that for every sub-family $\mathcal{S'}\subseteq\mathcal{S}$, the conflict graph $G_{\mathcal{S}'}$ of $(\mathcal{S}',w\upharpoonright_{\mathcal{S}'})$ equals the sub-graph of $G_{\mathcal{S}}$ induced by the set of vertices that represent $\mathcal{S}'$. In particular, we obtain a weight-preserving one-to-one correspondence between solutions to the weighted $k$-Set Packing problem and the Maximum Weight Independent Set problem (MWIS) in $G_{\mathcal{S}}$.\footnote{A vertex set in a graph is called independent if none of its vertices share an edge.} While there cannot be an $\mathcal{O}(n^{1-\epsilon})$-approximation for the general Maximum Weight Independent Set problem for any $\epsilon > 0$ unless $P=NP$~\cite{InapproxIndependentSet}, the conflict graphs of instances of the (weighted) $k$-Set Packing problem bear a certain structure that we can exploit: They are \emph{$k+1$-claw free}.
 
 For $d\geq 1$, a $d$-claw is defined to be a star consisting of $d+1$ vertices, one \emph{center node} and $d$ \emph{talons} connected to it. We call a graph $G$ \emph{$d$-claw free} if none of its induced sub-graphs form a $d$-claw. This is equivalent to stating that any vertex in $G$ can have at most $d-1$ neighbors in any independent set in $G$. To see that the conflict graph $G_{\mathcal{S}}$ that corresponds to an instance $(\mathcal{S},w)$ of the weighted $k$-Set Packing problem is $k+1$-claw free, we observe that a set corresponding to the center node of a $k+1$-claw in $G_\mathcal{S}$ would need to have a non-empty intersection with each of the $k+1$ pairwise disjoint sets corresponding to the talons, contradicting the fact that its cardinality is bounded by $k$.
 \subsection{Berman's algorithm SquareImp\label{SecSquareImp}}
 We now discuss Berman's algorithm SquareImp~\cite{Berman} and its analysis our results build upon. SquareImp yields a $\frac{k+1}{2}+\epsilon$-approximation not only for the weighted $k$-Set Packing problem, but also for the more general MWIS in $k+1$-claw free graphs. SquareImp starts with the empty solution and iteratively applies \emph{claw-shaped improvements} until no more exist. Claw-shaped improvements are local improvements with respect to the \emph{squared weight function} that either consist of a single vertex, or form the \emph{set of talons of a claw} in the given graph (see Definition~\ref{DefLocalImprAndClawShaped}).
\begin{definition}
 	Let $(G,w)$ be an instance of the MWIS in $k+1$-claw free graphs and let $A\subseteq V(G)$ be independent.
 	We call an independent set $X$ a \emph{local improvement} of $A$ if $w^2(X) > w^2(N(X,A))$, where
 	$N(X,A)=(X\cap A)\cup\{u\in A:\exists v\in X: \{u,v\}\in E(G)\}$.
 	
 	We call a local improvement $X$ \emph{claw-shaped} if $|X|=1$ and $N(X,A)=\emptyset$ or if there is $v\in A$ such that $\{v\}\cup X$ induces a $|X|$-claw in $G$ centered at $v$. We say that \emph{no claw improves $A$} to state that there is no claw-shaped improvement w.r.t.\ $A$.\label{DefLocalImprAndClawShaped}
 \end{definition}
%Using Definition~\ref{DefLocalImprAndClawShaped}, Berman's algorithm SquareImp can now be formulated as in Algorithm~\ref{AlgoSquareImp}.
%\begin{algorithm}[t]
%	\DontPrintSemicolon
%	\KwIn{$k+1$-claw free graph $G=(V,E)$, $w:V\rightarrow\mathbb{Q}_{>0}$\;}
%	\KwOut{an independent set $A\subseteq V$\;}
%	$A\gets \emptyset$\;
%	\While{there exists a claw-shaped improvement $X$}
%	{$A\gets A\setminus N(X, A)\cup X$\;}
%	\Return $A$\;
%	\caption{SquareImp \cite{Berman}}\label{AlgoSquareImp}	
%\end{algorithm}

The main idea of the analysis of SquareImp presented in \cite{Berman} is to charge the vertices in  the solution $A$ returned by SquareImp for preventing adjacent vertices in an optimum solution $A^*$ from being included into $A$. (Observe that by positivity of the weight function, SquareImp returns a maximal independent set.) More precisely, Berman shows how to spread the weight of the vertices in $A^*$ among their neighbors in $A$ in such a way that no vertex in $A$ receives more than $\frac{k+1}{2}$ times its own weight. The suggested weight distribution proceeds in two steps:\\ First, each vertex $u\in A^*$ invokes costs of $\frac{w(v)}{2}$ at each $v\in N(u,A)$. In the second step, each vertex in $u$ sends the remaining amount of $w(u)-\frac{w(N(u,A))}{2}$ to a heaviest neighbor it possesses in $A$, which is captured by the following definition of \emph{charges}:
\begin{definition}[charges \cite{Berman}]
	For each $u\in A^*$, pick a vertex $v\in N(u,A)$ of maximum weight and call it $n(u)$. Define a map $\mathrm{charge}:A^*\times A\rightarrow \mathbb{R}$ via
\[\chrg{u}{v}:=\begin{cases}
	w(u)-\frac{1}{2}\cdot w(N(u,A)) &,\text{ if }v=n(u)\\
	0&,\text{ otherwise }
	\end{cases}.\] \label{DefCharges}
\end{definition}
The fact that $G$ is $k+1$-claw free directly implies Proposition~\ref{PropNeighborhoodsdminus1}. All proofs for this section are deferred to appendix~\ref{AppendixAnalysisSquareImp}.
\begin{proposition}
	Let $G$ be $k+1$-claw free, and let $A^*$ be independent. Then $|N(v,A^*)|\leq k$ for any $v\in V(G)$.\label{PropNeighborhoodsdminus1}
\end{proposition}
In particular, the total weight $v\in A$ receives in the first step is bounded by $\frac{k}{2}\cdot w(v)$.

In order to bound the total amount a vertex from $A$ has to pay for in the second step, we need to exploit the fact that when the algorithm terminates, there is no claw-shaped improvement. To this end, we introduce the notion of the contribution, which measures how much adding a vertex $u$ to the set of talons of a claw centered at $v$ contributes towards making that claw improving.
\begin{definition}
For $u\in A^*\cup V\setminus A$ and $v\in A$, define \[\mathrm{contr}(u,v):=\begin{cases}\max\left\{0,\frac{w^2(u)-w^2(N(u,A)\setminus\{v\})}{w(v)}\right\} &, \text{ if $v\in N(u,A)$}\\
0 &,\text{ else}
\end{cases}.\]
\end{definition}
The fact that there is no claw-shaped improvement yields Proposition~\ref{PropUpperBoundContr}.
\begin{proposition}
	For each $v\in A$, we have $\sum_{u\in A^*}\mathrm{contr}(u,v)\leq w(v)$.\label{PropUpperBoundContr}
\end{proposition}
Moreover,we can bound the charges send from $u\in A^*$ to $v\in A$ by half of the respective contribution.
\begin{lemma}
For all $u\in A^*$ and $v\in A$, we have $2\cdot\chrg{u}{v}\leq \contr{u}{v}$.\label{LemPropPositiveCharges}
\end{lemma}
Combining Proposition~\ref{PropNeighborhoodsdminus1}, Proposition~\ref{PropUpperBoundContr} and Lemma~\ref{LemPropPositiveCharges} results in Theorem~\ref{TheoApproxFactor}.
\begin{theorem}[\cite{Berman}]
	Let $G=(V,E)$ be $k+1$-claw free and $w:V\rightarrow\mathbb{Q}_{>0}$. Let further $A^*$ be an independent set in $G$ of maximum weight and let $A$ be independent in $G$ with the property that there is no claw-shaped improvement with respect to $A$.
	Then
	\begin{align*}
	w(A^*)&\leq\frac{k+1}{2}\cdot w(A)-\frac{1}{2}\cdot \sum_{v\in A} (k-|N(v,A^*)|)\cdot w(v)- \frac{1}{2}\cdot\sum_{u\in A^*}\left(\sum_{v\in A} \contr{u}{v} - 2\cdot \chrg{u}{n(u)}\right)\\&\leq \frac{k+1}{2}\cdot w(A).
	\end{align*}\label{TheoApproxFactor}
\end{theorem}

\subsection{A tight example\label{SecTightExample}}
\begin{figure}
	\begin{tikzpicture}[yscale = 0.5, mynode/.style = {circle, draw = black, thick, fill = none, inner sep = 0mm, minimum size = 4mm}, neighbor/.style = {circle, draw = black, thick, fill = red!70!black, inner sep = 0mm, minimum size = 4mm}]
	\node at (-2,2){part of $A$};
	\node at (-2,0){$N(v,A^*)$};
	\node[mynode, label=above:{$1$}] (1) at (0,2) {};
		\node[mynode, label=above:{$1$}] (2) at (2,2) {};
		\node[mynode, label=above:{$1$}] (3) at (4,2) {$v$};
		\node[mynode, label=above:{$1$}] (4) at (6,2) {};
		\node[mynode, label=above:{$1$}] (5) at (8,2) {};
		\node[mynode, fill, label=below:{$1$}] (13) at (0,0) {};
		\node[mynode, fill, label=below:{$1$}] (23) at (2,0) {};
		\node[mynode, fill, label=below:{$1$}] (33) at (4,0) {};
		\node[mynode, fill, label=below:{$1$}] (43) at (6,0) {};
		\node[mynode, fill, label=below:{$1$}] (53) at (8,0) {};

		\draw (5)--(53);
		\draw (3)--(53);
	\draw (1)--(13);
	\draw (2)--(23);
	\draw (3)--(33);
	\draw (4)--(43);
	\draw (3)--(13);
	\draw (3)--(23);
	\draw (3)--(43);
	\end{tikzpicture}
	\caption{The local structure of Berman's tight example.\label{Fig:TightExample}}
\end{figure}
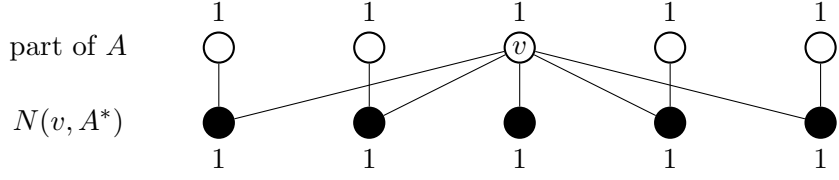

In his paper, Berman also provides an example showing that his analysis of SquareImp is best possible. The example features unit weights and locally exhibits the following structure: Every vertex $v\in A$ has one neighbor in $A^*$ of degree one to $A$, and $k-1$ further neighbors of degree two to $A$. Their second neighbors in $A$ are pairwise distinct. See Figure~\ref{Fig:TightExample} for an illustration. It is not hard to see that in this instance, we cannot find a claw-shaped improvement because $A$ is maximal and for each subset $T$ of the neighbors of a vertex $v\in A$, we have $|N(T,A)|\geq |T|$. Moreover, counting degrees from both sides yields $w(A^*)=|A^*|=\frac{k+1}{2}\cdot |A|=\frac{k+1}{2}\cdot w(A)$. By taking a closer look at the proofs of Lemma~\ref{LemPropPositiveCharges} and Theorem~\ref{TheoApproxFactor}, one can observe that this is essentially the only tight example. More precisely, one can show that every instance in which the weight of an optimum solution $A^*$ is by exactly $\frac{k+1}{2}$ larger than the weight of an independent set $A$ that no claw improves, exhibits the following properties: Within each connected component of $G[A\cup A^*]$, all weights are equal, and moreover, the neighborhood of each vertex in $A$ bears exactly the same structure as depicted in Figure~\ref{Fig:TightExample}.
The high-level idea of the prior works \cite{NeuwohnerLipics} and \cite{neuwohner2021limits} as well as of the current paper is to show that close-to-tight instances, in which the analysis of SquareImp is worse than what we are aiming at, retain enough of the structure of the unweighted worst case example to allow us to employ techniques from the unweighted setting, where the best known approximation guarantee is by a factor of $\frac{2}{3}$ lower than for general weights.
\subsection{Circular improvements}
 As Neuwohner's algorithm \emph{LogImp}\cite{neuwohner2021limits}, we make use of the notion of a \emph{circular improvement}, a certain type of local improvement of logarithmic size. It is inspired by the type of local improvements considered in the state-of-the-art work on the unweighted $k$-Set Packing problem~\cite{FurerYu}, as well as the specific structure of the tight example discussed in the previous section.
  The backbone of a circular improvement is given by a cycle $C$ of logarithmic size in an auxiliary graph, where each vertex $u$ that is not contained in our current solution $A$ with at least two neighbors in $A$ induces an edge between its two heaviest ones $n(u)$ and $n_2(u)$.\footnote{Here, the map $n:V\rightarrow A$ is chosen in accordance with the definition of charges, and $n_2$ is defined in such a way that it maps each vertex in $V$ with at least two neighbors in $A$ to one of maximum weight in $N(u,A)\setminus\{n(u)\}$.} Denote the set of vertices from $V\setminus A$ that induce the edges of the cycle by $U$. Our circular improvement $X$ is now of the form $X=U\dot{\cup}\dot{\bigcup}_{v\in V(C)}Y_v$, where, for every $v\in V(C)$, $Y_v\subseteq N(v,V\setminus A)$ can be thought of as a set of vertices that have positive contributions to $v$. Note that $V(C)\subseteq N(U,A)$ will be contained in the neighborhood of our local improvement anyways, so adding the vertices from the sets $Y_v$ can only help us. In order to ensure that the sets $Y_v$ are pairwise disjoint, we further require that $n(u)=v$ for every $u\in Y_v$.
 We now give the formal definition:
  \begin{definition}[Circular improvement]
 	Let $\kappa\in(0,1)$ with $\frac{1}{\kappa}\in\mathbb{N}^+$. Let $(G,w)$ be an instance of the MWIS in $k+1$-claw free graphs, let $A$ be a maximal independent set and let two fixed maps \begin{itemize} \item $n:V\rightarrow A$ mapping $u$ to an element of $N(u,A)$ of maximum weight and \item $n_2:\{u\in V: |N(u,A)|\geq 2\}\rightarrow A$ mapping $u$ to an element of $N(u,A)\setminus\{n(u)\}$ of maximum weight\end{itemize} be given. We call an independent set $X\subseteq V\setminus A$ a \emph{circular improvement} if there exists\\ $U\subseteq X\cap\{u\in V\setminus A: |N(u,A)|\geq 2\}$ with $|U|\leq \frac{8}{\kappa}\cdot \log(|V|)$ such that 
 	\begin{enumerate}
 		\item $C:=\left(\bigcup_{u\in U}\{n(u), n_2(u)\}, \{e_u=\{n(u), n_2(u)\}, u\in U\}\right)$ is a cycle, where we view the edge set as a multi-set and consider two parallel edges as the edge set of a cycle of length $2$.
 		\item $X=U\dot{\cup}\dot{\bigcup}_{v\in V(C)} Y_v$, where $Y_v:=\{x\in X\setminus U: n(x)=v\}$ for $v\in V(C)\subseteq A$. (Note that as we require $X$ to be independent and $G$ is $k+1$-claw free, we automatically have $|Y_v|\leq k$ for all $v\in V(C)$, which is explicitly stated as part of the definition in \cite{neuwohner2021limits}.)
 		\item For each $u\in U$, we have
 		\begin{align*}&\phantom{=}w^2(u)+\frac{1}{2}\cdot w^2(Y_{n(u)}) +\frac{1}{2}\cdot w^2(Y_{n_2(u)})\\&> \frac{w^2(n(u))+w^2(n_2(u))}{2}+w^2(N(u,A)\setminus\{n(u),n_2(u)\})\\\phantom{=}&\quad+\frac{1}{2}\cdot\sum_{x\in Y_{n(u)}} w^2(N(x,A)\setminus\{n(u)\})+\frac{1}{2}\cdot\sum_{x\in Y_{n_2(u)}} w^2(N(x,A)\setminus\{n_2(u)\}).\end{align*}
 	\end{enumerate}\label{DefCircular}
 \end{definition}
The third constraint is used to ensure that $X$ indeed defines a local improvement by reducing it to a property that can be checked locally for each edge. The latter is necessary to be able to search for circular improvements in polynomial time for conflict graphs of $k$-Set Packing instances, which is shown to be possible in \cite{neuwohner2021limits}.

 In more detail, the third constraint tells us that for each vertex $u$ that induces an edge, we can use the squared weight of $u$, together with half of the contributions of $Y_{n(u)}$ to $n(u)$ and $Y_{n_2(u)}$ to $n_2(u)$, respectively, to cover for half of the squared weights of $n(u)$ and $n_2(u)$, plus the squared weight of those neighbors of $u$ in $A$ other than $n(u)$ and $n_2(u)$. As each vertex in $V(C)$ has two incident edges, summing up these (strict) inequalities tells us that every circular improvement defines a local improvement.

  To get some intuition why this notion might be helpful, we point out that for $k\geq 4$, the auxiliary graph we consider for our tight example, when restricted to edges induced by $A^*$, is $(k-1)$-regular with $k-1\geq 3$, as every vertex in $A$ features $k-1$ neighbors in $A^*$ that have degree two to $A$ and, hence, induce edges. In particular, a result by Berman and F\"urer about the girth of sufficiently dense graphs~\cite{berman1994approximating} implies the existence of a cycle of logarithmically bounded size. We can further choose the sets $Y_v$ to consist of the unique vertex $u\in A^*$ with $N(u,A)=\{v\}$. In doing so, we have, on average, $2\cdot \frac{1}{2}+1=2$ vertices to cover for each vertex $v$ of our cycle, meaning that we indeed obtain a local improvement. Moreover, the latter property is preserved when locally perturbing weights by a factor of, say, $1\pm\epsilon$. This indicates that a similar approach might also work for instances where the analysis of SquareImp is only close to being tight. 
 \section{Outline of our contribution and the key technical ideas\label{SecOutline}}
In this section, we provide an overview of the key steps and ideas towards better approximation guarantees for the weighted $k$-Set Packing that asymptotically beat the $\frac{k}{2}$-barrier established in \cite{neuwohner2021limits} for algorithms that rely on local improvements of logarithmically bounded size, but also yield visible improvements for small values of $k$ (cf. Theorem~\ref{TheoSimpleMainTheorem}). 

To reach this goal, we would like to build upon Berman's algorithm SquareImp and its analysis because this allows us to exploit the nice structure close-to-tight instances exhibit. Recall that the analysis presented in section~\ref{SecSquareImp} applies to any independent set $A$ for which no claw-shaped improvement exists. Thus, it appears natural to study a local improvement algorithm which, in each iteration, searches for different types of local improvements (each with respect to the squared weight function to make sure that we do not cycle) including claw-shaped ones, until no more exist. In this situation, the lower bound result in \cite{neuwohner2021limits} implies that if we aim at approximation guarantees below $\frac{k}{2}$, then we need to search for local improvements of more than logarithmic size. However, we have seen in section~\ref{SecTightExample} that the worst case instances for the analysis of SquareImp are actually (basically) unweighted. In addition, we know that the best known approximation factor for the unweighted case is by a factor of $\frac{2}{3}$ smaller than the approximation guarantee Berman's analysis yields. Hence, it seems natural to try to obtain local improvements by simply dropping weights and running a black box algorithm for the unweighted setting. In the following, as we find it more convenient terminology-wise, we will phrase most of our ideas and results in terms of the MWIS in $k+1$-claw free graphs, instead of the weighted $k$-Set Packing problem. In doing so, we will, however, assume all of our input instances to be conflict graphs of explicitly given instances of the weighted $k$-Set Packing problem. In particular, induced sub-graphs of our input graphs will be conflict graphs of the respective sub-instances of the (weighted) $k$-Set Packing problem, and whenever we say that we apply a black box approximation for the Maximum (Cardinality) Independent Set problem (MIS) to them, we actually mean that implicitly, we cast them back into a $k$-Set Packing instance and run the respective algorithm for unweighted $k$-Set Packing. To shorten notation, we use \texttt{MIS} to refer to the state-of-the-art $\frac{k+1+\tilde{\epsilon}}{3}$-approximation algorithm for unweighted $k$-Set Packing by F\"urer and Yu~\cite{FurerYu}, where we can choose $\tilde{\epsilon}>0$ arbitrarily small. 

 Having settled the notation, we would like to motivate our approach by first showing how to make it work in the case in which no claw improves our current solution $A$, and Berman's analysis is tight, that is, we have $w(A^*)=\frac{k+1}{2}\cdot w(A)$ for any optimum solution $A^*$. As pointed out in section~\ref{SecTightExample}, this results in the very restricted setting where weights are equal within each connected component of $G[A\cup A^*]$. In this situation, for each weight $x$, we define the vertex set $V^x$ to consist of the set $A^x$ of all vertices from $A$ of weight $x$, together with all vertices $u\in V\setminus A$ such that $w(u)=x$ and every neighbor of $u$ in $A$ has weight $x$ as well. We claim that if we apply \texttt{MIS} to each of the instances corresponding to the sub-graphs $G[V^x]$, where $x$ occurs among the vertex weights, then for at least one of them, we obtain a local improvement with respect to the squared weight function. To see this, first observe that by our assumption, all vertices from $A^*$ of weight $x$ must be contained in $V^x$ since all of their neighbors in $A$ bear the same weight. Hence, as $w(A^*)=\frac{k+1}{2}\cdot w(A)$, there must be a weight $x_0$ with the property that $|A^*\cap V^{x_0}|\geq \frac{k+1}{2}\cdot|A^{x_0}|$. In addition, we know that $A^*\cap V^{x_0}$ defines an independent set in $G[V^{x_0}]$, meaning that \texttt{MIS} will return an independent set $Y$ of cardinality at least \[ |Y|\geq\frac{3}{k+1+\tilde{\epsilon}}\cdot|A^*\cap V^{x_0}|\geq \frac{3}{k+1+\tilde{\epsilon}}\cdot\frac{k+1}{2}\cdot |A^{x_0}| > |A^{x_0}|\] for $\tilde{\epsilon}$ chosen sufficiently small. By construction of $V^{x_0}$, we know that $N(Y,A)\subseteq A^{x_0}$ and, in particular, $w^2(Y)= x_0^2\cdot |Y| > x_0^2\cdot |A^{x_0}|= w^2(A^{x_0})\geq w^2(N(Y,A))$, so $Y$ yields a local improvement.
 We would like to generalize this approach in order to prove Theorem~\ref{TheoSimpleMainTheorem}, which tells us that for every $k\geq 4$\footnote{It will become apparent in the following why we exclude the case $k=3$.}, we can improve on the guarantee of $\frac{k+1}{2}$ by at least $0.0002$ and, in addition, obtain an asymptotic improvement over the lower bound of $\frac{k}{2}$ from \cite{neuwohner2021limits}. We restate it here for convenience.
\TheoSimpleMainTheorem*
To establish Theorem~\ref{TheoSimpleMainTheorem}, we have to deal with the case where the ratio between the weight of an optimum solution $A^*$ and our current solution $A$ may be smaller than $\frac{k+1}{2}$, but is still larger than the guarantee we are targeting.\footnote{Note that if $\frac{w(A^*)}{w(A)}$ is already smaller than what we are aiming at, there is nothing left to show.} In a slight abuse of notation, we will refer to these triples consisting of an instance, $A$ and $A^*$ as \emph{close-to-tight instances} simply. In order to tackle arbitrary close-to-tight instances, we first show that we retain a lot of the structure from our motivating example in that they are ``locally unweighted'' in a certain sense. In doing so, in section~\ref{SecUnweightedSubInstances}, we first follow the argumentation in \cite{neuwohner2021limits} (but we refine it in section~\ref{SecTechnicalDetails} to obtain significantly better constants). Then, section~\ref{SecLocalImprovementOutline} explains how to adjust our construction of sub-instances to which we apply the algorithm \texttt{MIS} from the motivating example to this more general setting. Finally, section~\ref{SecTechnicalDetails} comments on some technical tricks that we employ in order to improve our guarantees.

In the following, we fix an instance $(G,w)$ of the MWIS in $k+1$-claw free graphs and an independent set $A^*$ of maximum weight. In addition, we use $A$ to refer to the current solution of our algorithm, which should have the property that no claw improves $A$, but $(G,w,A,A^*)$ is close-to-tight.
\subsection{Locally unweighted sub-instances\label{SecUnweightedSubInstances}}
Following \cite{neuwohner2021limits}, we introduce the concept of \emph{helpful vertices} (see Definition~\ref{DefHelpful}). It refers to those vertices from $V\setminus A$ that are eligible to appear as neighbors of a certain vertex $v\in A$ in the sub-instances we would like to apply the algorithm $\texttt{MIS}$ to. Thus, helpful vertices help us to find a local improvement. As we would like the mentioned sub-instances to be as close to the unweighted setting as possible, helpful vertices are defined to mimic the two types of neighbors that appear in Berman's worst case example. More precisely, we fix a constant $\epsilon > 0$ and say that a vertex $u\in V\setminus A$ is helpful for its heaviest neighbor $n(u)$ if $w(u)$ and $w(n(u))$ differ only by a factor of $1+\epsilon$, and $w(N(u,A)\setminus\{n(u)\})\leq \epsilon\cdot w(u)$. In addition, if $u$ possesses at least two neighbors in $A$, we call $u$ helpful for its two heaviest ones $n(u)$ and $n_2(u)$ if the weights of $u$, $n(u)$ and $n_2(u)$ differ only by a factor of $1+\epsilon$, and $w(N(u,A)\setminus\{n(u), n_2(u)\})\leq \epsilon\cdot w(u)$. The vertices $u$ for which the first case applies can be thought of as having ``degree $1$'' to $A$, regarding vertices of weight $\mathcal{O}(\epsilon)\cdot w(u)$ as ``almost not existent''. The vertices for which the second set of conditions applies correspond to those of degree $2$ to $A$ in the unweighted setting.

In order to be able to construct a local improvement by applying \texttt{MIS}, we need to see that the number of helpful vertices is large enough such that even after factoring in the losses due to locally differing weights, as well as the fact that \texttt{MIS} may only recover a $\frac{3}{k+1+\tilde{\epsilon}}$-fraction of the helpful vertices, we still obtain an improvement over $A$. We do this by explaining how vertices in $A$ with few helpful neighbors improve our approximation guarantee.

 In doing so, we adopt some of the terminology from \cite{neuwohner2021limits} and say that each vertex $v\in A$ has $k-|N(v,A^*)|$ \emph{missing neighbors} (Def.~\ref{DefMissing}), where the term missing alludes to the fact that the worst case bound of $\frac{k+1}{2}$ in the analysis of SquareImp is only attained if every vertex in $A$ possesses the maximum number of $k$ neighbors in $A^*$. Looking at Theorem~\ref{TheoApproxFactor}, we can see that we in fact gain $\frac{w(v)}{2}$ in our bound on $w(A^*)$ for every missing neighbor of a vertex $v\in A$.
 
  In addition to missing neighbors, we also need to deal with vertices in $N(v,A^*)$ that are present, but do not turn out to be helpful for $v$, e.g.\ because their weight deviates too much from the weight of $v$. The idea of our analysis is to make each vertex $u\in A^*$ recompense each one of its neighbors $v\in N(u,A)$ for which $u$ is not helpful by improving our bound on $w(A^*)$ by some constant fraction of $w(v)$. This gives rise of the notion of the \emph{support} $\mathrm{supp}(u)$ of a vertex $u\in A^*$, which consists of all vertices in $N(u,A)$ that $u$ has to reimburse (see Def.~\ref{DefSupport}). In order to do so, we allow $u$ to spend the slack we have in the inequality $\frac{1}{2}\cdot \left(\sum_{v\in A} \contr{u}{v}-2\cdot\chrg{u}{n(u)}\right)\geq 0$, which appears in our bound on $w(A^*)$ in Theorem~\ref{TheoApproxFactor}. If we choose the support of $u$ to consist of all neighbors of $u$ in $A$ that $u$ is not helpful for, then we can show that we can compensate every such vertex with an $\Omega(\epsilon^2)$ fraction of its own weight (see~\cite{neuwohner2021limits}). This is sufficient to show that we either gain $\Omega(\epsilon^2)\cdot k\cdot w(A)$ in the approximation guarantee, which is (up to some technical details that we will comment on in section~\ref{SecTechnicalDetails}) the statement of Lemma~\ref{LemGainFromLowDegrees}, or we have \begin{equation}
  \sum_{v\in A} w(v)\cdot \text{\# helpful neighbors of $v$} > 2\cdot\frac{(1+\epsilon)^3}{1-\epsilon^2}\cdot\frac{k+1+\tilde{\epsilon}}{3}\cdot w(A). \footnote{The factor $2$ comes from the fact that we may count each helpful vertex twice (once for $n(u)$ and once for $n_2(u)$), and the factor $\frac{(1+\epsilon)^3}{1-\epsilon^2}$ covers for the local weight differences.}\label{EqHelpfulNeighbors}
  \end{equation} We will outline in the next section how to obtain a local improvement by applying the algorithm \texttt{MIS} if the latter holds. This then shows how we can obtain an $\Omega(k)$ improvement in the approximation guarantee, where the constants may, however, be rather tiny. To remedy this issue, section~\ref{SecTechnicalDetails} comments on how to tweak the definition of the support and refine our analysis in order to also obtain a visible improvement in the approximation guarantee for small values of $k$.

\subsection{Obtaining a local improvement\label{SecLocalImprovementOutline}}
In this section, we explain how to obtain a local improvement via an application of our black box algorithm \texttt{MIS} for the unweighted problem, provided the weighted sum over the number of helpful neighbors a vertex $v\in A$ has in $A^*$ is sufficiently large. See Lemma~\ref{LemImprovementNextIter} for the precise statement. Similar to what we did in the motivating example, we would like to derive the existence of a certain sub-instance $G[V']$ with the property that $|A^*\cap V'|>\frac{(1+\epsilon)^2}{1-\epsilon^2}\cdot\frac{k+1+\tilde{\epsilon}}{3}\cdot | A\cap V'|$. This means that \texttt{MIS} is bound to produce an independent set of cardinality larger than $\frac{(1+\epsilon)^2}{1-\epsilon^2}\cdot|V'\cap A|$  in $G[V']$. For our sub-instances, we only consider vertices from $V\setminus A$ that are helpful for some of their neighbors in $A$ as for these, we retain enough structural information about our instance when dropping weights. To bound the number of helpful vertices from $A^*$ in a certain sub-instance, we observe that as each vertex $u\in A^*$ is helpful for at most two vertices in $A$, it is sufficient to show that \begin{equation}\sum_{v\in A\cap V'}(\text{\# vertices in $A^*\cap V'$ that are helpful for $v$})>2\cdot\frac{(1+\epsilon)^2}{1-\epsilon^2}\cdot\frac{k+1+\tilde{\epsilon}}{3}\cdot |A\cap V'|.\label{EqNumHelpfulNeighborsSubInstance}\end{equation} To reach this goal, we need to exploit the bound \eqref{EqHelpfulNeighbors} and the fact that the weights of vertices in $A$ and their helpful neighbors only differ by a factor of at most $1+\epsilon$. We first point out that these properties do not suffice to derive a bound on the ratio between the cardinalities of $A^*$ and $A$. The reason for this is that it could happen that there are a few vertices in $A$ of very large weight that each have $k$ helpful neighbors in $A^*$ (which, hence, also have a very high weight) and dominate the weighted sum, whereas there is a huge number of vertices in $A$ with very small weight and no helpful neighbors at all. However, if we look at sub-instances that only contain vertices the weight of which lies above a certain threshold, it is not hard to see that in the previous example, we can find such a sub-instance in which the number of vertices from $A^*$ is much larger than the number of vertices from $A$. As it turns out, this is not a coincidence, but in fact, we can show that there always exists a weight threshold $L$ such that if we restrict to the set of vertices of weight at least $L$, then the corresponding sub-instance meets \eqref{EqNumHelpfulNeighborsSubInstance}. The intuition behind this may be that either, the high numbers of helpful neighbors are accumulated at vertices from $A$ of high weight, or the numbers of helpful neighbors are stretched out evenly, showing that the number of vertices from $A^*$ in our sub-instance is sufficiently large compared to the respective number of vertices from $A$.
As a consequence, if we apply \texttt{MIS} to all of the sub-instances $G[V_{\geq L}]$ containing vertices from $A$ and helpful vertices from $V\setminus A$ that bear at least a certain weight $L$, for at least one of them, \texttt{MIS} yields an independent set $X$ such that $|X|>\frac{(1+\epsilon)^2}{1-\epsilon^2}\cdot |A\cap V_{\geq L}|$. Ideally, we would now like to conclude that $X$ yields a local improvement. Unfortunately, we can run into a problem that is complementary to the one we previously encountered: If all of the vertices in $X$ feature a weight very close to the lower bound $L$, whereas $A$ contains a lot of vertices of high weight that are adjacent to a few vertices in $X$, $X$ does not need to define a local improvement. However, if instead, we look at subsets $X^{\leq U}$ of $X$ that contain all vertices in $X$ the weight of which is upper bounded by $U$, a similar argument as before shows that $(1-\epsilon^2)\cdot w^2(X)> w^2(N(X,A\cap V_{\geq L}))$. If during the sub-instance construction, we additionally make sure that we only include helpful vertices $u\in V\setminus A$ with the property that $w(N(u,A\setminus V_{\geq L})) \leq \epsilon\cdot w(u)$, this is sufficient to derive a local improvement.
\subsection{Technical details to improve the analysis\label{SecTechnicalDetails}}
In this section, we comment on some further technicalities we need to take care of in order to obtain improvements in a reasonable order of magnitude also for small values of $k$. They all originate from the fact that if we define the support of $u\in A^*$ to contain all vertices in $N(u,A)$ that $u$ is not helpful for, we can only guarantee a reimbursement in the order of $\mathcal{O}(\epsilon^2)$ times the weight of each supported vertex. See section~\ref{SecAnalysis} for the details. Luckily, it turns out that if we exclude what we refer to as \emph{special neighbors} (Def.~\ref{DefSpecial}) from the support of $u$, then we can ensure a reimbursement in the order of $\epsilon$ times the weight of each supported vertex (cf.~Lemma~\ref{LemProfitFromBadNeighbors}). We say that $u\in A^*$ and $v\in A$ are \emph{special neighbors} if $u$ is not helpful for $v$, $v=n(u)$ and $\contr{u}{v}>\frac{5}{8}\cdot w(v)$. By Proposition~\ref{PropUpperBoundContr}, we know that in every iteration where our algorithm does not find a claw-shaped improvement, every vertex $v\in A$ can have at most one special neighbor. In particular, for each $v\in A$, we gain $\Omega(\epsilon)\cdot (k-1-\text{\#helpful neighbors of $v$})\cdot w(v)$ in the analysis (cf. Lemma~\ref{LemGainFromLowDegrees}). For sufficiently large values of $k$, we can now, essentially, proceed in the same way as before, only getting a slightly worse dependence of the improvement on $k$. For $k=3,4,5$, however, we run into a problem: Recall that in order to be able to produce local improvements by applying \texttt{MIS}, we needed the property \eqref{EqHelpfulNeighbors}. But for $k\leq 5$, we have $\frac{2}{3}\cdot (k+1)\geq k-1$, so even if all non-special neighbors of the vertices in $A$ are helpful, this is not enough. To resolve this problem (at least for $k\geq 4$), circular improvements come into play. The key observation is that if we really have one special neighbor for each vertex in $A$, they form ideal candidates for the sets $Y_v$ we need for a circular improvement (meaning that we let $Y_v$ consist of the special neighbor only). Denote the set of vertices in $A$ that feature a special neighbor in $A^*$ by $A'$. We can show that if $w(A')$ is larger than a certain threshold, the conditions we need in order to ensure the existence of a circular improvement are weaker than \eqref{EqHelpfulNeighbors} (cf. Lemma~\ref{LemImprovementAprimeLarge}). Combining Lemma~\ref{LemGainFromLowDegrees}, Lemma~\ref{LemImprovementNextIter} and Lemma~\ref{LemImprovementAprimeLarge} yields our main technical theorem (Theorem~\ref{TheoMainTheorem}), which exposes the precise dependence of the approximation guarantee we obtain on our choices of constants. For $k\geq 4$, by plugging in the right constants (see Table~\ref{TableEpsXi}), we obtain Theorem~\ref{TheoSimpleMainTheorem}. For $k=3$ unfortunately, we cannot make this refined approach work, but need to settle for the much weaker constants we can achieve by supporting all non-helpful vertex. We defer the details to section~\ref{SecD4}.
\subsection{The relation between weighted and unweighted \boldmath$k$-Set Packing\unboldmath}
We conclude this overview of our contribution by pointing out that our approach does not only yield improved approximation guarantees for weighted $k$-Set Packing for $k\geq 4$ and allows us to asymptotically beat the threshold guarantee of $\frac{k}{2}$ established in \cite{neuwohner2021limits}. It is also the first work to tie the approximation guarantee that we can obtain in the weighted setting to the best one achievable in the unweighted case. More precisely, with a very similar analysis as the one we will present in the following to prove Theorem~\ref{TheoSimpleMainTheorem}, we can also establish the following result, which is proven in appendix~\ref{SecRelation}.
\begin{restatable}{theorem}{TheoConstantFactor}
 For any constant $\sigma\in (0,1)$, there exists a constant $\tau\in(0,1)$ with the following property: If there are $k_0\in\mathbb{N}_{\geq 3}$ and a polynomial time $1+\tau\cdot(k-1)$ approximation algorithm for the unweighted $k$-Set Packing problem for $k\geq k_0$, then there is a polynomial time $1+\sigma\cdot(k-1)$-approximation algorithm for the weighted $k$-Set Packing problem for $k\geq k_0$. \label{TheoConstantFactor}\end{restatable} 
 The rest of this paper is organized as follows:
In section~\ref{SecAlgo}, we formally introduce our new algorithm and show that it can be implemented to run in polynomial time. The analysis of the performance ratio is carried out in section~\ref{SecAnalysis}, which also gives a more detailed overview of the approximation guarantees we can prove for small values of $k$. Section~\ref{SecD4} explain in more detail why a different approach is needed for the case $k=3$. Finally, section~\ref{SecConclusion} concludes our paper.
\section{Our algorithm\label{SecAlgo}}
In this section, we introduce our new algorithm for the weighted $k$-Set Packing for $k\geq 4$. For convenience, we will phrase most of our algorithm and its analysis in terms of the more general MWIS in $k+1$-claw free graphs. The only points where we actually need the additional structure of an instance of the (weighted) $k$-Set Packing problem is to search for circular improvements in polynomial time (see~\cite{neuwohner2021limits}) and to approximate the MIS on induced sub-graphs of the input graph by applying an algorithm for the unweighted $k$-Set Packing problem. (Recall that for a given instance $(\mathcal{S},w)$ of the weighted $k$-Set Packing problem, the induced sub-graph $G_{\mathcal{S}}[V']$ equals the conflict graph of the $k$-Set Packing instance that is given by the sub-collection $\mathcal{S}'$ consisting of those sets corresponding to the vertices in $V'$. In particular, approximating the MIS in $G_{\mathcal{S}}[V']$ is equivalent to approximating the unweighted $k$-Set Packing problem on $\mathcal{S}'$.)
As in section~\ref{SecOutline}, we use \texttt{MIS} to refer to the state-of-the-art approximation algorithm for the unweighted $k$-Set Packing by F\"urer and Yu~\cite{FurerYu} and denote its approximation guarantee by \begin{equation}\rho:=\frac{k+1+\tilde{\epsilon}}{3}> 1,\label{EqBoundsRho}\end{equation} where we will choose $\tilde{\epsilon}>0$ sufficiently small to suit our purposes. Whenever we apply \texttt{MIS} to an induced sub-graph of the conflict graph of an instance of the weighted $k$-Set Packing problem, it is understood that we implicitly run \texttt{MIS} on the sub-instance of the (unweighted) $k$-Set Packing problem that corresponds to the induced sub-graph.

In order to formulate our algorithm, we introduce a constant $\epsilon> 0$, which can be thought of as a threshold for declaring relative weight differences as small. Its precise values (depending on $k$) will be given in section~\ref{SecAnalysis}. We can now formally define the notion of helpful neighbors of vertices in $A$. The reader may think of these as vertices from $V\setminus A$ that would be adjacent to $v$ in an unweighted approximation of our instance.
\begin{definition}[helpful vertex]
	Let $v\in A$. We say that a vertex $u\in N(v,V\setminus A)$ is \emph{helpful for $v$} if 
	\begin{enumerate}
			\item \begin{enumerate}
			\item $v= n(u)$ and
			\item $w(n(u))\leq (1+\epsilon)\cdot w(u)$ and
			\item $w(N(u,A)\setminus\{n(u)\})\leq \epsilon\cdot w(u)$, or
		\end{enumerate}
		\item \begin{enumerate}
			\item $|N(u,A)|\geq 2$ and 
			\item $v\in\{n(u),n_2(u)\}$ and
			\item $(1+\epsilon)^{-1}\cdot w(n(u))\leq w(n_2(u))\leq w(n(u))\leq (1+\epsilon)\cdot w(u)$ and
			\item$w(N(u,A)\setminus\{n(u),n_2(u)\})\leq \epsilon\cdot w(u)$. \end{enumerate}
	\end{enumerate} For $u\in V\setminus A$, we define $\mathrm{help}(u):=\{v\in A: \text{$u$ is helpful for $v$}\}$ and for $v\in A$, we set $\mathrm{help}(v):=\{u\in V\setminus A: \text{$u$ is helpful for $v$}\}$.\label{DefHelpful}
\end{definition}
We point out that in contrast to the intuition we gave earlier on, we actually do not impose an upper bound on the weight of helpful vertices $u$, requiring their weight to be bounded by $(1+\epsilon)\cdot w(v)$ for every vertex $v$ they are helpful for. While such an additional constraint would not hurt our analysis, it turns out to be unnecessary, so we omit it in order to keep the already quite tedious calculations a little shorter. The intuitive reason why we can drop the upper bound on the weight of helpful vertices is that increasing their weight only makes it easier for us to find a local improvement.
 
 We are now prepared to present our new algorithm, Algorithm~\ref{LocalImprovementAlgo}. Given an instance $(\mathcal{S},w)$ of the weighted $k$-Set Packing problem, Algorithm~\ref{LocalImprovementAlgo} first constructs its conflict graph $G=G_{\mathcal{S}}$ and initializes the independent set $A$ in $G$ it maintains to be the empty set. Now, it iteratively calls the sub-routine \texttt{RunIteration} (see Algorithm~\ref{AlgoRunIteration}), which checks whether a local improvement of one of four possible types exists. If \texttt{RunIteration} finds such a local improvement, it updates $A$ accordingly and returns \texttt{true}. Otherwise, it returns \texttt{false}. As soon as \texttt{RunInteration} does not manage to find a local improvement anymore, Algorithm~\ref{LocalImprovementAlgo} terminates and returns the collection of sets that corresponds to the current solution $A$.
 
 We now take a closer look at what is happening within one iteration of Algorithm~\ref{LocalImprovementAlgo} (i.e.\ one call to \texttt{RunIteration}). We start by checking for the existence of a local improvement of size at most $3$ in line~\ref{LineXSmallIf}. Note that we have defined our notion of local improvements with respect to the squared weight function (see Definition~\ref{DefLocalImprAndClawShaped}). The reason why we check for local improvements of size at most $3$ is rather technical and will become clear in our analysis of the case where many vertices from $A$ feature a special neighbor, which is presented in appendix~\ref{AppendixManySpecialNeighbors}. If we find such a local improvement, then we update $A$ accordingly in line~\ref{LineXSmall} and return \texttt{true}. Otherwise, we proceed by checking whether a claw-shaped improvement exists in line~\ref{LineClawShapedIf}. Again, if we are successful, we update $A$ and return \texttt{true}. In case we do not succeed, we continue by searching for a circular improvement in line~\ref{LineCircularIf}. Note that the definition of a circular improvement (Definition~\ref{DefCircular}) depends on the parameter $\kappa$ that controls our bound on the length of the underlying cycle. We will explain how to choose $\kappa$ when stating our main technical theorem (Theorem~\ref{TheoMainTheorem}).  Once more, if we find an improvement, we update $A$ and return \texttt{true}. If we also do not manage to find a circular improvement, we try to obtain a local improvement by applying the black box algorithm \texttt{MIS} for the unweighted setting. To this end, we define $V_{help}$ to consist of all vertices in $V\setminus A$ that are helpful for at least one of their neighbors. Intuitively, this means that their neighborhood bears a structure that nicely resembles the unweighted setting. Next, we loop over all possible choices (among the set of vertex weights) of a lower weight threshold $L$, and define $A_{\geq L}$ to consist of all vertices from $A$ of weight at least $L$. Moreover, we define $V_{\geq L}$ to consist of all vertices from $A_{\geq L}$, together with all of the vertices $u\in V_{help}$ with the property that $w(u)\geq L$ and moreover, all vertices $u$ is helpful for are contained in $A_{\geq L}$. The idea behind this is that for $u\in V_{help}$, the vertices in $\mathrm{help}(u)$ each have a weight that cannot be much larger than the weight of $u$ and, moreover, make up all of the weight of $N(u,A)$ except for maybe some $\epsilon$-fraction of $w(u)$. In particular, the neighborhood of $u$ in $G[V_{\geq L}]$ provides a reasonable estimate of whether or not it might be a good idea to include $u$ in a local improvement. We point out that forcing $L\in\{w(v),v\in V\}$ does not restrict the range of (non-empty) sets $V_{\geq L}$ we may encounter, as opposed to choosing $L\in \mathbb{R}$ arbitrarily. This is because whenever $V_{\geq L}$ is non-empty, raising $L$ to the minimum weight among the vertices in $V_{\geq L}\supseteq A_{\geq L}$ does not change the sets $A_{\geq L}$ and $V_{\geq L}$. 
 
For each threshold $L$, we now apply \texttt{MIS} to the induced sub-graph $G[V_{\geq L}]$, which, as discussed earlier, is the conflict graph of the sub-collection of $\mathcal{S}$ consisting of the sets represented by the vertices in $V_{\geq L}$. We denote the output of $\texttt{MIS}$ by $\bar{X}$ and define $X:=\bar{X}\setminus A$. Note that any independent set $Y\subseteq V$ defines a local improvement w.r.t.\ $A$ if and only if $Y\setminus A$ does. The reason that we remove the intersection with $A$ is mainly because it makes the analysis more convenient. 
 
  Finally, we loop over all possible choices (again among the set of vertex weights) of an upper weight threshold $U$ and define $X^{\leq U}$ to consist of all vertices from $X$ that are of weight at most $U$, that is, we sweep $X$ from low to high weights and consider the initial segments of vertices we encounter.\footnote{Note that it is actually unnecessary to also check thresholds $U<L$ because for these, $X^{\leq U}$ will be empty and, thus, no local improvement. However, to simplify the presentation, we omit the constraint $U\geq L$.} Again, restricting $U$ to values in $\{w(v),v\in V\}$ does not change the collection of non-empty sets $X^{\leq U}$ we may encounter. For each of the sets $X^{\leq U}$, we check whether it constitutes a local improvement. If yes, we update $A$ and return \texttt{true}. Otherwise, if none of the sets $X^{\leq U}$ yields a local improvement, we return \texttt{false}, causing Algorithm~\ref{LocalImprovementAlgo} to terminate.
 \begin{algorithm}[t]
 	\SetArgSty{textrm}
 		\DontPrintSemicolon
\KwIn{an instance $(\mathcal{S},w)$ of the weighted $k$-Set Packing problem}
\KwOut{a disjoint sub-collection of the sets in $\mathcal{S}$\;}
$G:=(V,E)\gets G_{\mathcal{S}}$\label{LineGS}\;
$A\gets\emptyset$ \;
 	$\texttt{continue}\gets$\textbf{true}\;
 	\While{\texttt{continue}}{
 	$\texttt{continue}\gets\texttt{RunIteration}(G, A, w)$\;	}
 	\Return the collection of sets corresponding to A\;
 	\caption{Local improvement algorithm}\label{LocalImprovementAlgo}
 \end{algorithm}
\begin{algorithm}[t]
	\SetArgSty{textrm}
 		\DontPrintSemicolon
\KwIn{the conflict graph $G=(V,E)=G_{\mathcal{S}}$ of a $k$-Set Packing instance $\mathcal{S}$,\; a positive weight function $w:V\rightarrow\mathbb{Q}_{>0}$,\; $A\subseteq V$ independent\;}
\KwOut{Whether a local improvement was found.\;}
\If{$\exists$ local improvement $X$ such that $|X|\leq 3$\label{LineXSmallIf}}{
	$A\gets (A\setminus N(X,A))\cup X$\label{LineXSmall}\;
	\textbf{return} \texttt{true}\;}
\If{$\exists$ claw-shaped improvement $X$\label{LineClawShapedIf}}{
$A\gets (A\setminus N(X,A))\cup X$\label{LineXClaw-Shaped}\;
\textbf{return} \texttt{true}\;}
\If{$\exists$ circular improvement $X$\label{LineCircularIf}}{
	$A\gets (A\setminus N(X,A))\cup X$\label{LineXCircular}\;
	\textbf{return} \texttt{true}\;}
$V_{help}\gets \{u\in V\setminus A: \mathrm{help}(u)\neq\emptyset\}$\label{LineVHelp}\;
\For{$L\in \{w(v),v\in V\}$}
{
	$A_{\geq L}\gets \{v\in A: w(v)\geq L\}$\;
	$V_{\geq L}\gets A_{\geq L}\cup\{u\in V_{help}: w(u)\geq L\wedge \mathrm{help}(u)\subseteq A_{\geq L}\}$\;
	$\bar{X}\gets \mathrm{MIS}(G[V_{\geq L}])$\label{LineBarX}\;
	$X\gets\bar{X}\setminus A$\label{LineXUnweighted}\;
	\For{$U\in \{w(v),v\in V\}$}
	{
		$X^{\leq U}:=\{u\in X: w(u)\leq U\}$\label{LineXleqU}\;
		\If{$w^2(X^{\leq U})>w^2(N(X^{\leq U}, A))$}{
			$A\gets (A\setminus N(X^{\leq U}, A))\cup X^{\leq U}$\;
			\textbf{return} \texttt{true}\;
		}
	}
}
\textbf{return} \texttt{false}\;
\caption{$\texttt{RunIteration}(G, A, w)$}\label{AlgoRunIteration}
\end{algorithm}
\subsection{Correctness of Algorithm~\ref{LocalImprovementAlgo}}
After having outlined the algorithm we propose, we would like to quickly convince ourselves that it is correct, meaning that it terminates and, when doing so, returns a feasible solution to the weighted $k$-Set Packing problem.
\begin{proposition}
	Algorithm~\ref{LocalImprovementAlgo} terminates and returns a disjoint sub-collection of sets.
\end{proposition}
\begin{proof}
	We first observe that Algorithm~\ref{LocalImprovementAlgo} is guaranteed to terminate because no set $A$ can be attained twice, given that $w^2(A)$ strictly increases in each iteration of the while-loop except the last one, and there are only finitely many possibilities.
	
	In order to show that  Algorithm~\ref{LocalImprovementAlgo} returns a disjoint sub-collection of sets, it suffices to show that throughout the course of the algorithm, $A$ is an independent set in the conflict graph $G=G_{\mathcal{S}}$ representing the input instance $(\mathcal{S},w)$.
	This is clear initially because $\emptyset$ is independent and none of our update steps performed in lines~\ref{LineXSmall},~\ref{LineXClaw-Shaped} and~\ref{LineXCircular} can harm this invariant since the respective set $X$ is always independent by definition. Hence, it remains to see that each of the sets $X^{\leq U}$ we consider in line~\ref{LineXleqU} is independent. By definition of \texttt{MIS}, its output $\bar{X}$ (line~\ref{LineBarX}) constitutes an independent set in the induced sub-graph $G[V_{\geq L}]$, and, hence, also in $G$. As a subset of $X\subseteq \bar{X}$, each of the sets $X^{\leq U}$ is independent, too.
\end{proof}
\subsection{Obtaining a polynomial running time}
Next, we show how to obtain a polynomial running time. In doing so, Lemma~\ref{LemRuntimeIt} bounds the running time of each iteration, while Lemma~\ref{LemPolyNumIt} shows how to achieve a polynomial number of iterations by scaling and truncating the weight function as explained in \cite{ChandraHalldorsson} and \cite{Berman}. Note that as in prior works~\cite{Berman},~\cite{FurerYu}, $k$ is considered a constant for the runtime analysis.

\begin{lemma} $\texttt{RunIteration}(G,A,w)$ can be implemented to run in polynomial time.\label{LemRuntimeIt}
\end{lemma}
\begin{proof}First, observe that we can check for a local improvement of size at most $3$ and for a claw-shaped improvement in polynomial time $\mathcal{O}(|V|^{k}\cdot(|V|+|E|))$, assuming $k\geq 3$. (Note that $|V|=|\mathcal{S}|$ and $|E|\leq |V|^2$.) Moreover, \cite{neuwohner2021limits} shows how to check for the existence of a circular improvement in polynomial time assuming the underlying structure of an instance of the weighted $k$-Set Packing problem, which is given in our setting. For every fixed choice of $\tilde{\epsilon}> 0$ (recall~\eqref{EqBoundsRho}), \texttt{MIS} runs in polynomial time and we call it $|V|$-times, which is polynomial. Finally, we consider a polynomial number of pairs of thresholds $L$ and $U$, and can, for each of them, compute $X^{\leq U}$ and check whether it constitutes a local improvement in time $\mathcal{O}(|V|+|E|)$.  \end{proof}
\begin{lemma}
	Let $r>0$ such that Algorithm~\ref{LocalImprovementAlgo} is an $r$-approximation for the weighted $k$-Set Packing problem and let $\delta > 0$. Then we can, given an instance $(\mathcal{S},w)$, compute, in polynomial time, a weight function $w'$ such that running Algorithm~\ref{LocalImprovementAlgo} on $(\mathcal{S},w')$  takes at most $k^2\cdot (\delta^{-1}+2)^2\cdot |\mathcal{S}|^2+1$ iterations, and yields a $(1+\delta)\cdot r$-approximation to the original problem.\label{LemPolyNumIt}
\end{lemma}
\begin{proof}Let $N:=\lceil \delta^{-1}+1\rceil$. We first apply the greedy algorithm to compute a solution $A'$. The greedy algorithm considers the sets in order of decreasing weight and picks every set that does not intersect an already chosen one. It is known to yield an approximation guarantee of $k$ (see for example \cite{ChandraHalldorsson}). We re-scale the weight function $w$ such that $w(A')=N\cdot |\mathcal{S}|$ holds. Note that this operation preserves the property that every disjoint sub-collection of sets is of weight at most $k\cdot w(A')$. Then, we delete sets $v$ of truncated weight $\lfloor w(v) \rfloor = 0$ and run Algorithm~\ref{LocalImprovementAlgo} with the integral weight function $w':=\lfloor w\rfloor$. In doing so, we know that $\lfloor w\rfloor^2(A)$ equals zero initially and must increase by at least one within each call to \texttt{RunIteration}, except for the last one. On the other hand, at each point, we have \[\lfloor w\rfloor ^2(A)\leq w^2(A)\leq (w(A))^2\leq k^2\cdot w^2(A')=k^2\cdot N^2\cdot |\mathcal{S}|^2, \] which bounds the total number of iterations by $k^2\cdot N^2\cdot |\mathcal{S}|^2+1$. Finally, if $A$ denotes the solution Algorithm~\ref{LocalImprovementAlgo} returns and $A^*$ is a disjoint sub-collection of maximum weight with respect to the original respectively the scaled, but un-truncated weight function $w$, we know that \[r\cdot w(A)\geq r\cdot\lfloor w\rfloor (A)\geq\lfloor w\rfloor(A^*)\geq w(A^*)-|A^*|\geq w(A^*)-|\mathcal{S}|=w(A^*) -\frac{w(A')}{N}\geq\frac{N-1}{N}\cdot w(A^*), \]so the approximation ratio increases by a factor of at most $\frac{N}{N-1}=1+\frac{1}{N-1}\leq 1+\delta$.\end{proof}
 \section{Analysis of the Performance Ratio\label{SecAnalysis}}
 In this section, we analyze the approximation guarantee of Algorithm~\ref{LocalImprovementAlgo} and prove our main theorem, Theorem~\ref{TheoSimpleMainTheorem}, which we restate here again.
 \TheoSimpleMainTheorem* 
 In contrast to the previous papers by Neuwohner, where the improvement in the approximation guarantee was less than $6\cdot 10^{-7}$~\cite{NeuwohnerLipics} or only took effect for values of $k\geq 200,000$~\cite{neuwohner2021limits}, we can bring the order of magnitude of the improvement to a more notable scale. In addition, our result implies the first asymptotic improvement over the lower bound of $\frac{k}{2}$ that was established for algorithms searching for local improvements with respect to some fixed power of the weight function and of logarithmically bounded size~\cite{neuwohner2021limits}. Note that the previous state-of-the-art algorithms for weighted $k$-Set Packing~\cite{Berman},~\cite{NeuwohnerLipics} and~\cite{neuwohner2021limits} all fit into the latter scheme.
 
 For the analysis, for each value of $k\geq 4$, we fix the constant $\epsilon$ introduced in the previous section, as well as an additional parameter $\xi$, which bounds the gain from non-helpful vertices. More precisely, $\xi$ will be chosen in such a way that for every $u\in A^*$, we gain $\xi\cdot w(\mathrm{supp}(u))$ in our bound on $w(A^*)$. Our choices of $\epsilon$ and $\xi$ need satisfy a bunch of inequalities that pop up during our analysis and are listed in appendix~\ref{AppendixInequalities}. The precise values we choose for these constants can be read from Table~\ref{TableEpsXi}.  
 \begin{table}[t]
 	\begin{tabular}{|r|r|r|r|}
 		\hline
 		$k$ & $\epsilon$ & $\xi$ & approximation guarantee\\\hline
 		$4$ & $0.01422$ & $0.001764$ & $2.4998$ \\
 		$5$ & $0.02674$ & $0.003298$ & $2.9990$ \\
 		$6$ & $0.03466$ & $0.004258$ & $3.4980$ \\
 		$7$ & $0.04013$ & $0.004917$ & $3.9968$ \\
 		$8$ & $0.04415$ & $0.005399$ & $4.4955$ \\
 		$9$ & $0.04723$ & $0.005767$ & $4.9941$ \\
 		$10$ & $0.04966$ & $0.006057$ & $5.4928$ \\
 		$11$ & $0.05164$ & $0.006292$ & $5.9914$ \\
 		$12$ & $0.05328$ & $0.006487$ & $6.4899$ \\
 		$13$ & $0.05465$ & $0.006649$ & $6.9885$ \\
 		$\geq 14$ & $0.084$ & $0.01$ & $0.4986\cdot (k+1) + 0.0208$\\
 		\hline 
 	\end{tabular}
 	\caption{Choices of $\epsilon$ and $\xi$ and resulting approximation guarantees for different values of $k$.}\label{TableEpsXi}
 \end{table}
Theorem~\ref{TheoSimpleMainTheorem} is a direct consequence of our main technical theorem, Theorem~\ref{TheoMainTheorem}, which states the precise dependence of the approximation guarantee we obtain on the parameters $\epsilon$ and $\xi$. Plugging in the values listed in the second and third column of Table~\ref{TableEpsXi} yields the approximation guarantees displayed in the last column, which are less or equal to \[\min\{0.5\cdot (k+1)-0.0002, 0.4986\cdot (k+1) + 0.0208\}\] as for $k\geq 14$, we have \[0.4986\cdot (k+1)+0.0208\leq 0.5\cdot (k+1)-0.0014\cdot 15+0.0208=0.5\cdot (k+1)-0.0002.\]  We point out that in fact, all of the approximation guarantees given for $k=4,\dots,13$ as well as the coefficient $0.4986$ and the constant term $0.0208$ have been strictly rounded up, meaning that the actual approximation ratios we obtain are strictly smaller. In particular, by choosing the constant $\delta$ from Lemma~\ref{LemPolyNumIt} sufficiently small, we can actually also achieve the guarantees from Table~\ref{TableEpsXi} in polynomial time (and not only up to some arbitrarily small additive error).
\begin{restatable}{theorem}{TheoMainTheorem}
 	For $k\geq 4$, Algorithm~\ref{LocalImprovementAlgo} yields an approximation guarantee of \[\frac{k+1}{2}-\xi\cdot\left(k-1-\frac{1}{k}\cdot \frac{(2+\kappa)\cdot (1+\epsilon)^2}{2+\epsilon}-\frac{k-1}{k}\cdot \frac{2\cdot\rho\cdot(1+\epsilon)^3}{1-\epsilon^2}\right),\] 
  where $\rho:=\frac{k+1+\tilde{\epsilon}}{3}$ is the approximation guarantee of \texttt{MIS} and $\kappa>0$ is the constant from Definition~\ref{DefCircular}. We choose $\kappa=\frac{1}{\lceil\epsilon^{-1}\rceil}$ and $\tilde{\epsilon}=\epsilon$. For $\epsilon:=0.084$ and $\xi:=0.01$, we can bound the approximation guarantee by $0.4986\cdot (k+1)+0.0208.$ \label{TheoMainTheorem}
 \end{restatable}
We remark that as we can choose $\kappa > 0$ and $\tilde{\epsilon}>0$ arbitrarily small, we can get slightly better constants by making them even smaller. However, these further improvements are comparably tiny, so to simplify calculations, we decided to choose them (essentially) equal to $\epsilon$. 
We further point out that the last bound of $0.4986\cdot (k+1)+0.0208$  yields the lowest asymptotic growth of the approximation guarantee achievable by our analysis (up to the previous point and the fact that the constants are rounded, of course). However, especially for small values of $k$, this need not be optimum, and in particular, for $k\leq 13$, $0.4986\cdot (k+1)+0.0208$ is not even smaller than $\frac{k+1}{2}$. For this reason, Table~\ref{TableEpsXi} displays (approximately) optimum choices of $\epsilon$ and $\xi$ and the resulting approximation guarantees for $k\leq 13$.

The remainder of this section is dedicated to the proof of Theorem~\ref{TheoMainTheorem}. We fix an instance $(\mathcal{S},w)$ of the weighted $k$-Set Packing problem and denote its conflict graph by $G=(V,E)$. Vertex weights on $V=\mathcal{S}$ are given by $w$. We further fix an independent set $A^*$ in $G$ of maximum weight and denote the solution found by Algorithm~\ref{LocalImprovementAlgo} by $A$. Observe that by positivity of the weight function, $A$ must be a maximal independent set, as a single vertex without any neighbors in $A$ would certainly yield a claw-shaped improvement. 

For our analysis, we re-use the notation introduced in section~\ref{SecSquareImp}, as well as most of the analysis of the algorithm SquareImp presented there. Observe that when Algorithm~\ref{LocalImprovementAlgo} terminates, no claw improves $A$, so all of the results from section~\ref{SecSquareImp} also apply to our setting. In particular, we can employ Theorem~\ref{TheoApproxFactor} to bound the weight of $A^*$. This motivates Definition~\ref{DefMissing} because Theorem~\ref{TheoApproxFactor} tells us that every missing neighbor a vertex $v\in A$ features improves our upper bound on $w(A^*)$ by $\frac{w(v)}{2}$.
 \begin{definition}[missing neighbor]
 	For $v\in A$ with $|N(v,A^*)|<k$, we say that $v$ has $k-|N(v,A^*)|$ \emph{missing neighbors}.\label{DefMissing}
 \end{definition}
Next, as explained in section~\ref{SecTechnicalDetails}, we introduce the concept of special neighbors, which we need to exclude in order to make sure that $u\in A^*$ can reimburse all of its other neighbors it is not helpful for with an $\Omega(\epsilon)$-fraction of their weight.

To see that excluding certain vertices is really necessary, consider, for example, a vertex $u^*\in A^*$ of weight $1-2\cdot\epsilon$, that has precisely one neighbor $v^*=n(u^*)$ in $A$, which is of weight $1$. Then $u^*$ is not helpful for $v^*$ since their weights differ by more than a factor of $1+\epsilon$. However, we can easily calculate
\begin{align*}&\phantom{=}\sum_{v\in A}\contr{u^*}{v}-2\cdot\chrg{u^*}{n(u^*)}=\contr{u^*}{n(u^*)}-2\cdot\chrg{u^*}{n(u^*)}\\&=\frac{w^2(u^*)}{w(v^*)}-(2\cdot w(u^*)-w(v^*))=(1-2\cdot\epsilon)^2-(2\cdot(1-2\cdot\epsilon)-1)=4\cdot\epsilon^2.\end{align*} This shows that if we wanted to reimburse all vertices for which $u^*$ is not helpful, we could only do so with some $\mathcal{O}(\epsilon^2)$ fraction of their weight, which dooms the order of magnitude of improvements in the approximation guarantee we can hope for to be very small, at least for small values for $k$.
\begin{definition}[special neighbors]
We say that $u\in A^*$ and $v\in A$ are \emph{special neighbors} if we have $v=n(u)$, $v\not\in\mathrm{help}(u)$ and $\contr{u}{v}> \frac{5}{8}\cdot w(v)$.

Let $A'$ consist of all vertices in $A\setminus A^*$ that have at least one special neighbor. Note that by Proposition~\ref{PropUpperBoundContr}, each $v\in A$ can have at most one special neighbor. Denote the unique special neighbor of $v\in A'$ by $t(v)$.
\label{DefSpecial}
\end{definition}
Now, we are ready to define the support of a vertex $u\in A^*$, which contains all of the vertices in $A$ it is supposed to compensate for not being helpful for them.
\begin{definition}[support]
	For $u\in A^*$, we define the support of $u$ to be \[\mathrm{supp}(u):=\{v\in N(u,A)\setminus\mathrm{help}(u): \text{ $u$ is not special for $v$}\}.\] \label{DefSupport}
\end{definition}
We point out that every vertex $v\in A\cap A^*$ is special for itself and in particular, its support is empty (see proof of Corollary~\ref{CorOneNeighbor}).\\
Our analysis now proceeds as follows. First, we show that if our instance is close-to-tight, then the weighted sum of the numbers of helpful neighbors the vertices in $A$ possess needs to be large. To see this, we prove that for every vertex $v\in A$, every missing neighbors of $v$ as well as every vertex $u\in A^*$ that supports $v$ improves our analysis by at least $\xi\cdot w(v)$.

On the other hand, we also show that whenever Algorithm~\ref{LocalImprovementAlgo} terminates, the weighted sum of the numbers of helpful neighbors of the vertices in $A$ can in fact \emph{not} be large because this would imply that Algorithm~\ref{LocalImprovementAlgo}, or more precisely the sub-routine \texttt{RunIteration}, would still be able to find a local improvement, which contradicts the termination criterion of our algorithm.
\subsection{Missing and supportive neighbors improve the analysis}
Lemma~\ref{LemProfitFromBadNeighbors} tells us that for every $u\in A^*$, the slack in the inequality \[\frac{1}{2}\cdot\left(\sum_{v\in N(u,A)}\contr{u}{v}-2\cdot \chrg{u}{n(u)}\right)\geq 0\] (see Theorem~\ref{TheoApproxFactor}) suffices to reimburse each of the neighbors its supports with a $\xi$-fraction of its weight. We will later choose $\xi=\frac{\epsilon}{4\cdot (2+\epsilon)} > \frac{\epsilon}{10}$, cf.\ appendix~\ref{AppendixInequalities}.
\begin{lemma}
	Let $u\in A^*$. Then \[\sum_{v\in N(u,A)}\contr{u}{v}-2\cdot \chrg{u}{n(u)}\geq 2\cdot\xi\cdot w(\mathrm{supp}(u)).\] \label{LemProfitFromBadNeighbors}
	\end{lemma}
 For better readability, we defer the (unfortunately rather tedious) proof to appendix~\ref{AppendixProfitBadNeighbors}.
By combining Lemma~\ref{LemProfitFromBadNeighbors} with Theorem~\ref{TheoApproxFactor}, we obtain Lemma~\ref{LemGainFromLowDegrees}, which makes it explicit how our approximation guarantee improves depending on $\sum_{v\in A} w(v)\cdot |\mathrm{help}(u)\cap A^*|$, the weighted sum of the number of helpful neighbors the vertices in $A$ have in $A^*$. Recall that $A'$ was defined as the set of all vertices in $A\setminus A^*$ that feature a special neighbor. Whereas for a vertex $v\in A\setminus A'$, we gain $\xi\cdot w(v)$ for every missing or non-helpful neighbor of $v$, for $v\in A'$, we gain the respective amount for each such neighbor except for the special one.
\begin{lemma}
\begin{align*}w(A^*)&\leq \frac{k+1}{2}\cdot w(A)-\xi\cdot\sum_{v\in A'}(k-1-|\mathrm{help}(v)\cap A^*|)\cdot w(v)-\xi\cdot\sum_{v\in A\setminus A'}(k-|\mathrm{help}(v)\cap A^*|)\cdot w(v)\\
&=\frac{k+1}{2}\cdot w(A)-\xi\cdot\sum_{v\in A}(k-|\mathrm{help}(v)\cap A^*|)\cdot w(v)+\xi\cdot w(A')\end{align*}\label{LemGainFromLowDegrees}
\end{lemma}
\begin{proof}
By Theorem~\ref{TheoApproxFactor}, we know that 
\begin{align}w(A^*)\leq\frac{k+1}{2}\cdot w(A)&-\frac{1}{2}\cdot \sum_{v\in A} (k-|N(v,A^*)|)\cdot w(v)\notag\\&- \frac{1}{2}\cdot\sum_{u\in A^*}\left(\sum_{v\in A} \contr{u}{v} - 2\cdot \chrg{u}{n(u)}\right).\label{EqLemSquareImp}\end{align}
We calculate
\begin{align*}
&\phantom{=}\frac{1}{2}\cdot\sum_{u\in A^*}\left(\sum_{v\in A} \contr{u}{v} - 2\cdot \chrg{u}{n(u)}\right)
\stackrel{Lem.~\ref{LemProfitFromBadNeighbors}}{\geq} \sum_{u\in A^*} \xi\cdot w(\mathrm{supp}(u))\\&=\sum_{u\in A^*}\sum_{v\in\mathrm{supp}(u)}\xi\cdot w(v)=\sum_{v\in A} |\{u\in A^*:v\in\mathrm{supp}(u)\}|\cdot \xi\cdot w(v).
\end{align*}
Plugging this into \eqref{EqLemSquareImp} yields
\[w(A^*)\leq \frac{k+1}{2}\cdot w(A)-\sum_{v\in A}\left((k-|N(v,A^*)|)\cdot \frac{1}{2}\cdot w(v)+|\{u\in A^*:v\in\mathrm{supp}(u)\}|\cdot \xi\cdot w(v)\right),\]
and to verify the statement of the lemma, it suffices to show that
\begin{enumerate}
	\item $(k-|N(v,A^*)|)\cdot \frac{1}{2}\cdot w(v)+|\{u\in A^*:v\in\mathrm{supp}(u)\}|\cdot \xi\cdot w(v)\geq (k-1-|\mathrm{help}(v)\cap A^*|)\cdot \xi\cdot w(v)$ for all $v\in A'$ and
	\item $(k-|N(v,A^*)|)\cdot \frac{1}{2}\cdot w(v)+|\{u\in A^*:v\in\mathrm{supp}(u)\}|\cdot \xi\cdot w(v)\geq (k-|\mathrm{help}(v)\cap A^*|)\cdot \xi\cdot w(v)$ for all $v\in A\setminus A'$.
\end{enumerate}
For $v\in A\cap A^*$, we have $|N(v,A^*)|=1$ and, hence, obtain
\begin{align*}&\phantom{=}(k-|N(v,A^*)|)\cdot \frac{1}{2}\cdot w(v)+|\{u\in A^*:v\in\mathrm{supp}(u)\}|\cdot \xi\cdot w(v)\geq \frac{k-1}{2}\cdot w(v)\\&\stackrel{(*)}{\geq} \xi\cdot k\cdot w(v)\geq (k-|\mathrm{help}(v)\cap A^*|)\cdot \xi\cdot w(v),\end{align*}
where the inequality marked $(*)$ follows from the facts that $k\geq 3$ and $\xi \leq \frac{1}{4}$  by \eqref{EqXi3}.

 Next, we consider the case that $v\in A\setminus A^*$.
By Definition~\ref{DefHelpful}, Definition~\ref{DefSpecial} and Definition~\ref{DefSupport}, we know that \[N(v,A^*)=(\mathrm{help}(v)\cap A^*)\dot{\cup}\{u\in N(v,A^*): \text{$u$ is special for $v$}\}\dot{\cup}\{u\in N(v,A^*): v\in\mathrm{supp}(u)\}.\]This tells us that
\begin{align*}
&\phantom{=}(k-|N(v,A^*)|)\cdot \frac{1}{2}\cdot w(v)+|\{u\in A^*:v\in\mathrm{supp}(u)\}|\cdot \xi\cdot w(v)\\&\geq (k-|N(v,A^*)|)\cdot \frac{1}{2}\cdot w(v)\\&\phantom{=}\quad+(|N(v,A^*)|-|\{u\in N(v,A^*):\text{$u$ is special for $v$}\}|-|\mathrm{help}(v)\cap A^*|)\cdot \xi\cdot w(v)\\
&\stackrel{\eqref{EqXi3}}{\geq} (k-|\{u\in N(v,A^*):\text{$u$ is special for $v$}\}|-|\mathrm{help}(v)\cap A^*|)\cdot \xi\cdot w(v).
\end{align*}
By definition of $A'$ and Proposition~\ref{PropUpperBoundContr}, we further know that every $v\in A'$ has exactly one special neighbor, whereas no vertex in $A\setminus(A'\cup A^*)$ can have a special neighbor. This concludes the proof.
\end{proof}
\subsection{Many helpful neighbors imply a local improvement}
In this section, we show that $\sum_{v\in A} |\mathrm{help}(v)\cap A^*|\cdot w(v)$ must be small when  Algorithm~\ref{LocalImprovementAlgo} terminates because otherwise, the last call to \texttt{RunIteration} (Algorithm~\ref{AlgoRunIteration}) would have produced a local improvement, contradicting the termination criterion of Algorithm~\ref{LocalImprovementAlgo}. We first observe that in the last call to \texttt{RunIteration} before Algorithm~\ref{LocalImprovementAlgo} terminates, the solution $A$ is not modified anymore. In particular, this means that $\sum_{v\in A} |\mathrm{help}(v)\cap A^*|\cdot w(v)$ does not change during the last iteration.  As a consequence, it suffices to prove that in the beginning of the last iteration, $\sum_{v\in A} |\mathrm{help}(v)\cap A^*|\cdot w(v)$ must have been small. For that purpose, Lemma~\ref{LemImprovementNextIter} tells us that we must have $\sum_{v\in A}|\mathrm{help}(v)\cap A^*|\cdot w(v) \leq \frac{2\cdot\rho\cdot(1+\epsilon)^3}{1-\epsilon^2}\cdot w(A)$ because otherwise, one of the sets $X^{\leq U}$ we considered in line~\ref{LineXleqU} would have been a local improvement. If $w(A')$ is small, then combining this bound with Lemma~\ref{LemGainFromLowDegrees} suffices to prove Theorem~\ref{TheoMainTheorem}. However, if $w(A')$ is large, then we need a better bound to compensate for the large $\xi\cdot w(A')$ term in Lemma~\ref{LemGainFromLowDegrees}. Such a bound is provided by Lemma~\ref{LemImprovementAprimeLarge}, showing that we must have \[\sum_{v\in A'}|\mathrm{help}(v)\cap A^*|\cdot w(v) \leq \frac{(2+\kappa)\cdot (1+\epsilon)^2}{(2+\epsilon)}\cdot w(A),\]  because otherwise, we would again have found a local improvement in the next iteration.
\begin{lemma}
	If at the beginning of a call to \texttt{RunIteration}, we have \[\sum_{v\in A}|\mathrm{help}(v)\cap A^*|\cdot w(v)> \frac{2\cdot\rho\cdot(1+\epsilon)^3}{1-\epsilon^2}\cdot w(A),\] then within this call, we find a local improvement.\label{LemImprovementNextIter}
\end{lemma}
The proof can be found in appendix~\ref{AppendixImprovementNextIter}.
\begin{lemma}If at the beginning of a call to \texttt{RunIteration}, we have  \[\sum_{v\in A'}|\mathrm{help}(v)\cap A^*|\cdot w(v) > \frac{(2+\kappa)\cdot (1+\epsilon)^2}{(2+\epsilon)}\cdot w(A),\] then within this call, we find a local improvement of size at most $3$, a claw-shaped improvement, or a circular improvement. \label{LemImprovementAprimeLarge}
\end{lemma}
The proof can be found in appendix~\ref{AppendixManySpecialNeighbors}.
We have now assembled all ingredients to prove our main technical theorem. For easier readability, we state it once again.
\TheoMainTheorem*
\begin{proof}
	Let \[\gamma:= 1+\frac{(2+\kappa)\cdot (1+\epsilon)^2}{(2+\epsilon)\cdot k}-\frac{2\cdot\rho\cdot(1+\epsilon)^3}{(1-\epsilon^2)\cdot k}.\]
	If $w(A')\leq \gamma\cdot w(A)$, we apply Lemma~\ref{LemImprovementNextIter}, which tells us that we must have
	\[\sum_{v\in A}|\mathrm{help}(v)\cap A^*|\cdot w(v)\leq \frac{2\cdot\rho\cdot(1+\epsilon)^3}{1-\epsilon^2}\cdot w(A).\]
	Plugging this into the bound from Lemma~\ref{LemGainFromLowDegrees} yields
	\begin{align*}
	w(A^*)&\leq \frac{k+1}{2}\cdot w(A)-\xi\cdot k\cdot w(A)+\xi\cdot \sum_{v\in A}|\mathrm{help}(v)\cap A^*|\cdot w(v)+\xi\cdot w(A')\\
	&\leq \frac{k+1}{2}\cdot w(A)-\xi\cdot k\cdot w(A)+\xi\cdot \frac{2\cdot\rho\cdot(1+\epsilon)^3}{1-\epsilon^2}\cdot w(A)\\&\quad+\xi\cdot \left(1+\frac{(2+\kappa)\cdot (1+\epsilon)^2}{(2+\epsilon)\cdot k}-\frac{2\cdot\rho\cdot(1+\epsilon)^3}{(1-\epsilon^2)\cdot k}\right)\cdot w(A)\\
	&=\left(\frac{k+1}{2}-\xi\cdot (k-1)+\xi\cdot\frac{1}{k}\cdot\frac{(2+\kappa)\cdot (1+\epsilon)^2}{2+\epsilon}+\xi\cdot\frac{k-1}{k}\cdot \frac{2\cdot\rho\cdot(1+\epsilon)^3}{1-\epsilon^2}\right)\cdot w(A)\\
	&=\left(\frac{k+1}{2}-\xi\cdot \left((k-1)-\frac{1}{k}\cdot\frac{(2+\kappa)\cdot (1+\epsilon)^2}{2+\epsilon}-\frac{k-1}{k}\cdot \frac{2\cdot\rho\cdot(1+\epsilon)^3}{1-\epsilon^2}\right)\right)\cdot w(A).
	\end{align*}
	If $w(A')>\gamma\cdot w(A)$, we make use of Lemma~\ref{LemImprovementAprimeLarge}, which yields
	\[\sum_{v\in A'}|\mathrm{help}(v)\cap A^*|\cdot w(v) \leq \frac{(2+\kappa)\cdot (1+\epsilon)^2}{(2+\epsilon)}\cdot w(A).\]
	Now, Lemma~\ref{LemGainFromLowDegrees} tells us that
	\begingroup\allowdisplaybreaks
	\begin{align*}
	w(A^*)&\leq \frac{k+1}{2}\cdot w(A)-\xi\cdot\sum_{v\in A'}(k-1-|\mathrm{help}(v)\cap A^*|)\cdot w(v)-\xi\cdot\sum_{v\in A\setminus A'}(k-|\mathrm{help}(v)\cap A^*|)\cdot w(v)\\
	&\leq \frac{k+1}{2}\cdot w(A)-\xi\cdot\sum_{v\in A'}(k-1-|\mathrm{help}(v)\cap A^*|)\cdot w(v)\\
	&=\frac{k+1}{2}\cdot w(A)-\xi\cdot (k-1)\cdot w(A')+\xi\cdot \sum_{v\in A'} |\mathrm{help}(v)\cap A^*|\cdot w(v)\\
	&\leq \frac{k+1}{2}\cdot w(A)-\xi\cdot (k-1)\cdot \left(1+\frac{(2+\kappa)\cdot (1+\epsilon)^2}{(2+\epsilon)\cdot k}-\frac{2\cdot\rho\cdot(1+\epsilon)^3}{(1-\epsilon^2)\cdot k}\right)\cdot w(A)\\&\quad+\xi\cdot \frac{(2+\kappa)\cdot (1+\epsilon)^2}{(2+\epsilon)}\cdot w(A)\\
	&=\bigg(\frac{k+1}{2}-\xi\cdot (k-1)-\xi\cdot\frac{k-1}{k}\cdot\frac{(2+\kappa)\cdot (1+\epsilon)^2}{2+\epsilon}+\xi\cdot\frac{(2+\kappa)\cdot (1+\epsilon)^2}{2+\epsilon}\\&\quad\quad+\xi\cdot\frac{k-1}{k}\cdot  \frac{2\cdot\rho\cdot(1+\epsilon)^3}{1-\epsilon^2}\bigg)\cdot w(A)\\
	&=\left(\frac{k+1}{2}-\xi\cdot \left((k-1)-\frac{1}{k}\cdot\frac{(2+\kappa)\cdot (1+\epsilon)^2}{2+\epsilon}-\frac{k-1}{k}\cdot \frac{2\cdot\rho\cdot(1+\epsilon)^3}{1-\epsilon^2}\right)\right)\cdot w(A).
	\end{align*}\endgroup
	When choosing $\kappa\leq \epsilon$, we obtain
	 \[\frac{(2+\kappa)\cdot (1+\epsilon)^2}{2+\epsilon}\leq (1+\epsilon)^2 <\frac{2\cdot\rho\cdot(1+\epsilon)^3}{1-\epsilon^2},\]
	 which yields
	 \begin{align*}
	 &\frac{k+1}{2}-\xi\cdot \left((k-1)-\frac{1}{k}\cdot\frac{(2+\kappa)\cdot (1+\epsilon)^2}{2+\epsilon}-\frac{k-1}{k}\cdot \frac{2\cdot\rho\cdot(1+\epsilon)^3}{1-\epsilon^2}\right)\\\leq \ & \frac{k+1}{2}-\xi\cdot \left((k-1)-\frac{2\cdot\rho\cdot(1+\epsilon)^3}{1-\epsilon^2}\right).
	 \end{align*}
	 Plugging $\epsilon:=0.084$, $\xi:=0.01$ and $\rho:=\frac{k+1+\epsilon}{3}$ into the right-hand-side yields an upper bound of $0.4986\cdot (k+1) + 0.0208$.
\end{proof}

Note that we can only find parameters $\epsilon$ and $\kappa$ for which \[k-1-\frac{1}{k}\cdot \frac{(2+\kappa)\cdot (1+\epsilon)^2}{2+\epsilon}-\frac{k-1}{k}\cdot \frac{2\cdot\rho\cdot(1+\epsilon)^3}{1-\epsilon^2}\] is positive for $k\geq 4$.
\section{The case \boldmath$k=3$\unboldmath\label{SecD4}}
\begin{figure}[t]
	\centering
	\begin{tikzpicture}[scale = 3,rednode/.style = {circle, draw = red, thick, fill = red, inner sep = 0.5mm, minimum size = 4mm}, bluenode/.style = {circle, draw = blue, thick, fill = blue, inner sep = 0.5mm, minimum size = 4mm}, yellownode/.style = {circle, draw = yellow!50!orange, thick, fill = yellow!50!orange, inner sep = 0.5mm, minimum size = 4mm},greennode/.style = {circle, draw = green!70!black, thick, fill = green!70!black, inner sep = 0.5mm, minimum size = 4mm}]
	\node[rednode] (V1) at ({cos(0)}, {sin(0)}) {};
	\node[bluenode] (V2) at ({cos(22.5)}, {sin(22.5)}) {};
	\node[yellownode]  (V3) at ({cos(45)}, {sin(45)}) {};
	\node[greennode] (B3) at ({1.4*cos(45)}, {1.4*sin(45)}) {};
	\draw[thick] (V3)--(B3);
	\node[bluenode] (V4) at ({cos(67.5)}, {sin(67.5)}) {};
	\node[rednode] (V5) at ({cos(90)}, {sin(90)}) {};
	\node[bluenode] (V6) at ({cos(112.5)}, {sin(112.5)}) {};
	\node[yellownode] (V7) at ({cos(135)}, {sin(135)}) {};
	\node[greennode] (B7) at ({1.4*cos(135)}, {1.4*sin(135)}) {};
	\draw[thick] (V7)--(B7);
	\node[bluenode] (V8) at ({cos(157.5)}, {sin(157.5)}) {};
	\node[rednode] (V9) at ({cos(180)}, {sin(180)}) {};
	\node[bluenode] (V10) at ({cos(202.5)}, {sin(202.5)}) {};
	\node[yellownode] (V11) at ({cos(225)}, {sin(225)}) {};
	\node[greennode] (B11) at ({1.4*cos(225)}, {1.4*sin(225)}) {};
	\draw[thick] (V11)--(B11);
	\node[bluenode] (V12) at ({cos(247.5)}, {sin(247.5)}) {};
	\node[rednode] (V13) at ({cos(270)}, {sin(270)}) {};
	\node[bluenode] (V14) at ({cos(292.5)}, {sin(292.5)}) {};
	\node[yellownode] (V15) at ({cos(315)}, {sin(315)}) {};
	\node[greennode] (B15) at ({1.4*cos(315)}, {1.4*sin(315)}) {};
	\draw[thick] (V15)--(B15);
	\node[bluenode] (V16) at ({cos(337.5)}, {sin(337.5)}) {};
	\draw[thick] (V1)--(V2)--(V3)--(V4)--(V5)--(V6)--(V7)--(V8)--(V9)--(V10)--(V11)--(V12)--(V13)--(V14)--(V15)--(V16)--(V1);
	\draw [bend left = 30, thick] (V1) to node[midway, bluenode] {} (V9);
	\draw [bend right = 30, thick] (V5) to node[midway, bluenode] {} (V13);	
	\end{tikzpicture}
	\caption{The "difficult instance" for the case $k=3$ with parameter $m=2$.}\label{FigInstanceD4}
\end{figure}
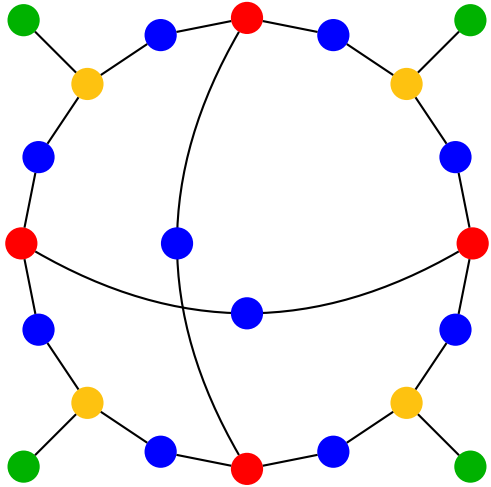

 In order to keep the paper at a reasonable length, we have decided against dealing with the case $k=3$, which requires a slightly different approach, in full details. However, as it seems a little unsatisfying to only present an improved approach for $k\geq 4$, given that the $k$-Set Packing problem is $NP$-hard for $k=3$ already, we would like to shortly point out the reason why a slightly modified proof is needed for $k=3$, and give an idea of how to conduct it.\\
To understand the problem we are facing for $k=3$, recall that for $k\geq 4$, we obtain an improvement in the approximation guarantee by combining Lemma~\ref{LemGainFromLowDegrees} with the bounds on \[\sum_{v\in A} |\mathrm{help}(v)\cap A^*|\cdot w(v)\text { and }\sum_{v\in A'} |\mathrm{help}(v)\cap A^*|\cdot w(v)\] Lemma~\ref{LemImprovementNextIter} and Lemma~\ref{LemImprovementAprimeLarge} provide. We show that for $k=3$, in case $w(A')=0.5\cdot w(A)$, none of these bounds can yield an improvement over the approximation guarantee of $\frac{k+1}{2}=2$ attained by SquareImp (without scaling and truncating the weight function). Indeed, given that we have $\frac{2\cdot\rho\cdot (1+\epsilon)^3}{1-\epsilon^2} > \frac{2\cdot (k+1)}{3}=\frac{8}{3}$, applying Lemma~\ref{LemImprovementNextIter} cannot exclude the case where \[\sum_{v\in A} |\mathrm{help}(v)\cap A^*| = \frac{8}{3}\cdot w(A).\] But plugging this into Lemma~\ref{LemGainFromLowDegrees} only results in a guarantee of \begin{align*}&\phantom{=}2\cdot w(A)-\xi\cdot 3\cdot w(A)+\xi\cdot\sum_{v\in A} |\mathrm{help}(v)\cap A^*|\cdot w(v) +\xi\cdot w(A')\\
&=2\cdot w(A)-\xi\cdot 3\cdot w(A)+\xi\cdot \frac{8}{3}\cdot w(A)+\xi\cdot\frac{1}{2}\cdot w(A)\\
&=2\cdot w(A)+\frac{\xi}{6}\cdot w(A)>2\cdot w(A). \end{align*}
On the other hand, Lemma~\ref{LemImprovementAprimeLarge} only tells us that
\[\sum_{v\in A'}|\mathrm{help}(v)\cap A^*|\leq \frac{(2+\kappa)\cdot (1+\epsilon)^2}{2+\epsilon}\cdot w(A)= 2\cdot\frac{(2+\kappa)\cdot (1+\epsilon)^2}{2+\epsilon}\cdot w(A'),\] and given that $2\cdot\frac{(2+\kappa)\cdot (1+\epsilon)^2}{2+\epsilon} > 2$ and any vertex in $A'$ can have at most $k-1 = 2$ helpful neighbors anyways, this bound is of no use, either. We conclude this section by presenting an example in which the mentioned problems occur. To this end, we first observe that any simple connected graph of maximum degree $3$ on at least $3$ vertices corresponds to the conflict graph of the instance of the weighted $3$-Set Packing problem we obtain by associating each vertex with its set of incident edges. Now, consider a graph that is constructed as follows:
Let $m\in\mathbb{N}_{>0}$, take a cycle of length $8m$, and color its vertices with the colors red, blue, yellow and blue in an alternating manner. Now, as the half-perimeter  of the cycle is divisible by $4$, each red vertex is opposed by another red vertex. For each pair of opposite red vertices, add another blue vertex and connect it to both red ones. All of the previous vertices receive a weight of $1$. Moreover, for each yellow vertex, add a green vertex of weight $1-2\cdot \epsilon$ and connect it to the yellow vertex (see Figure~\ref{FigInstanceD4}).
 Let $A^*$ consist of all blue and green vertices, and let $A$ contain the red and the yellow vertices. $A^*$ is an independent set of maximum weight, as it contains every second vertex from the cycle, and all other vertices. Moreover, $A$ is a maximal independent set, which might have been picked by Algorithm~\ref{LocalImprovementAlgo} vertex by vertex. We can further show that, after having done so, Algorithm~\ref{LocalImprovementAlgo} will terminate: 
 
  We first observe that there is no local improvement of size at most $3$: As $A$ is a maximal independent set, there is no improvement of size $1$, and as there is no vertex in $A$ with two neighbors from $A^*$ that are not adjacent to any other vertex in $A$, there is no improvement of size $2$. Finally, as there are neither two vertices in $A$ that share more than one common neighbor, nor two yellow vertices with a common neighbor, there is no local improvement of size $3$. 
  The fact that no local improvement of size at most $3$ exists tells us in particular that no claw-shaped improvement exists. 
  
  To see that there is no circular improvement, we observe that in the underlying cycle, every second vertex must be yellow as only yellow vertices feature neighbors with a positive contribution to them. But this means that the only candidate for this cycle is the cycle on all vertices from $A$ that we obtain by shortcutting the blue vertices that have a red and a yellow neighbor. For $m$ large enough, the size of this cycle is not logarithmically bounded in the number of vertices. Finally, for all $L$ for which the vertex set $V_{\geq L}$ is non-empty, it consists of all yellow, red and blue, but none of the green vertices (since they are not contained in $V_{help}$), so $\frac{|A^*\cap V_{\geq L}|}{|A\cap A^*|}=\frac{4m + m}{4m}=\frac{5}{4}<\frac{4}{3}$, which does not suffice to guarantee that we obtain an improvement by applying \texttt{MIS}, which could simply output $A$ again.
  
Now, we show that applying our analysis to this instance cannot produce an approximation ratio better than $\frac{k+1}{2}$. According to Definition~\ref{DefSpecial}, $A'$ consists of all yellow vertices, and for each yellow vertex $v$, its special neighbor is the green vertex $t(v)$ that is only connected to $v$. $A\setminus A'$ is the set of red vertices. In this example, every vertex from $A'$ has exactly $2=k-1$ helpful neighbors, whereas every vertex from $A\setminus A'$ features $k=3$ helpful neighbors. Hence, Lemma~\ref{LemGainFromLowDegrees} cannot be used to prove any approximation guarantee better than $\frac{k+1}{2}=2$.
  
Once way to overcome this issue is to simply re-define the support of a vertex $u\in A^*$ to contain all of the vertices it is not helpful for. This yields (for $u\in A^*\setminus A$ and a potentially much smaller choice of $\epsilon$), $\sum_{v\in A}\contr{u}{v}-2\cdot \chrg{u}{n(u)}\geq \Omega(\epsilon^2)\cdot w(\mathrm{supp}(u))$ (see~\cite{neuwohner2021limits}). Given that in case every vertex in $A$ had $k=3$ helpful neighbors, we would have $\frac{|A^*|}{|A|}\geq \frac{3}{2}>\frac{4}{3}=\frac{k+1}{3}$, we now have enough slack to obtain an improved approximation guarantee, provided there is neither a claw-shaped improvement, nor a local improvement resulting from an application of \texttt{MIS} as in Algorithm~\ref{AlgoRunIteration}. However, these considerations only result in a comparably small improvement, and given that Neuwohner~\cite{NeuwohnerLipics} has already shown how to obtain an approximation guarantee of $\frac{k+1}{2}- \frac{1}{63,700,993}$, we did not consider it worthwhile to carry out the exact calculation.

Another aspect that we would, however, like to point out is the fact that in the given example, even though we could not prove it with our approach, the ratio $\frac{w(A^*)}{w(A)}=\frac{4m+2m\cdot (1-2\cdot \epsilon)+m}{4m}<\frac{7}{4}<2$ is actually significantly smaller than the guarantee of $2$ implied by the analysis of SquareImp. The reason for this seems to be that the example is almost unweighted, and does not feature a local improvement of size $3$, which can be shown to exist in Berman's tight example from section~\ref{SecTightExample}. Thus, even though it cannot result in an approximation guarantee less than $\frac{3}{2}$ as shown in \cite{neuwohner2021limits}, it might be interesting to investigate how close to the lower bound of $\frac{k}{2}$ one can get (for $k=3$ or other small values of $k$ where the improvement over $\frac{k+1}{2}$ is still rather small) by considering local improvements of constant size (presumably with respect to the squared weight function).
  
\section{Conclusion\label{SecConclusion}}
In this paper, we have seen how to asymptotically beat the lower bound of $\frac{k}{2}$, which is the best any local improvement algorithm for the weighted $k$-Set Packing problem that searches for local improvements of logarithmically bounded size with respect to a fixed power of the weight function can achieve \cite{neuwohner2021limits}. We believe that this is a quite interesting result, given that the previous state-of-the-art works on the weighted $k$-Set Packing have been relying on such an approach, dooming them to produce no guarantees better than $\frac{k}{2}$. Moreover, in contrast to previous works that could either only guarantee an improvement over Berman's long-standing result of $\frac{k+1}{2}+\epsilon$ in a very small order of magnitude ($<6\cdot 10^{-7}$, \cite{NeuwohnerLipics}) or for extremely large values of $k$ ($k\geq 200,000$, \cite{neuwohner2021limits}), we can guarantee an improvement of at least $0.0002$ for every $k\geq 4$.

  Finally, we are the first ones to tie the approximation guarantees achievable for the weighted $k$-Set Packing problem to those for the unweighted setting in such a way that any improvement for the unweighted case (almost) immediately gives rise to an improved approximation guarantee for general weights. We hope that this result will also refuel interest in the unweighted $k$-Set Packing, perhaps ultimately resulting in improvements on that front.

\appendix
\section{Proofs omitted from the Analysis of SquareImp\label{AppendixAnalysisSquareImp}}
\begin{proof}[Proof of Proposition~\ref{PropNeighborhoodsdminus1}]
	If $v\in A^*$, then $N(v,A^*)=\{v\}$ since $A^*$ is independent. Hence, $|N(v,A^*)|=1\leq k$. If $v\in V(G)\setminus A^*$, then $N(v,A^*)$ constitutes the set of talons of a claw centered at $v$, provided it is non-empty, which again yields $|N(v,A^*)|\leq k$.
\end{proof}
 \begin{proof}[Proof of Prosition~\ref{PropUpperBoundContr}]
 	If $v\in A^*$, the statement is true because $N(v,A^*)=N(v,A)=\{v\}$ and $\mathrm{contr}(v,v)=w(v)$ in this case.\\
 	If $v\not\in A^*$, let $T$ denote the set of vertices sending strictly positive contributions to $v$. Then $T$ constitutes the set of talons of a claw centered at $v$ and we have $\sum_{u\in A^*}\contr{u}{v}=\sum_{u\in T}\mathrm{contr}(u,v)$ by non-negativity of the contribution. Consequently, $\sum_{u\in A^*}\contr{u}{v}>w(v)$ would imply that
 	\[w^2(T) = \sum_{u\in T} w^2(u) > w^2(v)+\sum_{u\in T} w^2(N(u,A)\setminus\{v\})\geq w^2(N(T,A)).\] But this would mean that $T$ constitutes a claw-shaped improvement, a contradiction.
 \end{proof}
\begin{proof}[Proof of Lemma~\ref{LemPropPositiveCharges}]
	If $v\neq n(u)$, the claim follows by non-negativity of the contribution. For $v=n(u)\in N(u,A)$, we obtain \begin{align*} w^2(N(u,A)\setminus\{v\})&=\sum_{x\in N(u,A)\setminus\{v\}}w^2(x)\notag\\&\leq \sum_{x\in N(u,A)\setminus\{v\}} w(x)\cdot \max_{y\in N(u,A)} w(y)\\&=w(N(u,A)\setminus\{v\})\cdot w(v)\\&=(w(N(u,A))-w(v))\cdot w(v).\end{align*} From this, we get
	\begin{align*} 2\cdot\chrg{u}{v}\cdot w(v) &= (2\cdot w(u)-w(N(u,A)))\cdot w(v)\\ &= 2\cdot w(u)\cdot w(v)-w(N(u,A))\cdot w(v)\\ &\leq w^2(u)+w^2(v)-w(N(u,A))\cdot w(v)\label{EqRelevantTight2}\\ &=w^2(u)-(w(N(u,A))-w(v))\cdot w(v)\\&\leq w^2(u)-w^2(N(u,A)\setminus\{v\})\\
	&\leq \contr{u}{v}\cdot w(v)\end{align*} and division by $w(v)> 0$ yields the claim.
\end{proof}
\begin{proof}[Proof of Theorem~\ref{TheoApproxFactor}]
	We have
	\begingroup\allowdisplaybreaks
	\begin{align*}
	w(A^*)&=\sum_{u\in A^*} \frac{w(N(u,A))}{2}+w(u)-\frac{w(N(u,A))}{2}\\
	&=\sum_{u\in A^*}\sum_{v\in N(u,A)} \frac{w(v)}{2}+\sum_{u\in A^*}\chrg{u}{n(u)}\\
	&=\frac{1}{2}\cdot\sum_{v\in A}|N(u,A^*)|\cdot w(v)\\&\quad+\frac{1}{2}\cdot \sum_{u\in A^*}\left(\sum_{v\in A} \contr{u}{v}-\left(\sum_{v\in A} \contr{u}{v}-2\cdot\chrg{u}{n(u)}\right)\right)\\
	&= \frac{k}{2}\cdot w(A)-\frac{1}{2}\cdot\sum_{v\in A}(k-|N(u,A^*)|)\cdot w(v)\\
	&\quad +\frac{1}{2}\cdot \sum_{v\in A}\sum_{u\in A^*} \contr{u}{v} -\frac{1}{2}\cdot \sum_{u\in A^*}\left(\sum_{v\in A} \contr{u}{v}-2\cdot\chrg{u}{n(u)}\right)\\
	&\stackrel{Prop.~\ref{PropUpperBoundContr}}{\leq} \frac{k+1}{2}\cdot w(A)-\frac{1}{2}\cdot\sum_{v\in A}(k-|N(v,A^*)|)\cdot w(v)\\
	&\quad  -\frac{1}{2}\cdot \sum_{u\in A^*}\left(\sum_{v\in A} \contr{u}{v}-2\cdot\chrg{u}{n(u)}\right)\\
	&\stackrel{\contr{-}{-}\geq 0}{\leq} \frac{k+1}{2}\cdot w(A)-\frac{1}{2}\cdot\sum_{v\in A}\underbrace{(k-|N(v,A^*)|)}_{\geq 0 \text{ by Prop.~\ref{PropNeighborhoodsdminus1}}}\cdot w(v)\\
	&\qquad  -\frac{1}{2}\cdot \sum_{u\in A^*}\left( \underbrace{\contr{u}{n(u)}-2\cdot\chrg{u}{n(u)}}_{\geq 0 \text{ by Lem.~\ref{LemPropPositiveCharges}}}\right)\\
	&\leq \frac{k+1}{2}\cdot w(A).
	\end{align*}\endgroup
\end{proof}
%\begin{proof}[Proof of Lemma~\ref{LemBoundCharges}] Assume for a contradiction that \[\sum_{u\in A^*: \chrg{u}{v}>0} \chrg{u}{v}>\frac{w(v)}{2} \] for some $v\in A$. Then $v\not\in A^*$ since \[\{u\in A^*:\chrg{u}{v}>0\}=\{v\}=N(v,A)=N(v,A^*) \] and \[\sum_{u\in A^*: \chrg{u}{v}>0}\chrg{u}{v}=\chrg{v}{v}=\frac{w(v)}{2} \] otherwise. Hence, $T:=\{u\in A^*: \chrg{u}{v}>0\}$ forms the set of talons of a claw centered at $v$. By Lemma~\ref{LemPropPositiveCharges}, it satisfies  \[w^2(T)=\sum_{u\in T} w^2(u)>\sum_{u\in T} w^2(N(u,A)\setminus \{v\}) +w^2(v)\geq w^2(N(T,A)), \] contradicting the fact that no claw improves $w^2(A)$.\end{proof}
\section{Inequalities satisfied by our choice of constants\label{AppendixInequalities}}
\begin{equation}
\frac{3}{8}\cdot(1+\epsilon)^2+\epsilon^2\leq  \frac{11}{16}\cdot (1+\epsilon)^2 +\epsilon^2\leq\frac{3}{4}\cdot(1+\epsilon)^2+\epsilon^2 \leq 1\label{EqEps1}
\end{equation}
\begin{equation}
\frac{1-\epsilon}{2}\geq\frac{1-\epsilon\cdot(1+\epsilon)}{2}\geq \frac{1-2\cdot\epsilon\cdot (1+\epsilon)}{2}\geq\xi\label{EqXi1}
\end{equation}
\begin{equation}
\dfrac{\left(1-\sqrt{\frac{5}{8\cdot(1-\epsilon^2)}}\right)^2}{1+\epsilon\cdot \sqrt{\frac{5}{8\cdot(1-\epsilon^2)}}}\geq 2\cdot\xi\label{EqXi2}
\end{equation}
\begin{equation}
\min\left\{\frac{\epsilon}{2+\epsilon},\frac{1}{4}, \frac{1}{4\cdot (1+\epsilon)},\frac{\epsilon}{2\cdot (1+\epsilon)},\frac{\epsilon}{4+2\epsilon}\right\}\geq \frac{\epsilon}{4\cdot (2+\epsilon)}\geq \xi\label{EqXi3}
\end{equation}
\begin{equation}
\frac{\epsilon}{2\cdot(1+\epsilon)\cdot (2+\epsilon)}\geq\frac{\epsilon}{2\cdot (1+\epsilon)^2\cdot (2+\epsilon)}\geq \frac{\epsilon}{2\cdot (1+\epsilon)^3\cdot (2+\epsilon)}\geq \xi \label{EqXi4}
\end{equation}
\begin{equation}
\dfrac{\left(\frac{1}{3}-\epsilon\right)^2}{\left(\frac{4}{3}+2\epsilon\right)\cdot (2+\epsilon)}\geq \xi\label{EqXi5}
\end{equation}
\begin{equation}
\epsilon\cdot\left(\frac{2+(1+\epsilon)^{-1}}{2+\epsilon}-(1+\epsilon)\right)\geq 2\cdot \xi \label{EqXi6}
\end{equation}
\begin{figure}
	\begin{subfigure}[b]{0.49\textwidth}
		\centering
		\includegraphics[height=0.3\textheight]{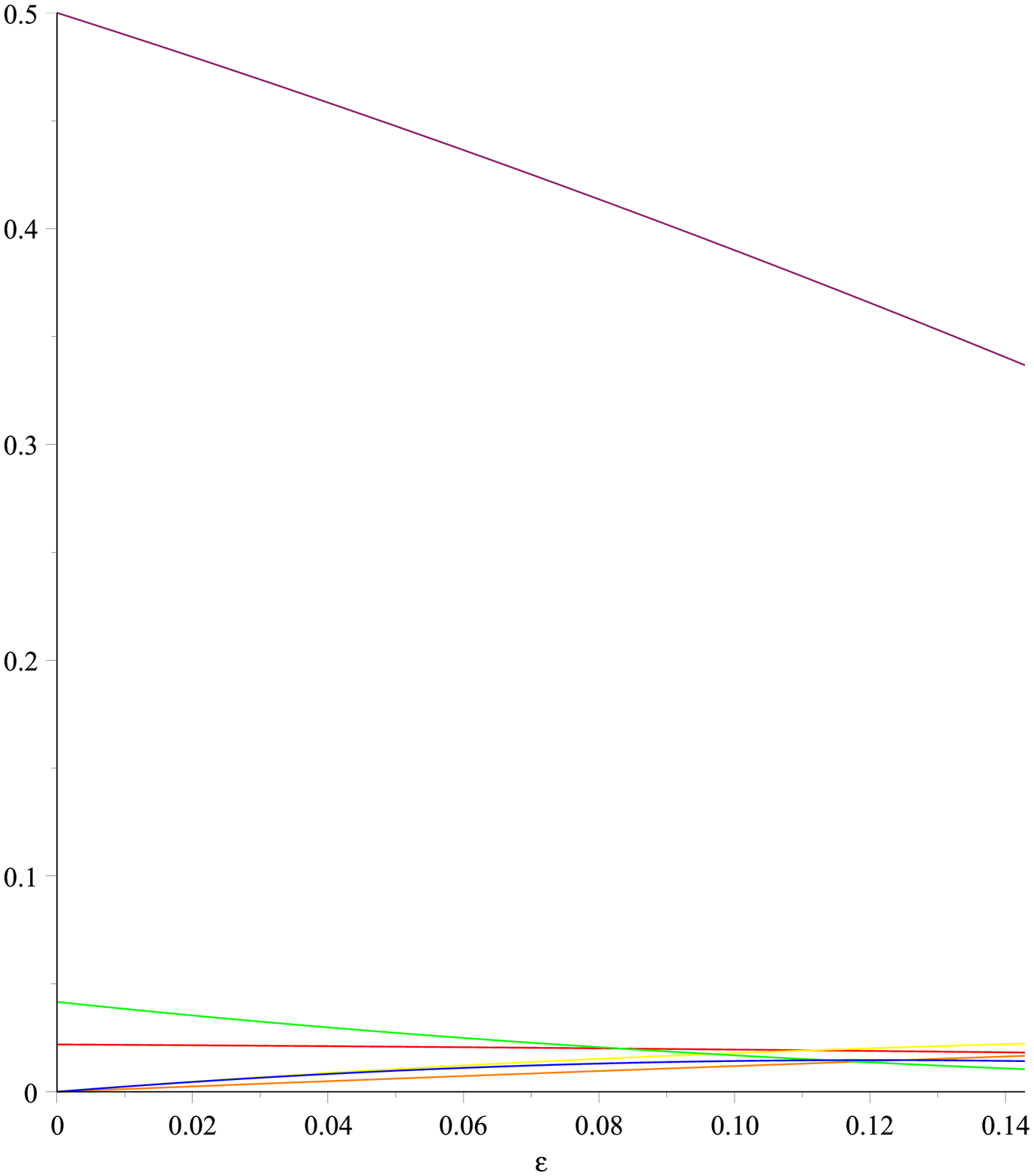}
		\caption{The functions given by \eqref{EqXi1}-\eqref{EqXi6} (colored in violet, red, orange, yellow, green and blue). }
		%TODO: function 1/4 is missing, same colors for both plots, gnuplot????
	\end{subfigure}
	\begin{subfigure}[b]{0.49\textwidth}
		\centering
		\includegraphics[height = 0.3\textheight]{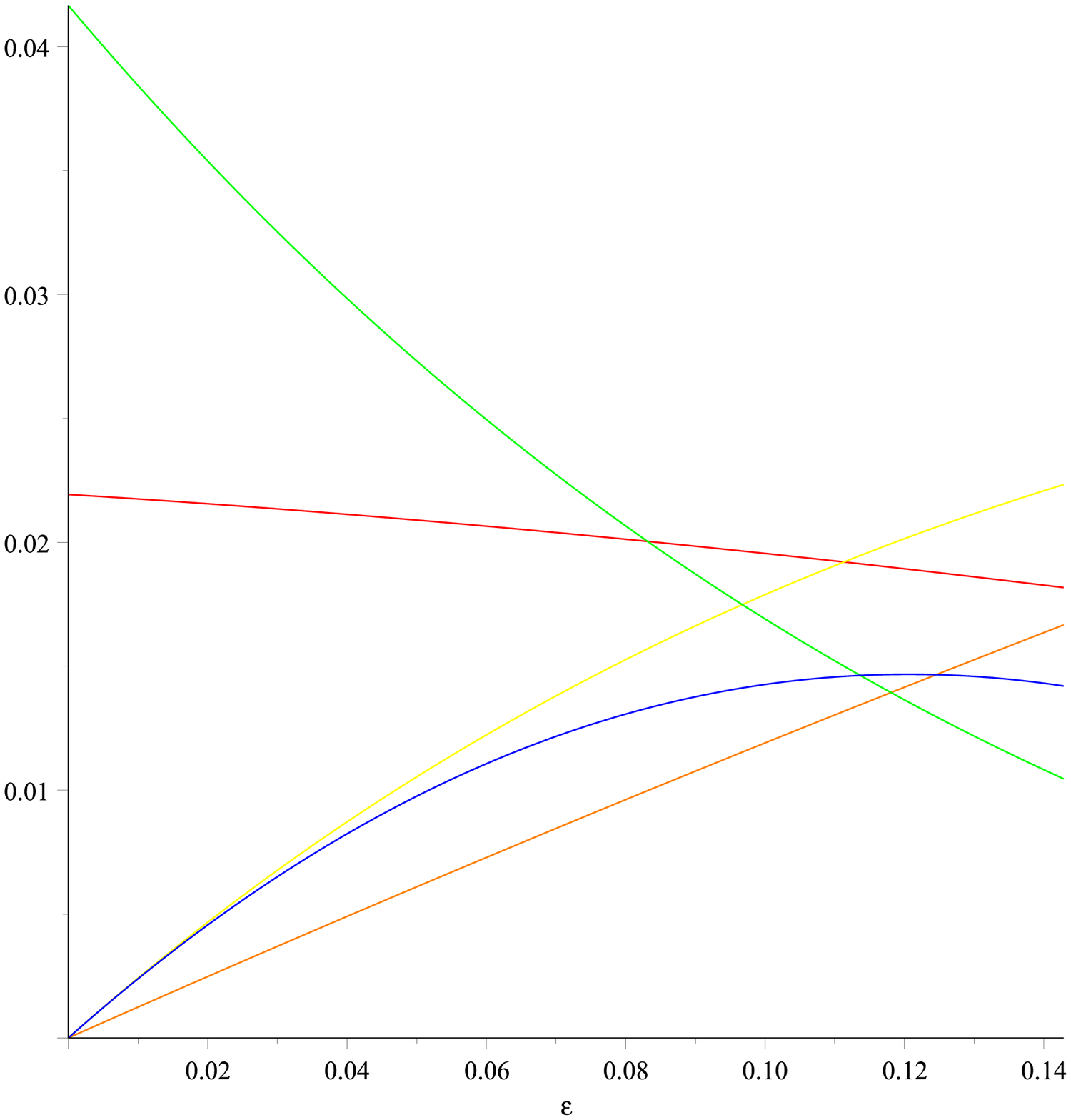}
		\caption{The functions given by \eqref{EqXi2}-\eqref{EqXi6} (colored in red, orange, yellow, green and blue). }
	\end{subfigure}
	\caption{The bounds on $\xi$ implied by \eqref{EqXi1}-\eqref{EqXi6}.} \label{FigPlotsXi}
\end{figure} 
To verify the correctness of the approximation guarantees claimed in Theorem~\ref{TheoMainTheorem} and Table~\ref{TableEpsXi}, it suffices to plug the values of $\epsilon$ and $\xi$ into the bounds provided in Theorem~\ref{TheoMainTheorem} (and its proof) and to check that each of the inequalities \eqref{EqEps1}-\eqref{EqXi6} holds. Instead of explicitly carrying our these calculations, which we do not believe to be particularly enlightening, we briefly explain how we came up with the values displayed in Table~\ref{TableEpsXi} so that the dedicated reader may reproduce them. First, we observe that \eqref{EqEps1} is, for $\epsilon \geq 0$, equivalent to $\epsilon\leq \frac{1}{7}$. Next, for a given value of $\epsilon\in \left(0,\frac{1}{7}\right]$, we compute the maximum value of $\xi$ subject to \eqref{EqXi1}-\eqref{EqXi6}. For this purpose, we regard the (smallest) left-hand-side of each of these constraints as a function in $\epsilon$ (where for \eqref{EqXi2} and \eqref{EqXi6}, we divide by $2$ first). It turns out that the maximum possible value of $\xi$ is attained for $\epsilon =1-\frac{\sqrt{7}}{3}\approx 0.1181$ at the intersection of the two functions given by \eqref{EqXi3} and \eqref{EqXi5} (see Figure~\ref{FigPlotsXi}). In particular, as the approximation guarantee stated in Theorem~\ref{TheoMainTheorem} is monotonically decreasing as a function of $\xi$ and monotonically increasing as a function of $\epsilon$, we can restrict ourselves to values of $\epsilon\in \left(0,1-\frac{\sqrt{7}}{3}\right]$. For $\epsilon\in\left(0,1-\frac{\sqrt{7}}{3}\right]$, \eqref{EqXi3} imposes the most restrictive constraint on $\xi$ (see Figure~\ref{FigPlotsXi}). Now, we can plug this, together with $\kappa:=\epsilon$ (actually, we want to choose $\kappa:=\frac{1}{\lceil \epsilon^{-1}\rceil}\leq \epsilon$, but this only gives a slightly better approximation guarantee) and $\rho:=\frac{k+1+\epsilon}{3}$ into the approximation guarantee from Theorem~\ref{TheoMainTheorem}, which then becomes a function in $k$ and $\epsilon$. For fixed $k$, we can, hence, (approximately) compute the optimum value of $\epsilon$, e.g. by using a computer algebra system. For general $k$, we employ the upper bound  \begin{align*}
&\frac{k+1}{2}-\xi\cdot \left((k-1)-\frac{1}{k}\cdot\frac{(2+\kappa)\cdot (1+\epsilon)^2}{2+\epsilon}-\frac{k-1}{k}\cdot \frac{2\cdot\rho\cdot(1+\epsilon)^3}{1-\epsilon^2}\right)\\\leq \ & \frac{k+1}{2}-\xi\cdot \left((k-1)-\frac{2\cdot\rho\cdot(1+\epsilon)^3}{1-\epsilon^2}\right),
\end{align*} which is asymptotically equal to the left-hand-side for $k\rightarrow\infty$. For a fixed choice of $\epsilon$, $\xi=\frac{\epsilon}{4\cdot (2+\epsilon)}$ and $\rho=\frac{k+1+\epsilon}{3}$, the right-hand-side becomes a linear function in $k$. To obtain the best asymptotic behavior, we must, hence, choose $\epsilon$ such that the coefficient of $k$ is (approximately) minimized. This results in values of $\epsilon = 0.084$ and $\xi= 0.01$, yielding an approximation guarantee of at most $0.4986\cdot (k+1) + 0.0208$.\\ We finally remark that since some of the estimates we employ in our analysis are rather crude, the approximation guarantees we obtain are most likely not the best ones achievable with our approach. However, they give an impression of the order of magnitudes of the improvements we can expect, while trying to keep the rather tedious and lengthy calculations to a minimum.
\section{Proof of Lemma~\ref{LemProfitFromBadNeighbors}\label{AppendixProfitBadNeighbors}}
In order to verify Lemma~\ref{LemProfitFromBadNeighbors}, we will show that for each $u\in A^*$, we have
\[\left(\sum_{v\in A}\contr{u}{v}-2\cdot\chrg{u}{n(u)}\right)\cdot w(n(u))\geq 2\cdot\xi\cdot w(n(u))\cdot w(\mathrm{supp}(u)).\]
As all weights are strictly positive, division by $w(n(u))$ then yields the desired statement.
We first show two auxiliary results. Lemma~\ref{LemProfitSmallRemainingLargeWeight} deals with the case where the weight of $n(u)$ is too large for $u$ to be helpful for any of its neighbors.
Lemma~\ref{LemBounds} assembles several bounds on $\left(\sum_{v\in N(u,A)}\contr{u}{v}-2\cdot\chrg{u}{n(u)}\right)\cdot w(n(u))$ that we will use in the main part of the proof of Lemma~\ref{LemProfitFromBadNeighbors}.
\begin{lemma}
	Let $u\in A^*$ such that \begin{itemize}
		\item $w(N(u,A)\setminus\{n(u)\})\leq \epsilon\cdot w(u)$, and
		\item  $(1+\epsilon)\cdot w(u) < w(n(u))$. \end{itemize}
	Then $\left(\sum_{v\in A}\contr{u}{v}-2\cdot\chrg{u}{n(u)}\right)\cdot w(n(u))\geq 2\cdot\xi\cdot w(n(u))\cdot w(\mathrm{supp}(u))$.\label{LemProfitSmallRemainingLargeWeight}
\end{lemma}
\begin{proof}
	If $\mathrm{supp}(u)=\emptyset$, we are done by Lemma~\ref{LemPropPositiveCharges} and non-negativity of the contribution, so assume that this is not the case.
	We calculate\begin{align}
	&\phantom{=}\sum_{v\in N(u,A)}\contr{u}{v}\cdot w(n(u))-2\cdot\chrg{u}{n(u)}\cdot w(n(u))\notag\\
	&\geq\contr{u}{n(u)}\cdot w(n(u))-(2\cdot w(u)-w(N(u,A)))\cdot w(n(u))\notag\\
	&\geq w^2(u)-w^2(N(u,A)\setminus\{n(u)\})-(2\cdot w(u)-w(N(u,A)))\cdot w(n(u))\notag\\
	&\geq w^2(u)-w(N(u,A)\setminus\{n(u)\})\cdot w(N(u,A)\setminus\{n(u)\})\notag\\&\phantom{=}-(2\cdot w(u)-w(N(u,A)\setminus\{n(u)\})-w(n(u)))\cdot w(n(u))\notag\\
	&=(w(u)-w(n(u)))^2+(w(n(u))-w(N(u,A)\setminus\{n(u)\}))\cdot w(N(u,A)\setminus\{n(u)\}).\label{EqBoundLemSmallRemaining}\end{align}
	Now, we distinguish two cases:\\
	\textbf{Case 1:} $w(u) > \sqrt{\frac{5}{8\cdot(1-\epsilon^2)}}\cdot w(n(u))$. As $w(N(u,A)\setminus\{n(u)\})\leq \epsilon\cdot w(u)$, we have \[\contr{u}{n(u)}\cdot w(n(u))\geq w^2(u)-w^2(N(u,A)\setminus\{n(u)\})\geq (1-\epsilon^2)\cdot w^2(u) > \frac{5}{8}\cdot w^2(n(u)),\] so $\contr{u}{n(u)}>\frac{5}{8}\cdot w(n(u))$. By Definition~\ref{DefHelpful}, Definition~\ref{DefSpecial} and Definition~\ref{DefSupport}, this tells us that $u$ is a special neighbor of $n(u)$ and,hence, $n(u)\not\in\mathrm{supp}(u)$.
	We get \begin{align*}
	&\phantom{=}\sum_{v\in N(u,A)}\contr{u}{v}\cdot w(n(u))-2\cdot\chrg{u}{n(u)}\cdot w(n(u))\\
	&\stackrel{\eqref{EqBoundLemSmallRemaining}}{\geq} (w(n(u))-w(N(u,A)\setminus\{n(u)\}))\cdot w(N(u,A)\setminus\{n(u)\})\\
	&\geq (w(n(u))-\epsilon\cdot w(u))\cdot w(N(u,A)\setminus\{n(u)\})\\
	&\geq (w(n(u))-\epsilon\cdot w(n(u)))\cdot w(N(u,A)\setminus\{n(u)\})\text{ $|$ $n(u)\not\in\mathrm{supp}(u)$}\\
	&\geq (1-\epsilon)\cdot w(n(u))\cdot w(\mathrm{supp}(u))\stackrel{\eqref{EqXi1}}{\geq}2\cdot\xi\cdot w(n(u))\cdot w(\mathrm{supp}(u)).
	\end{align*}
	\textbf{Case 2:} $w(u) \leq \sqrt{\frac{5}{8\cdot(1-\epsilon^2)}}\cdot w(n(u))$. Then
	\begin{align}w(\mathrm{supp}(u))&\leq w(N(u,A))\leq w(n(u))+ w(N(u,A)\setminus\{n(u)\})\notag\\&\leq w(n(u))+\epsilon\cdot w(u)\leq \left(1+\epsilon\cdot \sqrt{\frac{5}{8\cdot(1-\epsilon^2)}}\right)\cdot w(n(u)).\label{EqBoundSupp}\end{align}
	Moreover, by the assumptions of our lemma, we have \begin{align*}&\quad(w(n(u))-w(N(u,A)\setminus\{n(u)\}))\cdot w(N(u,A)\setminus\{n(u)\})\\&\geq ((1+\epsilon)\cdot w(u)-\epsilon\cdot w(u))\cdot w(N(u,A)\setminus\{n(u)\})\geq 0.\end{align*} Combining this observation with \eqref{EqBoundLemSmallRemaining} leads to
	\begin{align*}
	&\phantom{=}\sum_{v\in N(u,A)}\contr{u}{v}\cdot w(n(u))-2\cdot\chrg{u}{n(u)}\cdot w(n(u))\\
	&\geq(w(u)-w(n(u)))^2\\
	&\geq \left(1-\sqrt{\frac{5}{8\cdot(1-\epsilon^2)}}\right)^2\cdot w^2(n(u))\\
	&\stackrel{\eqref{EqBoundSupp}}{\geq} \frac{\left(1-\sqrt{\frac{5}{8\cdot(1-\epsilon^2)}}\right)^2}{1+\epsilon\cdot \sqrt{\frac{5}{8\cdot(1-\epsilon^2)}}}\cdot w(\mathrm{supp}(u))\cdot w(n(u))\\
		&\stackrel{\eqref{EqXi2}}{\geq} 2\cdot \xi\cdot w(n(u))\cdot w(\mathrm{supp}(u)).
	\end{align*}
\end{proof}
\begin{corollary}
	Let $u\in A^*$ such that $|N(u,A)|=1$. Then \[\left(\sum_{v\in A}\contr{u}{v}-2\cdot\chrg{u}{n(u)}\right)\cdot w(n(u))\geq 2\cdot\xi\cdot w(n(u))\cdot w(\mathrm{supp}(u)).\]\label{CorOneNeighbor}
\end{corollary}
\begin{proof}
	As $|N(u,A)|=1$, we have $N(u,A)=\{n(u)\}$. If $\mathrm{supp}(u)=\emptyset$, we are done by Lemma~\ref{LemPropPositiveCharges} and non-negativity of the contribution. In particular, this includes the case where $u\in A^*\cap A$ because in this case, $u\in A$ and $n(u)=u\not\in\mathrm{help}(u)\subseteq V\setminus A$. Hence, as further $N(u,A)=\{u\}=\{n(u)\}$, we have $\contr{u}{n(u)}=\contr{u}{u}=w(u)$ and $u$ is special for itself, implying that $\mathrm{supp}(u)\subseteq N(u,A)\setminus\{u\}=\emptyset$. Otherwise, if $\mathrm{supp}(u)\neq\emptyset$ and, in particular, $u\in A^*\setminus A$, we obtain $w(N(u,A)\setminus\{n(u)\})= w(\emptyset)=0\leq \epsilon\cdot w(u)$ and  $n(u)\in\mathrm{supp}(u)\subseteq N(u,A)\setminus\mathrm{help}(u)$. This implies $(1+\epsilon)\cdot w(n(u))< w(u)$ since all other conditions for $u$ being helpful for $n(u)$ are satisfied. Hence, we can apply Lemma~\ref{LemProfitSmallRemainingLargeWeight} to conclude the desired statement.	
\end{proof}
\begin{lemma}
	For each $u\in A^*$, we have\\
	\begin{equation}
	\left(\sum_{v\in N(u,A)}\contr{u}{v}-2\cdot\chrg{u}{n(u)}\right)\cdot w(n(u))\geq \left(w(N(u,A))-2\cdot w(u)\right)\cdot w(n(u)).\label{EqBound1}
	\end{equation}
	Moreover, for each $u\in A^*$ with $|N(u,A)|\geq 2$, we have the following:\\
	\begin{align}
	&\phantom{=}\left(\sum_{v\in N(u,A)}\contr{u}{v}-2\cdot\chrg{u}{n(u)}\right)\cdot w(n(u))\notag\\&\geq(w(u)-w(n(u)))^2 + (w(n(u))-w(n_2(u)))\cdot (w(N(u,A))-w(n(u)))\label{EqBound2}
	\end{align}
	\begin{align}
	&\phantom{=}\left(\sum_{v\in N(u,A)}\contr{u}{v}-2\cdot\chrg{u}{n(u)}\right)\cdot w(n(u))\notag\\
	&\geq 2\cdot w(u)\cdot(w(u)-w(n(u)))+w(n_2(u))\cdot (w(n(u))-w(n_2(u)))\notag\\
	&\phantom{=}+w(N(u,A)\setminus\{n(u),n_2(u)\})\cdot (w(n(u))-2\cdot w(N(u,A)\setminus\{n(u),n_2(u)\}))\label{EqBound3}
	\end{align}
	\begin{align}
	&\phantom{=}\left(\sum_{v\in N(u,A)}\contr{u}{v}-2\cdot\chrg{u}{n(u)}\right)\cdot w(n(u))\notag\\
	&\geq(w(u)-w(n(u)))^2\notag\\
	&+(w(n(u))-\min\{w(n_2(u)),w(N(u,A)\setminus\{n(u), n_2(u)\})\})\cdot w(N(u,A)\setminus\{n(u), n_2(u)\})\label{EqBound4}
	\end{align}\label{LemBounds}
\end{lemma}
\begin{proof}
	\eqref{EqBound1}:
	\[\sum_{v\in N(u,A)}\contr{u}{v}-2\cdot\chrg{u}{n(u)}\geq -2\cdot\chrg{u}{n(u)} = w(N(u,A))-2\cdot w(u)\]
	by non-negativity of the contribution.

	\eqref{EqBound2}:\begingroup\allowdisplaybreaks
	\begin{align*}
	&\phantom{=}\left(\sum_{v\in N(u,A)}\contr{u}{v}-2\cdot\chrg{u}{n(u)}\right)\cdot w(n(u))\\&\geq \contr{u}{n(u)}\cdot w(n(u))-2\cdot\chrg{u}{n(u)}\cdot w(n(u))\\
	&\geq w^2(u)-w^2(N(u,A)\setminus\{n(u)\})-(2\cdot w(u)-w(N(u,A)))\cdot w(n(u))\\
	&\geq w^2(u)-w(n_2(u))\cdot (w(N(u,A))-w(n(u)))-(2\cdot w(u)-w(N(u,A)))\cdot w(n(u))\\
	&= w^2(u)-w(n(u))\cdot (w(N(u,A))-w(n(u)))-(2\cdot w(u)-w(N(u,A)))\cdot w(n(u))\\&\phantom{=}+(w(n(u))-w(n_2(u)))\cdot (w(N(u,A))-w(n(u)))\\
	&=(w(u)-w(n(u)))^2 + (w(n(u))-w(n_2(u)))\cdot (w(N(u,A))-w(n(u))).
	\end{align*}\endgroup

	\eqref{EqBound3}:\begingroup\allowdisplaybreaks
	\begin{align*}
	&\phantom{=}\left(\sum_{v\in N(u,A)}\contr{u}{v}-2\cdot\chrg{u}{n(u)}\right)\cdot w(n(u))\\
	&\geq \contr{u}{n(u)}\cdot w(n(u))+\contr{u}{n_2(u)}\cdot w(n(u))-2\cdot\chrg{u}{n(u)}\cdot w(n(u))\\
	&\stackrel{\mathrm{contr}\geq 0}{\geq} \contr{u}{n(u)}\cdot w(n(u))+\contr{u}{n_2(u)}\cdot w(n_2(u))-2\cdot\chrg{u}{n(u)}\cdot w(n(u))\\
	&\geq w^2(u)-w^2(N(u,A)\setminus\{n(u)\}) + w^2(u)-w^2(N(u,A)\setminus\{n_2(u)\})\\&\phantom{=}-(2\cdot w(u)-w(N(u,A)))\cdot w(n(u))\\
	&= 2\cdot w^2(u)-2\cdot w^2(N(u,A)\setminus\{n(u),n_2(u)\})-w^2(n(u))-w^2(n_2(u))\\&\phantom{=}-(2\cdot w(u)-w(N(u,A)\setminus\{n(u),n_2(u)\})-w(n(u))-w(n_2(u)))\cdot w(n(u))\\
	&= 2\cdot w^2(u)-2\cdot w(u)\cdot w(n(u))-2\cdot w^2(N(u,A)\setminus\{n(u),n_2(u)\})\\&\phantom{=} +w(n(u))\cdot w(n_2(u))-w^2(n_2(u))+w(N(u,A)\setminus\{n(u),n_2(u)\})\cdot w(n(u))\\
	&\geq 2\cdot w(u)\cdot(w(u)-w(n(u)))+w(n_2(u))\cdot (w(n(u))-w(n_2(u)))\\&\phantom{=}+w(N(u,A)\setminus\{n(u),n_2(u)\})\cdot (w(n(u))-2\cdot w(N(u,A)\setminus\{n(u),n_2(u)\}))\end{align*}\endgroup

	\eqref{EqBound4}:\begingroup\allowdisplaybreaks
	\begin{align*}
	&\phantom{=}\left(\sum_{v\in N(u,A)}\contr{u}{v}-2\cdot\chrg{u}{n(u)}\right)\cdot w(n(u)) \\&\geq \contr{u}{n(u)}\cdot w(n(u))-2\cdot\chrg{u}{n(u)}\cdot w(n(u))\\
	&\geq w^2(u)-w^2(N(u,A)\setminus\{n(u)\})-(2\cdot w(u)-w(N(u,A)))\cdot w(n(u))\\
	&= w^2(u)-w^2(n_2(u))-w^2(N(u,A)\setminus\{n(u), n_2(u)\})\\
	&\phantom{=}-(2\cdot w(u)-w(N(u,A)\setminus\{n(u), n_2(u)\})-w(n(u))-w(n_2(u)))\cdot w(n(u))\\
	&= (w(u)-w(n(u)))^2+(w(n(u))-w(n_2(u))\cdot w(n_2(u))\\
	&\phantom{=}+w(n(u))\cdot w(N(u,A)\setminus\{n(u), n_2(u)\})-w^2(N(u,A)\setminus\{n(u), n_2(u)\})\\
	&\geq (w(u)-w(n(u)))^2+w(n(u))\cdot w(N(u,A)\setminus\{n(u), n_2(u)\})\\
	&\phantom{=}-(\max_{x\in N(u,A)\setminus\{n(u), n_2(u)\}} w(x))\cdot w(N(u,A)\setminus\{n(u), n_2(u)\})\\
	&\geq (w(u)-w(n(u)))^2+w(n(u))\cdot w(N(u,A)\setminus\{n(u), n_2(u)\})\\
	&\phantom{=}-\min\{w(n_2(u)),w(N(u,A)\setminus\{n(u), n_2(u)\})\}\cdot w(N(u,A)\setminus\{n(u), n_2(u)\})\\
	&=(w(u)-w(n(u)))^2\\
	&\phantom{=}+(w(n(u))-\min\{w(n_2(u)),w(N(u,A)\setminus\{n(u), n_2(u)\})\})\cdot w(N(u,A)\setminus\{n(u), n_2(u)\}) \end{align*}
	Here, the inequality \[\max_{x\in N(u,A)\setminus\{n(u), n_2(u)\}} w(x)\leq \min\{w(n_2(u)),w(N(u,A)\setminus\{n(u), n_2(u)\})\}\] follows since $n(u)$ and $n_2(u)$ are two vertices in $N(u,A)$ of maximum weight and by non-negativity of weights. Note that if we set $\max\emptyset :=-\infty$ and $-\infty\cdot 0:= 0$, then all of our calculations are also correct if $N(u,A)\setminus\{n(u),n_2(u)\}=\emptyset$.
\end{proof}\endgroup
Now, we are prepared to finally prove Lemma~\ref{LemProfitFromBadNeighbors}.
\begin{proof}[Proof of Lemma~\ref{LemProfitFromBadNeighbors}]
	By Corollary~\ref{CorOneNeighbor} and as $A$ is a maximal independent set, we can assume that $|N(u,A)|\geq 2$. Observe that this in particular implies $u\in A^*\setminus A$.\\
	\textbf{Case $1$:} $w(N(u,A))\geq (2+\epsilon)\cdot w(u)$.
	Then
	\begin{align*}&\phantom{=}\left(\sum_{v\in N(u,A)}\contr{u}{v}-2\cdot\chrg{u}{n(u)}\right)\cdot w(n(u))\\&\stackrel{\eqref{EqBound1}}{\geq}(w(N(u,A))-2\cdot w(u))\cdot w(n(u))=\left(1-\frac{2\cdot w(u)}{w(N(u,A))}\right)\cdot w(n(u))\cdot w(N(u,A))\\&\geq \left(1-\frac{2}{2+\epsilon}\right)\cdot w(n(u))\cdot w(\mathrm{supp}(u))=\frac{\epsilon}{2+\epsilon}\cdot w(n(u))\cdot w(\mathrm{supp}(u))\\&\stackrel{\eqref{EqXi3}}{\geq}2\cdot \xi\cdot w(n(u))\cdot w(\mathrm{supp}(u)).\end{align*}
	Hence, we can assume that \begin{equation}w(\mathrm{supp}(u))\leq w(N(u,A))< (2+\epsilon)\cdot w(u)\label{EqAssumptionWeightNeighborhood}\end{equation} for the remainder of the proof.

	\textbf{Case $2$:} $w(n(u))> (1+\epsilon)\cdot w(u)$.\\
	Then \[w(N(u,A)\setminus\{n(u)\})= w(N(u,A))-w(n(u)) < (2+\epsilon)\cdot w(u)-(1+\epsilon)\cdot w(u)= w(u).\]
	If $w(N(u,A))-w(n(u))\leq \epsilon\cdot w(u)$, then Lemma~\ref{LemProfitSmallRemainingLargeWeight} yields the desired statement. So assume that $w(N(u,A))-w(n(u))> \epsilon\cdot w(u)$.\\
	Then \eqref{EqBound2} tells us that
	\begin{align*}&\phantom{=}\left(\sum_{v\in N(u,A)}\contr{u}{v}-2\cdot\chrg{u}{n(u)}\right)\cdot w(n(u))\notag\\&\geq(w(n(u))-w(n_2(u)))\cdot (w(N(u,A))-w(n(u)))\notag\\&\geq (w(n(u))-w(n_2(u)))\cdot\max\{\epsilon\cdot w(u),w(n_2(u))\}.\end{align*} As a function of $w(n_2(u))$, the last term attains its minimum on the interval $[0,\epsilon\cdot w(u)]$ at $\epsilon\cdot w(u)$, and is concave on the interval $[\epsilon\cdot w(u), w(u)]$. (Recall that $w(n_2(u))\leq w(N(u,A)\setminus\{n(u)\})< w(u).$) Hence, we get
	\begin{align*}&\phantom{=}\left(\sum_{v\in N(u,A)}\contr{u}{v}-2\cdot\chrg{u}{n(u)}\right)\cdot w(n(u))\\&\geq(w(n(u))-w(n_2(u)))\cdot\max\{\epsilon\cdot w(u),w(n_2(u))\}\\&\geq \min\{(w(n(u))-\epsilon\cdot w(u))\cdot \epsilon\cdot w(u),(w(n(u))-w(u))\cdot w(u)\}
	\\&= w(u)\cdot w(n(u))\cdot \min\left\{\left(1-\epsilon\cdot\frac{w(u)}{w(n(u))}\right)\cdot\epsilon,1-\frac{w(u)}{w(n(u))} \right\}\\
	&\stackrel{(*)}{\geq} w(u)\cdot w(n(u))\cdot \min\left\{\left(1-\frac{\epsilon}{1+\epsilon}\right)\cdot\epsilon,1-\frac{1}{1+\epsilon} \right\}\\
	&= \frac{\epsilon}{1+\epsilon}\cdot w(u)\cdot w(n(u)) =\frac{\epsilon}{1+\epsilon}\cdot \frac{w(u)}{w(N(u,A))}\cdot w(n(u))\cdot w(N(u,A))\\
	&\stackrel{\eqref{EqAssumptionWeightNeighborhood}}{\geq} \frac{\epsilon}{(1+\epsilon)\cdot (2+\epsilon)}\cdot w(n(u))\cdot w(\mathrm{supp}(u))
	\stackrel{\eqref{EqXi4}}{\geq}2\cdot \xi\cdot w(n(u))\cdot w(\mathrm{supp}(u)).\end{align*} Here, the inequality marked $(*)$ follows from our case assumption that $w(n(u))> (1+\epsilon)\cdot w(u)$.\\\\
	\textbf{Case 3:} $(1+\epsilon)^{-1}\cdot w(u)\leq w(n(u))\leq (1+\epsilon)\cdot w(u)$ and $w(n_2(u))<(1+\epsilon)^{-1}\cdot w( n(u))$.\\
	\textbf{Case 3.1:} $w(N(u,A))-w(n(u))\leq \epsilon\cdot w(u)$\\ Then $u$ is helpful for $n(u)$ and $\mathrm{supp}(u)\subseteq N(u,A)\setminus\{n(u)\}$, implying that \begin{equation}w(\mathrm{supp}(u))\leq w(N(u,A))-w(n(u)).\label{Eq3p1supp}\end{equation} As a consequence,
	\begin{align*}&\phantom{=}\left(\sum_{v\in N(u,A)}\contr{u}{v}-2\cdot\chrg{u}{n(u)}\right)\cdot w(n(u))\\
	&\stackrel{\eqref{EqBound2}}{\geq} (w(n(u))-w(n_2(u)))\cdot (w(N(u,A))-w(n(u)))\\&\stackrel{\eqref{Eq3p1supp}}{\geq} (w(n(u))-w(n_2(u)))\cdot w(\mathrm{supp}(u))\geq (1-(1+\epsilon)^{-1})\cdot w(n(u))\cdot w(\mathrm{supp}(u))\\&=\frac{\epsilon}{1+\epsilon}\cdot w(n(u))\cdot w(\mathrm{supp}(u))\stackrel{\eqref{EqXi3}}{\geq}2\cdot\xi\cdot w(n(u))\cdot w(\mathrm{supp}(u)).\end{align*}
	\textbf{Case 3.2:} $w(N(u,A))-w(n(u)) > \epsilon\cdot w(u)$\\ We have \[w(N(u,A))-w(n(u)) > \epsilon\cdot w(u)\geq\frac{\epsilon}{1+\epsilon}\cdot w(n(u)).\]By the same reasoning as in case 2, using our case assumption $w(n_2(u))\leq (1+\epsilon)^{-1}\cdot w(n(u))$, we get
	\begin{align*} &\phantom{=}(w(n(u))-w(n_2(u)))\cdot (w(N(u,A))-w(n(u)))\\&\geq  (w(n(u))-w(n_2(u)))\cdot\max\left\{\frac{\epsilon}{1+\epsilon}\cdot w(n(u)), w(n_2(u))\right\}\\&\geq \min\bigg\{\left(w(n(u))-\frac{1}{1+\epsilon}\cdot w(n(u))\right)\cdot \frac{1}{1+\epsilon}\cdot w(n(u)),\\&\phantom{=}\left(w(n(u))-\frac{\epsilon}{1+\epsilon}\cdot w(n(u))\right)\cdot\frac{\epsilon}{1+\epsilon}\cdot w(n(u))\bigg\}\\&=\frac{\epsilon}{(1+\epsilon)^2}\cdot w^2(n(u)).\end{align*}
	As a consequence,\begin{align*}&\phantom{=}\left(\sum_{v\in N(u,A)}\contr{u}{v}-2\cdot\chrg{u}{n(u)}\right)\cdot w(n(u))\\&\stackrel{\eqref{EqBound2}}{\geq}(w(n(u))-w(n_2(u)))\cdot (w(N(u,A))-w(n(u)))\\&\geq \frac{\epsilon\cdot  w^2(n(u))}{(1+\epsilon)^2}\stackrel{(*)}{\geq} \frac{\epsilon}{(1+\epsilon)^3\cdot (2+\epsilon)}\cdot w(n(u))\cdot w(\mathrm{supp}(u))\\&\stackrel{\eqref{EqXi4}}{\geq}2\cdot \xi\cdot w(n(u))\cdot w(\mathrm{supp}(u)),\end{align*} where the inequality marked $(*)$ follows since \[w(\mathrm{supp}(u))\stackrel{\eqref{EqAssumptionWeightNeighborhood}}{\leq}(2+\epsilon)\cdot w(u)\leq (2+\epsilon)\cdot (1+\epsilon)\cdot w(n(u)).\]
	\textbf{Case 4:} $(1+\epsilon)^{-1}\cdot w(u)\leq  w(n(u))\leq (1+\epsilon)\cdot w(u)$ and $(1+\epsilon)^{-1}\cdot w(n(u))\leq w(n_2(u))$\\
	\textbf{Case 4.1:} $w(N(u,A)\setminus\{n(u), n_2(u)\})\leq \epsilon\cdot w(u)$:\\ Then $u$ is helpful for both $n(u)$ and $n_2(u)$ and we get $\mathrm{supp}(u)\subseteq N(u,A)\setminus\{n(u), n_2(u)\}$.\\
	\textbf{Case 4.1.1:} $w(n(u))\geq w(u)$.
	Now, if \[w(n(u))-w(n_2(u))\geq \frac{w(N(u,A)\setminus\{n(u), n_2(u)\})}{2},\] then
	\begin{align*}&\phantom{=}\left(\sum_{v\in N(u,A)}\contr{u}{v}-2\cdot\chrg{u}{n(u)}\right)\cdot w(n(u))\\&\stackrel{\eqref{EqBound2}}{\geq}(w(n(u))-w(n_2(u)))\cdot (w(N(u,A))-w(n(u)))\\&\geq(w(n(u))-w(n_2(u)))\cdot w(n_2(u))\\ &\geq\frac{w(N(u,A)\setminus\{n(u), n_2(u)\})}{2}\cdot (1+\epsilon)^{-1}\cdot w(n(u))\\&\geq \frac{1}{2\cdot (1+\epsilon)}\cdot w(n(u))\cdot w(\mathrm{supp}(u))\\&\stackrel{\eqref{EqXi3}}{\geq}2\cdot \xi\cdot w(n(u))\cdot w(\mathrm{supp}(u)) .\end{align*}
	On the other hand, if $w(n(u))-w(n_2(u)) < \frac{w(N(u,A)\setminus\{n(u), n_2(u)\})}{2}$, then 
	\begin{align*}w(N(u,A)) &= w(n(u))+w(n_2(u))+w(N(u,A)\setminus\{n(u), n_2(u)\}) \\
	&= 2\cdot w(n(u))-(w(n(u))-w(n_2(u)))+w(N(u,A)\setminus\{n(u), n_2(u)\}) \\
	&> 2\cdot w(n(u))-\frac{w(N(u,A)\setminus\{n(u), n_2(u)\})}{2}+w(N(u,A)\setminus\{n(u), n_2(u)\})\\&\geq 2\cdot w(u)+\frac{w(N(u,A)\setminus\{n(u), n_2(u)\})}{2}.\end{align*}
	This implies 
	\begin{align*} &\phantom{=}\left(\sum_{v\in N(u,A)}\contr{u}{v}-2\cdot\chrg{u}{n(u)}\right)\cdot w(n(u))\\
	&\stackrel{\eqref{EqBound1}}{\geq} (w(N(u,A))-2\cdot w(u))\cdot w(n(u))\\
	&\geq \frac{w(N(u,A)\setminus\{n(u), n_2(u)\})}{2}\cdot w(n(u))\geq \frac{1}{2}\cdot w(n(u))\cdot w(\mathrm{supp}(u))\\
	&\stackrel{\eqref{EqXi3}}{\geq}2\cdot \xi \cdot w(n(u))\cdot w(\mathrm{supp}(u)).\end{align*}
	\textbf{Case 4.1.2:} $w(n(u))<w(u)$.\\
	Then
	\begin{equation}
	w(n_2(u))\leq w(n(u))<w(u).\label{EqOrderOfWeights}
	\end{equation}
	Hence,\begingroup\allowdisplaybreaks
	\begin{align*}
	&\phantom{=}\left(\sum_{v\in N(u,A)}\contr{u}{v}-2\cdot\chrg{u}{n(u)}\right)\cdot w(n(u))\\
	&\stackrel{\eqref{EqBound3}}{\geq} 2\cdot w(u)\cdot(w(u)-w(n(u)))+w(n_2(u))\cdot (w(n(u))-w(n_2(u)))\\&\phantom{=}+w(N(u,A)\setminus\{n(u),n_2(u)\})\cdot (w(n(u))-2\cdot w(N(u,A)\setminus\{n(u),n_2(u)\}))\\
	&\stackrel{\eqref{EqOrderOfWeights}}{\geq} w(N(u,A)\setminus\{n(u),n_2(u)\})\cdot (w(n(u))-2\cdot w(N(u,A)\setminus\{n(u),n_2(u)\}))\\
	&\geq w(N(u,A)\setminus\{n(u),n_2(u)\})\cdot (w(n(u))-2\cdot \epsilon\cdot w(u))\\
	&\geq w(N(u,A)\setminus\{n(u),n_2(u)\})\cdot (w(n(u))-2\cdot \epsilon\cdot (1+\epsilon)\cdot w(n(u)))\\
	&\geq (1-2\cdot\epsilon\cdot (1+\epsilon))\cdot w(n(u))\cdot w(\mathrm{supp}(u))\\
	&\stackrel{\eqref{EqXi1}}{\geq} 2\cdot \xi\cdot w(n(u))\cdot w(\mathrm{supp}(u)).
	\end{align*}\endgroup
	\textbf{Case 4.2:} $w(N(u,A)\setminus\{n(u), n_2(u)\})> \epsilon\cdot w(u)$:\\
	We want to apply \eqref{EqBound4}.
	As $w(N(u,A))<(2+\epsilon)\cdot w(u)$ and \[(1+\epsilon)^{-1}\cdot w(n(u))\leq w(n_2(u))\leq w(n(u)),\] we have
	\begin{align*}&\phantom{=}w(n(u))-w(N(u,A)\setminus\{n(u), n_2(u)\})\\&\geq w(n(u))-((2+\epsilon)\cdot w(u)-w(n(u))-w(n_2(u)))\\&\geq \left(2+(1+\epsilon)^{-1}\right)\cdot w(n(u))-(2+\epsilon)\cdot w(u)\\&\geq \left(2+(1+\epsilon)^{-1}-(2+\epsilon)\cdot (1+\epsilon)\right)\cdot w(n(u)).\end{align*}
	Hence,
	\begin{align*} &\phantom{=}\left(\sum_{v\in N(u,A)}\contr{u}{v}-2\cdot\chrg{u}{n(u)}\right)\cdot w(n(u))\\
	&\stackrel{\eqref{EqBound4}}{\geq} (w(n(u))-w(N(u,A)\setminus\{n(u), n_2(u)\}))\cdot w(N(u,A)\setminus\{n(u), n_2(u)\})\\
	&\geq \left(2+(1+\epsilon)^{-1}-(2+\epsilon)\cdot(1+\epsilon)\right)\cdot w(n(u))\cdot \epsilon\cdot w(u)\\
	&\stackrel{\eqref{EqAssumptionWeightNeighborhood}}{\geq}\frac{2+(1+\epsilon)^{-1}-(2+\epsilon)\cdot (1+\epsilon)}{2+\epsilon}\cdot w(n(u))\cdot \epsilon\cdot w(\mathrm{supp}(u))\\
	&\stackrel{\eqref{EqXi6}}{\geq} 2\cdot\xi\cdot w(n(u))\cdot w(\mathrm{supp}(u)).
	\end{align*}
	\textbf{Case 5:} $w(n_2(u))\leq w(n(u))<(1+\epsilon)^{-1}\cdot w(u)$: \\
	\textbf{Case 5.1:} $2\cdot w(N(u,A)\setminus\{n(u),n_2(u)\})\leq w(n(u))$:\\
	We get 
	\begin{align*}
	&\phantom{=}\left(\sum_{v\in N(u,A)}\contr{u}{v}-2\cdot\chrg{u}{n(u)}\right)\cdot w(n(u))\\
	&\stackrel{\eqref{EqBound3}}{\geq} 2\cdot w(u)\cdot(w(u)-w(n(u)))+w(n_2(u))\cdot (w(n(u))-w(n_2(u)))\\&\phantom{=}+w(N(u,A)\setminus\{n(u),n_2(u)\})\cdot (w(n(u))-2\cdot w(N(u,A)\setminus\{n(u),n_2(u)\}))\\
	&\geq 2\cdot w(u)\cdot(w(u)-w(n(u))=2\cdot w(u)\cdot\left(\frac{w(u)}{w(n(u))}-1\right)\cdot w(n(u))\\&\geq 2\cdot w(u)\cdot w(n(u))\cdot(1+\epsilon-1)\stackrel{\eqref{EqAssumptionWeightNeighborhood}}{\geq} 2\cdot\frac{\epsilon}{2+\epsilon}\cdot w(n(u))\cdot w(\mathrm{supp}(u))\\
	& \stackrel{\eqref{EqXi3}}{\geq}2\cdot\xi\cdot w(n(u))\cdot w(\mathrm{supp}(u)).
	\end{align*}
	\textbf{Case 5.2:} $2\cdot w(N(u,A)\setminus\{n(u),n_2(u)\}) > w(n(u))$:\\
	We get
	\begin{align*}
	&\phantom{=}\left(\sum_{v\in N(u,A)}\contr{u}{v}-2\cdot\chrg{u}{n(u)}\right)\cdot w(n(u))\\
	&\stackrel{\eqref{EqBound4}}{\geq}(w(u)-w(n(u)))^2\\
	&\phantom{=}+(w(n(u))-\min\{w(n_2(u)),w(N(u,A)\setminus\{n(u), n_2(u)\})\})\cdot w(N(u,A)\setminus\{n(u), n_2(u)\}). \end{align*}
	Note that both summands are non-negative since all weights are positive and \mbox{$w(n_2(u))\leq w(n(u))$}.\\\\
	\textbf{Case 5.2.1:} $w(n(u))\leq \epsilon\cdot w(u)+\min\{w(n_2(u),w(N(u,A)\setminus\{n(u),n_2(u)\}))\}$:\\
	Then \begin{align*}3\cdot w(n(u))-2\cdot\epsilon\cdot w(u)&\leq w(n(u))+w(n_2(u))+w(N(u,A)\setminus\{n(u),n_2(u)\})\\&=w(N(u,A))\leq (2+\epsilon)\cdot w(u)\end{align*}
	and, hence, $w(n(u))\leq (\frac{2}{3}+\epsilon)\cdot w(u)$. As a consequence,
	\begin{align*}&\phantom{=}\left(\sum_{v\in N(u,A)}\contr{u}{v}-2\cdot\chrg{u}{n(u)}\right)\cdot w(n(u))\\
	&\stackrel{\eqref{EqBound4}}{\geq} (w(u)-w(n(u)))^2\\
	&\geq \left(\frac{1}{3}-\epsilon\right)^2\cdot (w(u))^2\geq \dfrac{\left(\frac{1}{3}-\epsilon\right)^2}{\frac{2}{3}+\epsilon}\cdot w(n(u))\cdot w(u)\\
	&\stackrel{\eqref{EqAssumptionWeightNeighborhood}}{\geq} \dfrac{\left(\frac{1}{3}-\epsilon\right)^2}{\left(\frac{2}{3}+\epsilon\right)\cdot (2+\epsilon)}\cdot w(n(u))\cdot w(\mathrm{supp}(u))\\
	&\stackrel{\eqref{EqXi5}}{\geq} 2\cdot \xi\cdot w(n(u))\cdot w(\mathrm{supp}(u)).
	\end{align*}
	\textbf{Case 5.2.2:} $w(n(u))> \epsilon\cdot w(u)+\min\{w(n_2(u),w(N(u,A)\setminus\{n(u),n_2(u)\}))\}$:\\
	Recall that by assumption on Case 5.2, we also have \[w(N(u,A)\setminus\{n(u),n_2(u)\}) > \frac{w(n(u))}{2}.\] Therefore, we obtain,
	\begin{align*} &\phantom{=}\left(\sum_{v\in N(u,A)}\contr{u}{v}-2\cdot\chrg{u}{n(u)}\right)\cdot w(n(u))\\
	&\stackrel{\eqref{EqBound4}}{\geq} (w(n(u))-\min\{w(n_2(u)),w(N(u,A)\setminus\{n(u), n_2(u)\})\})\cdot w(N(u,A)\setminus\{n(u), n_2(u)\})\\
	&\geq \epsilon\cdot w(u)\cdot \frac{w(n(u))}{2}\stackrel{\eqref{EqAssumptionWeightNeighborhood}}{\geq}\frac{\epsilon}{2\cdot(2+\epsilon)}\cdot w(n(u))\cdot w(\mathrm{supp}(u))\\
	&= \stackrel{\eqref{EqXi3}}{\geq} 2\cdot \xi\cdot w(n(u))\cdot w(\mathrm{supp}(u)).
	\end{align*}
\end{proof}
\section{Conversion between weighted and unweighted sums}
The following section contains two technical lemmata that allow us to translate back and forth between statements about weighted and unweighted sums of degrees as well as weights and sizes of neighborhoods in a setting where weights of neighboring vertices only differ by a (small) constant factor. Lemma~\ref{LemAverage} is applied in the proofs of Lemma~\ref{LemImprovementAprimeLarge}, Lemma~\ref{LemGoodUpperSweepSet} and Lemma~\ref{LemGoodUpperSweepSet1}, whereas we use Lemma~\ref{LemWeightedAverage} in the proofs of  Lemma~\ref{LemGoodLowerSweepSet} and Lemma~\ref{LemGoodLowerSweepSet2}.
\begin{lemma}
	Let $S$ be a finite set, $\varphi:S\rightarrow\mathbb{R}_{\geq 0}$, $\mu:S\rightarrow\mathbb{R}_{\geq 0}$ and $\eta > 0$ such that
	\[\sum_{s\in S}\varphi(s)\cdot \mu(s)> \eta\cdot \varphi(S).\] Let further $0<\lambda<1$. Then there exists $x\in (0,\infty)$ such that
	\[\sum_{s\in S: \lambda\cdot \varphi(s) \geq x} \mu(s)> \lambda\cdot\eta\cdot |\{s\in S:\varphi(s)\geq x\}|.\]\label{LemAverage}
\end{lemma}
\begin{proof}
	Assume towards a contradiction that there was no $x\in(0,\infty)$ with the desired property. We get
	\begingroup
	\allowdisplaybreaks
	\begin{align*}
	&\phantom{=}\sum_{s\in S}\varphi(s)\cdot \mu(s)=\sum_{s\in S}\lambda^{-1}\cdot \lambda\cdot \varphi(s)\cdot \mu(s)=\sum_{s\in S}\lambda^{-1}\cdot\int_{0}^{\lambda\cdot \varphi(s)}\mu(s)dx\\
	&=\sum_{s\in S}\lambda^{-1}\cdot\int_{0}^{\infty}\mu(s)\cdot\mathbb{1}_{\lambda\cdot \varphi(s)\geq x}dx=\lambda^{-1}\cdot \int_{0}^{\infty}\sum_{s\in S}\mu(s)\cdot\mathbb{1}_{\lambda\cdot \varphi(s)\geq x}dx\\
	&=\lambda^{-1}\cdot \int_{0}^{\infty}\sum_{s\in S:\lambda\cdot \varphi(s)\geq x}\mu(s)dx\leq \lambda^{-1}\cdot \int_{0}^{\infty}\lambda\cdot\eta\cdot |\{s\in S:\varphi(s)\geq x\}|dx\\
	&=\eta\cdot\int_{0}^{\infty}|\{s\in S:\varphi(s)\geq x\}|dx=\eta\cdot \int_{0}^{\infty}\sum_{s\in S}\mathbb{1}_{\varphi(s)\geq x}dx\\
	&=\eta\cdot\sum_{s\in S}\int_{0}^{\infty}\mathbb{1}_{\varphi(s)\geq x}dx=\eta\cdot\sum_{s\in S}\int_{0}^{\varphi(s)}1dx\\
	&=\eta\cdot \varphi(S)<\sum_{s\in S}\varphi(s)\cdot \mu(s),\end{align*}\endgroup a contradiction. Hence, there is $x\in(0,\infty)$ such that \[\sum_{s\in S: \lambda\cdot \varphi(s) \geq x} \mu(s)> \lambda\cdot\eta\cdot |\{s\in S:\varphi(s)\geq x\}|.\] 
\end{proof}
\begin{lemma}
	Let $S_1$ and $S_2$ be finite sets, $\varphi:S_1\cup S_2\rightarrow\mathbb{R}_{>0}$ and $\eta>0$ such that 
	\[|S_1|>\eta\cdot|S_2|.\] Let further $0<\lambda$. Then, there exists $x\in(0,\infty)$ such that 
	\[\sum_{s\in S_1: \varphi(s)\leq x} \varphi(s)> \lambda\cdot\eta\cdot\sum_{s\in S_2: \varphi(s)\leq \lambda^{-1}\cdot x} \varphi(s).\]\label{LemWeightedAverage}
\end{lemma}
\begin{proof}
	As $S_1$ and $S_2$ are finite and $S_1$ is not empty (since $|S_1|>\eta\cdot|S_2|\geq 0$), we can define $\Phi:=\max_{s\in S_1\cup S_2}\varphi(s)>0$. Moreover, let \begin{equation} x_0:=\inf\{x\geq 0:|\{s\in S_1: \varphi(s)\leq x\}|>\eta\cdot|\{s\in S_2:\varphi(s)\leq \lambda^{-1}\cdot x\}|.\label{DefX0}\end{equation} Observe that $x_0$ exists because for $x=\max\{1,\lambda\}\cdot\Phi$, the sets we obtain are just $S_1$ and $S_2$, which satisfy the given condition. Hence, we take the infimum over a non-empty set of values.\\ We want to prove that \[\sum_{s\in S_1: \varphi(s)\leq x_0} \varphi(s)> \lambda\cdot\eta\cdot\sum_{s\in S_2: \varphi(s)\leq \lambda^{-1}\cdot x_0} \varphi(s).\] First, note that by positivity of $\varphi$, we must have $x_0\geq \min\{\varphi(s):s\in S_1\}>0$ because otherwise, by definition of the infimum, there has to be $0\leq \beta<\min\{\varphi(s):s\in S_1\}$ for which \[0=|\{s\in S_1: \varphi(s)\leq \beta\}|>\eta\cdot|\{s\in S_2:\varphi(s)\leq \lambda^{-1}\cdot \beta\}|\geq 0,\] a contradiction.\\
	Next, we show that \[|\{s\in S_1: \varphi(s)\leq x_0\}|>\eta\cdot|\{s\in S_2:\varphi(s)\leq \lambda^{-1}\cdot x_0\}|:\]
	Let \[\bar{x}:=\min\{x > x_0:x\in\{\varphi(s):s\in S_1\}\cup\{\lambda\cdot\varphi(s):s\in S_2\}\},\] where $\min\emptyset:=\infty$.
	Then $\bar{x}>x_0$ and by definition of the infimum, there is $x_0\leq \beta<\bar{x}$ such that 
	\[|\{s\in S_1: \varphi(s)\leq \beta\}|>\eta\cdot|\{s\in S_2:\varphi(s)\leq \lambda^{-1}\cdot \beta\}|.\] But now, for each $s\in S_1$, we have \[\varphi(s)\leq x_0\Leftrightarrow\varphi(s)\leq\beta\] and for each $s\in S_2$, we have \[\varphi(s)\leq \lambda^{-1}\cdot x_0\Leftrightarrow\lambda\cdot\varphi(s)\leq x_0\Leftrightarrow\lambda\cdot\varphi(s)\leq \beta\Leftrightarrow\varphi(s)\leq \lambda^{-1}\cdot \beta.\] Hence, \[|\{s\in S_1: \varphi(s)\leq x_0\}|>\eta\cdot|\{s\in S_2:\varphi(s)\leq \lambda^{-1}\cdot x_0\}|.\]
	To simplify notation, let $S_1':=\{s\in S_1:\varphi(s)\leq x_0\}$ and $S_2':=\{s\in S_2:\varphi(s)\leq\lambda^{-1}\cdot x_0\}$. Note that by definition, $|S'_1|>\eta\cdot |S'_2|$ and in particular, $|S'_1|>0$. Observe that by minimality of $x_0$, for $0<x<x_0$, we have
	\begin{align}
	|\{s\in S'_1: \varphi(s)\leq x\}|&=|\{s\in S_1: \varphi(s)\leq x\}|\leq\eta\cdot|\{s\in S_2:\varphi(s)\leq \lambda^{-1}\cdot x\}|\notag\\&=\eta\cdot |\{s\in S'_2:\varphi(s)\leq \lambda^{-1}\cdot x\}| .\label{EqMinimalityX0}
	\end{align} 
	We compute
	\begingroup
	\allowdisplaybreaks
	\begin{align*} x_0\cdot |S'_1| &= \sum_{s\in S'_1} \varphi(s)+x_0-\varphi(s)= \varphi(S'_1)+\sum_{s\in S'_1} x_0-\varphi(s)\\
	&= \varphi(S'_1)+\sum_{s\in S'_1} \int_{\varphi(s)}^{x_0}1 dx=\varphi(S'_1)+\sum_{s\in S'_1} \int_{0}^{x_0}\mathbb{1}_{x\geq \varphi(s)} dx\\
	&=\varphi(S'_1)+ \int_{0}^{x_0}\sum_{s\in S'_1}\mathbb{1}_{x\geq \varphi(s)} dx=\varphi(S'_1)+\int_{0}^{x_0}|\{s\in S'_1: \varphi(s)\leq x\}| dx\\
	&\stackrel{\eqref{EqMinimalityX0}}{\leq}\varphi(S'_1)+\eta\int_{0}^{x_0}|\{s\in S'_2: \varphi(s)\leq \lambda^{-1}\cdot x\}| dx=\varphi(S'_1)+\eta\int_{0}^{x_0} \sum_{s\in S'_2}\mathbb{1}_{\varphi(s)\leq \lambda^{-1}\cdot x}dx\\
	&=\varphi(S'_1)+\eta\sum_{s\in S'_2}\int_{0}^{x_0} \mathbb{1}_{\varphi(s)\leq \lambda^{-1}\cdot x}dx=\varphi(S'_1)+\eta\sum_{s\in S'_2}\int_{0}^{x_0} \mathbb{1}_{\lambda\cdot\varphi(s)\leq  x}dx\\
	&=\varphi(S'_1)+\eta\sum_{s\in S'_2}\int_{\lambda\cdot \varphi(s)}^{x_0} 1 dx=\varphi(S'_1)+\eta\sum_{s\in S'_2} x_0-\lambda\cdot\varphi(s)\\
	&=\varphi(S'_1)+x_0\cdot\eta\cdot |S'_2|-\lambda\cdot\eta\cdot\varphi(S'_2).
	\end{align*}
	\endgroup
	This results in  \[\varphi(S'_1)\geq \lambda\cdot\eta\cdot\varphi(S'_2)+x_0\cdot(|S'_1|-\eta\cdot |S'_2|)> \lambda\cdot\eta\cdot\varphi(S'_2),\]
	where the last inequality follows since $|S'_1|>\eta\cdot|S'_2|$ and $x_0>0$. This finishes the proof.
\end{proof}
\section{Proof of Lemma~\ref{LemImprovementNextIter}\label{AppendixImprovementNextIter}}
In order to prove Lemma~\ref{LemImprovementNextIter}, it suffices to deal with the case in which no local improvement of size $\leq 3$, no claw-shaped improvement and no circular improvement exists since we are done otherwise. In this situation, we need to make sure that one of the sets $X^{\leq U}$ Algorithm~\ref{AlgoRunIteration} considers in line~\ref{LineXleqU} constitutes a local improvement. The proof of this statement consists of two main parts. First, Lemma~\ref{LemGoodUpperSweepSet} shows that there exists a weight $L$ with the property that $|A^*\cap V_{\geq L}|$ is sufficiently large compared to $|A_{\geq L}|$ so that \texttt{MIS}, applied to $G[V_{\geq L}]$, outputs a set $\bar{X}$ the cardinality of which is by some constant factor larger than the cardinality of $A_{\geq L}$. This factor is chosen in a way that it cuts us enough slack to cover for the fact that the weights in $X=\bar{X}\setminus A$ might be by a factor of $1+\epsilon$ smaller than the weights of their neighbors, and that there might be some further neighbors in $A\setminus A_{\geq L}$, the total (squared) weight of which is, however, bounded by $\epsilon^2\cdot w^2(X)$.
Under these assumptions, Lemma~\ref{LemGoodLowerSweepSet} tells us that one of the sets $X^{\leq U}$ we consider constitutes a local improvement.
\begin{lemma}
	If 	$\sum_{v\in A}|\mathrm{help}(v)\cap A^*|\cdot w(v)> \frac{2\cdot\rho\cdot(1+\epsilon)^3}{1-\epsilon^2}\cdot w(A)$ and Algorithm~\ref{AlgoRunIteration} does not find a local improvement of size at most $3$ (line~\ref{LineXSmall}), a claw-shaped improvement (line~\ref{LineXClaw-Shaped}) or a circular improvement (line~\ref{LineXCircular}), then there is $L\in \{w(v),v\in V\}$ such that 
	\[|A^*\cap V_{\geq L}|> \frac{\rho\cdot (1+\epsilon)^2}{1-\epsilon^2}\cdot |A_{\geq L}|.\]
	In particular, \texttt{MIS}, when applied to $G[V_{\geq L}]$, returns an independent set $\bar{X}$ of cardinality
	\[|\bar{X}| > \frac{(1+\epsilon)^2}{1-\epsilon^2}\cdot |A_{\geq L}|.\]
	\label{LemGoodUpperSweepSet}
\end{lemma}
\begin{lemma}
	Assume that the set $\bar{X}$ in line~\ref{LineBarX} of Algorithm~\ref{AlgoRunIteration} satisfies \[|\bar{X}| > \frac{(1+\epsilon)^2}{1-\epsilon^2}\cdot |A_{\geq L}|.\] Then there exists $U\in\{w(v),v\in V\}$ such that the set $X^{\leq U}$ defined in line~\ref{LineXleqU} yields a local improvement.
	\label{LemGoodLowerSweepSet}
\end{lemma}
Compliant with Algorithm~\ref{AlgoRunIteration}, we employ the following notation:
\begin{itemize}
	\item We denote the set of helpful vertices by $V_{help}:=\{u\in V\setminus A: \mathrm{help}(u)\neq\emptyset\}$.
	\item For $x\in\mathbb{R}_{\geq 0}$, we let $A_{\geq x}=\{v\in A: w(v)\geq x\}$ consist of all vertices in $A$ of weight at least $x$ and  define $V_{\geq x}:=A_{\geq x}\cup\{u\in V_{help}: w(u)\geq x \wedge \mathrm{help}(u)\subseteq A_{\geq x} \}$ to contain all vertices from $V_{help}$ with the property that these and all of the vertices in $A$ they are helpful for are of weight at least $x$. The intuition behind this definition is that we want to make sure that for every vertex $u\in V_{\geq x}\setminus A$ that might appear in our candidate local improvement $X$, $G[V_{\geq x}]$ captures enough information about $N(u,A)$ in that it contains all of $u$'s neighbors in $A$ that make up a significant fraction of $w(N(u,A))$. These are precisely those neighbors for which $u$ is helpful.
\end{itemize}
\begin{proof}[Proof of Lemma~\ref{LemGoodUpperSweepSet}]
	We first show that if the weight of $v\in A$ amounts to at least $(1+\epsilon)\cdot x$, then all helpful neighbors of $v$ are contained in $V_{\geq x}$.
	\begin{claim}
		For $v\in A$ with $(1+\epsilon)\cdot x\leq w(v)$, we have $\mathrm{help}(v)\cap A^*\subseteq V_{\geq x}$.\label{ClaimSameDegree}
	\end{claim}
	\begin{proof}
		First, by choice of $x$, we have $v\in A_{\geq x}$.
		Next, we know that for $u\in\mathrm{help}(v)$, one of the following applies: 
		\begin{itemize}
			\item $\mathrm{help}(u)=\{n(u)\}=\{v\}$ and $w(v)\leq (1+\epsilon)\cdot w(u)$. 
			\item $\mathrm{help}(u)=\{n(u), n_2(u)\}\ni v$ and 
			$(1+\epsilon)^{-1}\cdot w(n(u))\leq w(n_2(u))\leq w(n(u))\leq (1+\epsilon)\cdot w(u).$
		\end{itemize} 
		In either case, $u$ and all of the vertices $u$ is helpful for bear a weight of at least $\frac{w(v)}{1+\epsilon}\geq x$. This implies that $u\in V_{\geq x}$.\end{proof} 
	Next, we prove the existence of $L$ with the property that the sum of helpful neighbors the vertices in $A_{\geq L}$ have in $A^*\cap V_{\geq L}$ is large compared to the cardinality of $A_{\geq L}$.
	\begin{claim}There is $L\in\{w(v),v\in V\}$ such that \[\sum_{v\in A_{\geq L}} |\mathrm{help}(v)\cap A^*\cap V_{\geq L}|> \frac{2\cdot \rho\cdot (1+\epsilon)^2}{1-\epsilon^2}\cdot |A_{\geq L}|.\] \end{claim}
	\begin{proof}
		We want to apply Lemma~\ref{LemAverage}. To this end, set $S:=A$, $\mu(v):=|\mathrm{help}(v)\cap A^*|$, $\varphi(v):=w(v)$, $\lambda:=(1+\epsilon)^{-1}$ and $\eta:= \frac{2\cdot \rho\cdot (1+\epsilon)^3}{1-\epsilon^2}$. Then Lemma~\ref{LemAverage} tells us that there is $L\in (0,\infty)$ such that \begin{align*}&\phantom{=}\sum_{v\in A_{\geq L}} |\mathrm{help}(v)\cap A^*\cap V_{\geq L}|\stackrel{\small\text{Claim~\ref{ClaimSameDegree}}}{\geq} \sum_{v\in A_{\geq (1+\epsilon)\cdot L}}|\mathrm{help}(v)\cap A^*|\\&\stackrel{\small\text{Lem.~\ref{LemAverage}}}{>}\frac{2\cdot \rho\cdot (1+\epsilon)^3}{1-\epsilon^2}\cdot (1+\epsilon)^{-1}\cdot |A_{\geq L}|=\frac{2\cdot \rho\cdot (1+\epsilon)^2}{1-\epsilon^2}\cdot |A_{\geq L}|.\end{align*}
		The strict inequality tells us that $A_{\geq L}\neq \emptyset$ and by increasing $L$ to the minimum one among the vertex weights in $A_{\geq L}$, we may assume that $L\in\{w(v),v\in V\}$.
	\end{proof} From this, we calculate
	\begin{align*}
	2\cdot |A^*\cap V_{\geq L}|&\geq \sum_{u\in (A^*\setminus A)\cap V_{\geq L}} |\mathrm{help}(u)| = \sum_{v\in A_{\geq L}} |\mathrm{help}(v)\cap A^*\cap V_{\geq L}| \\&> \frac{2\cdot \rho\cdot (1+\epsilon)^2}{1-\epsilon^2}\cdot |A_{\geq L}|.
	\end{align*}
	Here, the first inequality follows since  $|\mathrm{help}(u)|\leq 2$ for all $u\in A^*\setminus A$ by definition. The following equation is implied by the facts that $\mathrm{help}(u)\subseteq A_{\geq L}$ for $u\in V_{\geq L}$, $\mathrm{help}(v)\subseteq V\setminus A$ for $v\in A$, and $v\in \mathrm{help}(u)\Leftrightarrow u\in \mathrm{help}(v)$ for $u\in V\setminus A$ and $v\in A$. Division by $2$ yields \[|A^*\cap V_{\geq L}|> \frac{\rho\cdot (1+\epsilon)^2}{1-\epsilon^2}\cdot |A_{\geq L}|,\] and as $A^*\cap V_{\geq L}$ is independent in $G$, we can conclude that the algorithm $\texttt{MIS}$, applied to $G[V_{\geq L}]$, finds an independent set $\bar{X}$  of size at least \[|\bar{X}|> \rho^{-1}\cdot \frac{\rho\cdot (1+\epsilon)^2}{1-\epsilon^2}\cdot |A_{\geq L}|=\frac{ (1+\epsilon)^2}{1-\epsilon^2}\cdot |A_{\geq L}|.\] Note that the strict inequality is inherited from the strict inequality on \mbox{$|A^*\cap V_{\geq L}|$} and $|A_{\geq L}|$. Further observe that as $G[V_{\geq L}]$ is an induced sub-graph of $G$, $\bar{X}$ is independent in $G$ as well.
\end{proof}
\begin{proof}[Proof of Lemma~\ref{LemGoodLowerSweepSet}] Define $X:=\bar{X}\setminus A$ and for $x\geq 0$, let $X^{\leq x}:=\{v\in X: w(v)\leq x\}$. We point out that it suffices to prove the existence of any $U\in (0,\infty)$ such that $X^{\leq U}$ constitutes a local improvement because this property will ensure that $X^{\leq U}\neq \emptyset$ and, hence, decreasing $U$ to the maximum weight in $X^{\leq U}$ will preserve the set $X^{\leq U}$, and ensure $U\in\{w(v),v\in V\}$.
	
	Our goal is to apply Lemma~\ref{LemWeightedAverage}. For this purpose, we have to derive a lower bound on the cardinality of $X$.
	As $\bar{X}\cap A\subseteq V_{\geq L}\cap A=A_{\geq L}$, we get \[|X|>\frac{(1+\epsilon)^2}{1-\epsilon^2}\cdot |A_{\geq L}|-|\bar{X}\cap A|=\frac{(1+\epsilon)^2}{1-\epsilon^2}\cdot |A_{\geq L}|-|\bar{X}\cap A_{\geq L}|\geq \frac{(1+\epsilon)^2}{1-\epsilon^2}\cdot |A_{\geq L}\setminus\bar{X}|,\] where the last inequality follows from the fact that $\frac{(1+\epsilon)^2}{1-\epsilon^2}>1$. Moreover, as $\bar{X}$ is independent in $G$, no vertex in $X$ is adjacent to a vertex in $A_{\geq L}\cap\bar{X}$, implying that $N(X,A_{\geq L})\subseteq A_{\geq L}\setminus\bar{X}$. Hence, we obtain
	\[|X|>\frac{(1+\epsilon)^2}{1-\epsilon^2}\cdot |N(X,A_{\geq L})|.\] 
	\begin{claim}There is $U\in (0,\infty)$ such that $w^2(X^{\leq U})> \frac{1}{1-\epsilon^2}\cdot w^2(N(X^{\leq U},A_{\geq L}))$.\end{claim}
	\begin{proof} We apply Lemma~\ref{LemWeightedAverage} with $S_1:=X$, $S_2:=N(X,A_{\geq L})$, $\varphi(s):=w^2(s)>0$ for $s\in S_1\cup S_2$, $\eta:= \frac{(1+\epsilon)^2}{1-\epsilon^2}$ and $\lambda:=(1+\epsilon)^{-2}$.
		In this setting, Lemma~\ref{LemWeightedAverage} tells us that there is $U\in (0,\infty)$ such that \begin{align*}w^2(X^{\leq U})&=w^2(\{u\in X: w^2(u)\leq U^2\})\\&>  \frac{1}{1-\epsilon^2}\cdot w^2(\{v\in N(X,A_{\geq L}): w^2(v)\leq (1+\epsilon)^2\cdot U^2\}).\end{align*}
		By definition of $V_{\geq L}$, for we know that for $u\in X\subseteq V_{\geq L}\setminus A\subseteq V_{help}$, any vertex in $N(u,A)$ is of weight at most $(1+\epsilon)\cdot w(u)$ since this holds for the vertices in $\mathrm{help}(u)$ and moreover, the total weight of $N(u,A)\setminus\mathrm{help}(u)$ is bounded by $\epsilon\cdot w(u)$. In particular, for $u\in X^{\leq U}$ and $v\in N(u,A)$, we have $w^2(v)\leq (1+\epsilon)^2\cdot w^2(u)\leq (1+\epsilon)^2\cdot U^2$. These facts imply that
		\begin{align}w^2(X^{\leq U})&> \frac{1}{1-\epsilon^2}\cdot w^2(\{v\in N(X,A_{\geq L}): w^2(v)\leq (1+\epsilon)^2\cdot U^2\})\notag\\&\geq\frac{1}{1-\epsilon^2}\cdot w^2(N(X^{\leq U},A_{\geq L})).\label{EqXleqUhelpfulNeighborhood}\end{align}\end{proof}
	To finally see that $X^{\leq U}$ constitutes a local improvement of the squared weight function, it remains to bound $w^2(N(X^{\leq U},A\setminus A_{\geq L}))$. As $X^{\leq U}\subseteq V_{\geq L}\setminus A\subseteq V_{help}$, we know that for $u\in X$, we have $\emptyset\neq\mathrm{help}(u)\subseteq A_{\geq L}$. By Definition~\ref{DefHelpful}, this implies that \[w(N(u,A\setminus A_{\geq L}))\leq w(N(u,A)\setminus\mathrm{help}(u))\leq \epsilon\cdot w(u),\] and, hence, \[w^2(N(u,A\setminus A_{\geq L}))\leq (w(N(u,A\setminus A_{\geq L})))^2\leq \epsilon^2\cdot w^2(u)\] for $u\in X^{\leq U}$. Consequently, we obtain \[w^2(N(X^{\leq U},A\setminus A_{\geq L}))\leq \sum_{u\in X^{\leq U}} w^2(N(u,A\setminus A_{\geq L}))\leq \epsilon^2\cdot w^2(X^{\leq U}).\] Combining this with \eqref{EqXleqUhelpfulNeighborhood} finally yields
	\begin{align*}w^2(X^{\leq U})&=(1-\epsilon^2)\cdot w^2(X^{\leq U})+\epsilon^2\cdot w^2(X^{\leq U}) \\&> w^2(N(X^{\leq U},A_{\geq L}))+w^2(N(X^{\leq U},A\setminus A_{\geq L})) \\&= w^2(N(X^{\leq U},A)),\end{align*} showing that $X^{\leq U}$ constitutes a local improvement.
\end{proof}
\section{Proof of Lemma~\ref{LemImprovementAprimeLarge}\label{AppendixManySpecialNeighbors}}
To prove Lemma~\ref{LemImprovementAprimeLarge}, we consider the auxiliary graph $J^*$ on the vertex set $A$ in which each vertex $u\in A^*\setminus A$ that is helpful for $n(u)$ and $n_2(u)$ induces an edge between them, provided at least one of them is contained in $A'$. It turns out that cycles of logarithmic size in $J^*$ yield circular improvements because for $v\in A'$, we can choose $Y_v:=\{t(v)\}$ (cf. Def.~\ref{DefSpecial}) to have a large contribution to $v$ and moreover, if $u$ is helpful for $n(u)$ and $n_2(u)$, then $2\cdot w^2(u)\gtrapprox w^2(n(u)) + w^2(n_2(u))$ and $w^2(N(u,A)\setminus\{n(u),n_2(u)\})$ is small.

In order to derive the existence of a cycle of logarithmically bounded size in $J^*$, by a result by Berman and F\"urer, it suffices to show that $J^*$ is sufficiently dense~\cite{berman1994approximating}. For this purpose, Lemma~\ref{LemNoSingleNeighborsInHstar} tells us that no vertex $v\in A'$ can have a helpful neighbor $u$ with the property that $\mathrm{help}(u)=\{v\}$. Moreover, Lemma~\ref{LemNoEdgesWithinAPrime} tells us that in $J^*$, we do not have any edges within $A'$. We use these two statements in the proof of Lemma~\ref{LemImprovementAprimeLarge} to obtain a good bound on the weighted sum of degrees in $J^*$, which we can then, by exploiting local similarity of weights, translate into a density result for a sub-graph of $J^*$.
\begin{lemma}
	If there exists $u\in A^*\setminus A$ such that $\mathrm{help}(u)=\{n(u)\}$ and $n(u)\in A'$, then $\{u, t(n(u))\}$ constitutes a local improvement.\label{LemNoSingleNeighborsInHstar}
\end{lemma}
\begin{proof}
	First, note that $u$ and $t(n(u))$ are distinct because $u$ is helpful for $n(u)$, but $t(n(u))$ is not. By the definition of helpful vertices (Definition~\ref{DefHelpful}) and by our choice of $t(n(u))$, we know that
	\begin{enumerate}
		\item $w(n(u))\leq (1+\epsilon)\cdot w(u)$,
		\item $w(N(u,A)\setminus\{n(u)\})\leq \epsilon\cdot w(u)$ and
		\item $\contr{t(n(u))}{n(u)}>\frac{5}{8}\cdot w(n(u))$, which, as $w(n(u))>0$, implies \begin{equation}
		w^2(t(n(u)))-w^2(N(t(n(u)),A)\setminus\{n(u)\}) > \frac{5}{8}\cdot w^2(n(u)).\label{EqContrSpecial}
		\end{equation}
	\end{enumerate}
	Let $X:=\{u, t(n(u))\}$. Then
	\begingroup\allowdisplaybreaks
	\begin{align*}
	w^2(N(X,A))&\leq w^2(n(u))+w^2(N(u,A)\setminus\{n(u)\})+w^2(N(t(n(u)),A)\setminus\{n(u)\})\\
	&\stackrel{\eqref{EqContrSpecial}}{<} w^2(n(u))+w^2(N(u,A)\setminus\{n(u)\})+w^2(t(n(u)))-\frac{5}{8}\cdot w^2(n(u))\\
	&\leq \frac{3}{8}\cdot w^2(n(u)) +(w(N(u,A)\setminus\{n(u)\}))^2+w^2(t(n(u)))\\
	&\leq \frac{3}{8}\cdot w^2(n(u))+\epsilon^2\cdot w^2(u)+ w^2(t(n(u)))\\
	&\leq \left(\frac{3\cdot (1+\epsilon)^2}{8}+\epsilon^2\right)\cdot w^2(u)+w^2(t(n(u)))\\
	&\stackrel{\eqref{EqEps1}}{\leq} w^2(u)+w^2(t(n(u)))=w^2(X).
	\end{align*}\endgroup
	Hence, $X$ is a local improvement.
\end{proof}
\begin{lemma}
	If there exists $u\in A^*\setminus A$ such that $n(u)$ and $n_2(u)$ are contained in $A'\cap\mathrm{help}(u)$, then $\{t(n(u)), u, t(n_2(u))\}$ is a local improvement.\label{LemNoEdgesWithinAPrime}
\end{lemma}
\begin{proof}
	By the definition of $\mathrm{help}(u)$, we know that \begin{enumerate}
		\item  $w(n_2(u))\leq w(n(u))\leq (1+\epsilon)\cdot w(u)$ and
		\item$w(N(u,A)\setminus\{n(u),n_2(u)\})\leq \epsilon\cdot w(u)$. \end{enumerate}
	Moreover, the definition of the vertices $t(n(u))$ and $t(n_2(u))$ tells us that they are distinct because $n(t(n(u)))=n(u)$ and $n(t(n_2(u)))=n_2(u)$, and different from $u$ since neither of them is helpful for $n(u)$ or $n_2(u)$, respectively. In addition to that, the definition of $t(n(u))$ and $t(n_2(u))$ tells us that
	$\contr{t(n(u))}{n(u)}>\frac{5}{8}\cdot w(n(u))$ and $\contr{t(n_2(u))}{n_2(u)}>\frac{5}{8}\cdot w(n_2(u))$, which, as $w(n(u))>0$ and $w(n_2(u))>0$, implies
	\begin{equation}
	w^2(t(n(u)))-w^2(N(t(n(u)),A)\setminus\{n(u)\}) > \frac{5}{8}\cdot w^2(n(u))\quad\text{and}\label{EqContrSpecial1}
	\end{equation}
	\begin{equation}
	w^2(t(n_2(u)))-w^2(N(t(n_2(u)),A)\setminus\{n_2(u)\}) > \frac{5}{8}\cdot w^2(n_2(u)).\label{EqContrSpecial2}
	\end{equation}
	Let $X:=\{t(n(u)), u, t(n_2(u))\}$. Then
	\begingroup\allowdisplaybreaks
	\begin{align*}
	w^2(N(X,A))&\leq w^2(n(u))+w^2(n_2(u))+w^2(N(u,A)\setminus\{n(u),n_2(u)\})\\&\phantom{=}+w^2(N(t(n(u)), A)\setminus \{n(u)\})+w^2(N(t(n_2(u)), A)\setminus \{n_2(u)\})\\
	&\stackrel{\eqref{EqContrSpecial1},\eqref{EqContrSpecial2}}{<}w^2(n(u))+w^2(n_2(u))+w^2(N(u,A)\setminus\{n(u),n_2(u)\})\\&\phantom{=}+w^2(t(n(u)))-\frac{5}{8}\cdot w^2(n(u))+w^2(t(n_2(u)))-\frac{5}{8}\cdot w^2(n_2(u))\\
	&\leq\frac{3}{8}\cdot w^2(n(u))+\frac{3}{8}\cdot w^2(n_2(u))+(w(N(u,A)\setminus\{n(u),n_2(u)\}))^2\\
	&\phantom{=}+w^2(t(n(u)))+w^2(t(n_2(u)))\\
	&\leq \frac{3}{8}\cdot w^2(n(u))+\frac{3}{8}\cdot w^2(n_2(u))+\epsilon^2\cdot w^2(u)+w^2(t(n(u)))+w^2(t(n_2(u)))\\
	&\leq \left(\frac{3\cdot(1+\epsilon)^2}{8}+\frac{3\cdot(1+\epsilon)^2}{8}+\epsilon^2\right)\cdot w^2(u)+w^2(t(n(u)))+w^2(t(n_2(u)))\\
&\stackrel{\eqref{EqEps1}}{\leq} w^2(u)+w^2(t(n(u)))+w^2(t(n_2(u)))=w^2(X).
	\end{align*}\endgroup
	Hence, $X$ constitutes a local improvement as claimed.
\end{proof}
\begin{proof}[Proof of Lemma~\ref{LemImprovementAprimeLarge}]
	If there exists a local improvement of size at most $3$ or a claw-shaped improvement, we are done, so assume that this is not the case. Consider the multi-graph $J^*$ given by
	\begin{itemize}
		\item $V(J^*)=A$ and
		\item $E(J^*):=\{e_u:=\{n(u),n_2(u)\}: u\in A^*\setminus A: \mathrm{help}(u)=e_u=\{n(u),n_2(u)\} \wedge e_u\cap A'\neq \emptyset\}$.
	\end{itemize} We want to show that $J^*$ contains a cycle of logarithmic size (and that such a cycle yields a circular improvement). To this end, first observe that by Lemma~\ref{LemNoSingleNeighborsInHstar}, we know that for each $v\in A'$, we have $|\delta_{J^*}(v)|=|\mathrm{help}(v)\cap A^*|$ because $v$ does not have any helpful neighbor $u$ such that $\mathrm{help}(u)=\{n(u)\}$ and as $A'\subseteq A\setminus A^*$, we have $\mathrm{help}(v)\cap A^*\subseteq A^*\setminus A$ for $v\in A'$. Additionally, Lemma~\ref{LemNoEdgesWithinAPrime} tells us that $A'$ is an independent set in $J^*$. As by construction, $A\setminus A'$ is independent in $J^*$, too, we can conclude that $J^*$ is bipartite with bipartitions $A'$ and $A\setminus A'$.\\
	Next, we would like to see that $J^*$ has a sub-graph that is dense enough to contain a cycle of logarithmic size by making use of Lemma~\ref{LemAverage}. To this end, we have to calculate the weighted sum of degrees in $J^*$. Note that by Definition~\ref{DefHelpful}, for each edge $\{n(u),n_2(u)\}\in E(J^*)$, we have $(1+\epsilon)^{-1}\cdot w(n(u))\leq w(n_2(u))\leq w(n(u))$. Throughout the following calculation, we denote edges of $J^*$ in such a way that the first vertex is from $A'$ and the second one is from $A\setminus A'$. We compute
	\begingroup\allowdisplaybreaks
	\begin{align*}
	&\phantom{=}\sum_{v\in A} w(v)\cdot |\delta_{J^*}(v)| =\sum_{\{x,y\}\in E(J^*)} w(x)+w(y)\geq \sum_{\{x,y\}\in E(J^*)} w(x)+(1+\epsilon)^{-1}\cdot w(x)\\
	&= (1+(1+\epsilon)^{-1})\cdot \sum_{\{x,y\}\in E(J^*)} w(x)=(1+(1+\epsilon)^{-1})\cdot \sum_{v\in A'} w(v)\cdot |\delta_{J^*}(v)|\\
	&=(1+(1+\epsilon)^{-1})\cdot \sum_{v\in A'} w(v)\cdot |\mathrm{help}(v)\cap A^*|>(1+(1+\epsilon)^{-1})\cdot \frac{(2+\kappa)\cdot (1+\epsilon)^2}{(2+\epsilon)}\cdot w(A)\\
	&=(2+\epsilon)\cdot \frac{(2+\kappa)\cdot (1+\epsilon)}{2+\epsilon}\cdot w(A)=(1+\epsilon)\cdot (2+\kappa)\cdot w(A).
	\end{align*}\endgroup
	Define $S:=A$, $\mu(v):=|\delta_{J^*}(v)|$, $\varphi(v):=w(v)$, $\eta:=(1+\epsilon)\cdot (2+\kappa)$ and $\lambda := (1+\epsilon)^{-1}$.
	Then Lemma~\ref{LemAverage} tells us that there is $x>0$ such that
	\[\sum_{v\in A: w(v)\geq (1+\epsilon)\cdot x}|\delta_{J^*}(v)| > (2+\kappa)\cdot |\{v\in A: w(v)\geq x\}|.\]
	As for $v\in A$ with $w(v)\geq (1+\epsilon)\cdot x$, we know that every neighbor of $v$ in $J^*$ has weight at least $(1+\epsilon)^{-1}\cdot (1+\epsilon)\cdot x=x$, we can infer that
	\[\sum_{v\in A: w(v)\geq x}|\delta_{J^*[\{v\in A: w(v)\geq  x\}]}(v)|\geq \sum_{v\in A: w(v)\geq (1+\epsilon)\cdot x}|\delta_{J^*}(v)| > (2+\kappa)\cdot |\{v\in A: w(v)\geq x\}|.\]
	In particular, as $J^*$ does not contain any loops, the strict inequality tells us that \[|\{v\in A: w(v)\geq  x\}|\geq 2.\] Moreover, we obtain
	\begin{align*}|E(J^*[\{v\in A: w(v)\geq  x\}])|&=\frac{1}{2}\!\!\cdot\sum_{v\in A: w(v)\geq \cdot x}|\delta_{J^*[\{v\in A: w(v)\geq  x\}]}(v)|>\left(1+\frac{\kappa}{2}\right)\cdot |\{v\in A: w(v)\geq x\}|.\end{align*}
	Now, Lemma $3.2$ from \cite{berman1994approximating} and the fact that $\frac{1}{\kappa}\in\mathbb{N}^+$ allow us to conclude that the sub-graph $J^*[\{v\in A: w(v)\geq  x\}]$ contains a cycle of length at most $\frac{8}{\kappa}\cdot\log(|A|)\leq\frac{8}{\kappa}\cdot\log(|V(G)|)$. Call this cycle $C$ and let $U$ be the set of vertices from $A^*\setminus A$ that induce the edges of $C$. We show that $X:=U\dot{\cup}\dot{\bigcup}_{v\in V(C)\cap A'}\{t(v)\}$ defines a circular improvement. Note that for $v\neq v'$, we have $t(v)\neq t(v')$ since $n(t(v))=v\neq v'=n(t(v'))$, and that for $u\in U$ and $v\in V(C)\cap A'$, we have $t(v)\neq u$. To this end, observe that $t(v)=u$ would imply $n(u)=n(t(v))=v$. But by definition of $J^*$, this yields $v\in\mathrm{help}(u)$, whereas $v\not\in\mathrm{help}(t(v))$, a contradiction. Hence, we obtain a disjoint union as claimed.\\
	To see that $X$ defines a local improvement/satisfies the third condition from Definition~\ref{DefCircular}, we need to show that for each edge $\{n(u),n_2(u)\}=\{v,z\}\in E(C)$ such that $v\in A'$ and $z\in A\setminus A'$, we have
	\[2\cdot w^2(u)+w^2(t(v)) > w^2(v)+w^2(z)+2\cdot w^2(N(u,A)\setminus\{v,z\})+w^2(N(t(v), A)\setminus\{v\}).\] Note that summing up these inequalities then yields 
	\[2\cdot w^2(X) > 2\cdot w^2(V(C))+2\cdot w^2(N(X,A)\setminus V(C)) \geq 2\cdot w^2(N(X,A)),\] implying that $X$ is a local improvement.
	Recall that as $\mathrm{help}(u)=\{n(u),n_2(u)\}=\{v,z\}$, we have
	\begin{equation}
	\max\{w(v),w(z)\}\leq (1+\epsilon)\cdot w(u)\label{EqMaxWeightVZ}
	\end{equation}
	and
	\begin{equation}
	w(N(u,A)\setminus\{v,z\})=w(N(u,A)\setminus\{n(u),n_2(u)\})\leq\epsilon\cdot w(u).\label{EqWeightRemainingNeighborsOfU}
	\end{equation}
	Besides, by definition of $t(v)$, we have $\contr{t(v)}{v}>\frac{5}{8}\cdot w(v)$,
	so \begin{equation}
	w^2(t(v))-w^2(N(t(v),A)\setminus\{v\}) > \frac{5}{8}\cdot w^2(v)\label{EqContrTV}.
	\end{equation}
	Using this, we obtain 
	\begin{align*}
	&\phantom{=}w^2(v)+w^2(z)+2\cdot w^2(N(u,A)\setminus\{v,z\})+w^2(N(t(v), A)\setminus\{v\})\\
	&\stackrel{\eqref{EqContrTV}}{<}w^2(v)+w^2(z)+2\cdot w^2(N(u,A)\setminus\{v,z\})+w^2(t(v))-\frac{5}{8}\cdot w^2(v)\\
	&\leq \frac{3}{8}\cdot w^2(v) +w^2(z)+ 2\cdot (w(N(u,A)\setminus\{v,z\}))^2 + w^2(t(v))\\
	&\stackrel{\eqref{EqMaxWeightVZ},\eqref{EqWeightRemainingNeighborsOfU}}{\leq} \frac{3\cdot(1+\epsilon)^2}{8}\cdot w^2(u)+(1+\epsilon)^2\cdot w^2(u)+2\cdot\epsilon^2\cdot w^2(u)+w^2(t(v))\\
	&=\left(\frac{11}{8}\cdot (1+\epsilon)^2+2\cdot\epsilon^2\right)\cdot w^2(u)+w^2(t(v))\\
	&\stackrel{\eqref{EqEps1}}{\leq}2\cdot w^2(u)+w^2(t(v)).
	\end{align*}
	This finishes the proof.
\end{proof}
\section{The relation between weighted and unweighted \boldmath$k$-Set Packing\unboldmath\label{SecRelation}}
In this section, we would like to shed some light on the more general relation between approximation guarantees for the weighted and the unweighted $k$-Set Packing problem. More precisely, we would like to prove Theorem~\ref{TheoConstantFactor}, which we restate again for easier readability.
\TheoConstantFactor*
We would first like to point out that Theorem~\ref{TheoConstantFactor} is not a direct consequence of our previous analysis. Even though the proof of Theorem~\ref{TheoMainTheorem} is oblivious to the approximation guarantee $\rho$ for the unweighted $k$-Set Packing problem we plug into it, our choice of the constraints in appendix~\ref{AppendixInequalities}, and thus, of the constants $\epsilon$ and $\xi$, is tailored to the case where $\rho=\frac{k+1+\epsilon}{3}$. In particular, the constraints in appendix~\ref{AppendixInequalities} only allow for a value of $\xi$ smaller than $0.02$ (see Figure~\ref{FigPlotsXi}), and by just applying Theorem~\ref{TheoMainTheorem}, we get stuck above an approximation guarantee of $\frac{k+1}{2}-0.02\cdot(k-1)$, which is not sufficient for our purposes.

Instead, we will basically repeat a coarse version of the previous analysis, in which we, essentially, first relax the constant $\epsilon$ sufficiently to obtain a large enough value of $\xi$, and then compute how small $\tau$ needs to be, compared to $\sigma$, to still obtain a local improvement by an application of MIS in case there are too many helpful vertices to obtain an improved guarantee. 

Note that we express the approximation guarantees as $1+\chi\cdot (k-1)$ with $\chi\in (0,1)$ instead of $\chi\cdot (k+1)$ to ensure they are not smaller than $1$.  For technical reasons, it is, however, more convenient to deal with approximation guarantees for the unweighted $k$-Set Packing problem that are of the form $\tilde{\tau}\cdot (k+1)$, whereas the desired guarantee for the weighted case remains of the form $1+\sigma\cdot (k-1)$.
The proof of Theorem~\ref{TheoConstantFactor} hence consists of two parts. First, we show the following modified theorem:
\begin{theorem}
 For any constant $\sigma\in (0,1)$, there exists a constant $\tilde{\tau}\in(0,1)$ with the following property: If there are $k_0\in\mathbb{N}_{\geq 3}$ and a polynomial time $\tilde{\tau}\cdot(k+1)$ approximation algorithm for the unweighted $k$-Set Packing problem for $k\geq k_0$, then there is a polynomial time $1+\sigma\cdot(k-1)$-approximation algorithm for the weighted $k$-Set Packing problem for $k\geq k_0$.\label{TheoModified}	
\end{theorem}
Then, we show that Theorem~\ref{TheoModified} implies Theorem~\ref{TheoConstantFactor} by proving the following lemma, which basically exploits the fact that there exists a constant $C>1$ such that unless $P=NP$, there is no $C$-approximation for both the weighted and the unweighted $k$-Set Packing problem for $k\geq 3$.
\begin{lemma}
	Unless $P=NP$, for every constant $\tilde{\tau}\in (0,1)$, there exists a constant $\tau\in(0,1)$ with the following property: If there are $k_0\in\mathbb{N}_{\geq 3}$ and a polynomial time $1+\tau\cdot (k-1)$ approximation algorithm for the weighted or unweighted $k$-Set Packing problem for $k\geq k_0$, then for any $k\geq k_0$, we have $1+\tau\cdot (k-1)\leq \tilde{\tau}\cdot (k+1)$.\label{LemTildeTau}
	\end{lemma}
Note that Lemma~\ref{LemTildeTau} also allows us to replace the $1+\sigma\cdot (k-1)$ term in Theorem~\ref{TheoConstantFactor} and Theorem~\ref{TheoModified} by $\tilde{\sigma}\cdot (k+1)$.
We now give a short proof of Theorem~\ref{TheoConstantFactor}, assuming that Theorem~\ref{TheoModified} and Lemma~\ref{LemTildeTau} hold.
\begin{proof}[Proof of Theorem~\ref{TheoConstantFactor}]
In case $P\neq NP$, combining Theorem~\ref{TheoModified} and Lemma~\ref{LemTildeTau} directly yields Theorem~\ref{TheoConstantFactor}. To see this, let $\sigma\in (0,1)$. By Theorem~\ref{TheoModified}, there is $\tilde{\tau}\in (0,1)$ such that for any $k_0\in\mathbb{N}_{\geq 3}$, the existence of a polynomial time $\tilde{\tau}\cdot (k+1)$-approximation for the unweighted $k$-Set Packing problem for $k\geq k_0$ implies a polynomial time $1+\sigma\cdot(k-1)$-approximation for weighted $k$-Set Packing.  Let $\tau\in (0,1)$ as implied by Lemma~\ref{LemTildeTau}. We claim that $\tau$ meets the requirements of Theorem~\ref{TheoConstantFactor}. Indeed, by Lemma~\ref{LemTildeTau}, if there exists $k_0\in\mathbb{N}_{\geq 3}$ such that a polynomial time $1+\tau\cdot (k-1)$-approximation for $k\geq k_0$ exists, then by our assumption that $P\neq NP$, we have $1+\tau\cdot (k-1)\leq \tilde{\tau}\cdot (k+1)$ for all $k\geq k_0$ and in particular, the $1+\tau\cdot (k-1)$ approximation algorithm for $k\geq k_0$ also constitutes a $\tilde{\tau}\cdot (k+1)$-approximation. By our choice of $\tilde{\tau}$, this implies a polynomial time $1+\sigma\cdot (k-1)$-approximation for the weighted $k$-Set Packing problem for $k\geq k_0$.

Now, we deal with the case $P=NP$. As the decision variant of the weighted $k$-Set Packing problem (decide whether there is a solution of total weight $\geq q\in\mathbb{Q}$) is in $NP$, it follows that if $P=NP$, there exists a polynomial time algorithm $\mathcal{A}$ for the decision variant of weighted $k$-Set Packing. We can apply a standard trick to also obtain a polynomial time algorithm for the (optimization variant of the) weighted $k$-Set Packing problem (assuming all weights to be rational). By first multiplying all weights with their common denominator and then performing binary search, we obtain an algorithm $\mathcal{B}$ that can determine the value of an optimum solution in polynomial time. Now, we can simply run the following algorithm: Let $q^*:=\mathcal{B}(\mathcal{S},w)$ be the value of an optimum solution and initialize $\mathcal{S'}:=\mathcal{S}$. Now, for each set $u\in\mathcal{S}$, check whether deleting $u$ from $\mathcal{S}'$ decreases the value of an optimum solution below $q^*$. If it does, we know that we need to keep $u$, otherwise, we can delete $u$ from $\mathcal{S}'$ permanently. After having traversed all sets in $\mathcal{S}$, $\mathcal{S}'$ is a disjoint sub-collection of $\mathcal{S}$ of maximum weight.
\end{proof} 
\subsection{Proof of Theorem~\ref{TheoModified}}
This section is dedicated to the proof of Theorem~\ref{TheoModified}. It goes down a very similar road as our previous analysis. In particular, we consider the same algorithm as before (Algorithm~\ref{LocalImprovementAlgo}), but we re-define the notion of helpful vertices (see Def.~\ref{ReDefHelpful}) and instead of the $\frac{k+1+\tilde{\epsilon}}{3}$-approximation by F\"urer and Yu~\cite{FurerYu}, \texttt{MIS} now denotes the $\tilde{\tau}\cdot (k+1)$-approximation for the unweighted $k$-Set Packing problem for $k\geq k_0$ from the statement of Theorem~\ref{TheoModified} (we will explain soon how to choose $\tilde{\tau}$, depending on $\sigma$). We point out that as now, we are not interested in optimizing constants, but only in an existential statement, we can actually omit the search for improvements of size at most $3$ and circular improvements from \texttt{RunIteration} (Algorithm~\ref{AlgoRunIteration}) since their only purpose was to boost the order of magnitude of the constant $\xi$ (see Lemma~\ref{LemGainFromLowDegrees}). In particular, we will ignore these steps during the subsequent analysis and only use the properties that when Algorithm~\ref{LocalImprovementAlgo} terminates, there is neither an improving claw, nor does one of the sets $X^{\leq U}$ we obtain in line~\ref{LineXleqU} of \texttt{RunIteration} yield a local improvement.

We will show the following lemma:
\begin{lemma}
Let $\sigma\in (0,1)$. Then there is $\tilde{\tau}\in (0,1)$ with the following property: If there are $k_0\in \mathbb{N}_{\geq 3}$ and a polynomial time $\tilde{\tau}\cdot (k+1)$-approximation algorithm $\mathcal{A}$ for the unweighted $k$-Set Packing problem for $k\geq k_0$, then running Algorithm~\ref{LocalImprovementAlgo} with the notion of helpful vertices defined according to Definition~\ref{ReDefHelpful} and $\texttt{MIS}:=\mathcal{A}$ yields a $1+\frac{\sigma}{2}\cdot (k-1)$-approximation for the weighted $k$-Set Packing problem for $k\geq k_0$.\label{LemModified}
\end{lemma} Combining Lemma~\ref{LemModified} with Lemma~\ref{LemRuntimeIt} and Lemma~\ref{LemPolyNumIt} tells us that we can also obtain a polynomial time $1+\sigma\cdot (k-1)$-approximation for the weighted $k$-Set Packing problem by scaling and truncating the weight function, which concludes the proof of Theorem~\ref{TheoModified}.

To prove Lemma~\ref{LemModified}, we first need to introduce three constants $\delta\in \left(0,\frac{1}{4}\right)$, $\alpha > 1$ and $\beta\in (0,1)$. $\frac{1}{2}-\delta$ will play the role of $\xi$ in our previous analysis (cf. Lemma~\ref{LemProfitFromBadNeighbors} and Lemma~\ref{LemProfitNew}), $\alpha$ will correspond to our threshold $1+\epsilon$ on the proximity of vertex weights, and $\beta$ will more or less assume the role of the threshold $\epsilon^2$ telling us that we should consider $w^2(N(u,A)\setminus\mathrm{help}(u))$ as small if it is bounded by $\beta\cdot w^2(u)$. In addition, we need a fourth constant $m\in\mathbb{N}$ that tells us for how many of its neighbors we can allow a vertex $u\in V\setminus A$ to be helpful. On the one hand, even if $w(N(u,A))\leq 2\cdot w(u)$, $N(u,A)$ may still contain $2\cdot \alpha$ vertices that are within a weight range of $[\alpha^{-1}\cdot w(u),\alpha\cdot w(u)]$ and we would like $u$ to be helpful for as many of these as possible in order to be able to guarantee a sizeable reimbursement for the remaining neighbors in $\mathrm{supp}(u)$. On the other hand, when lower bounding $|A^*\cap V_{\geq L}|$ by using a lower bound on $\sum_{v\in A_{\geq L}} |\mathrm{help}(v)\cap V_{\geq L}\cap A^*|$ (cf. proof of Lemma~\ref{LemGoodUpperSweepSet1}), we need a bound on the maximum number of vertices in $A$ a vertex in $A^*$ can be helpful for. Our value of $m$ will provide a trade-off between these two aspects.

Fix $\sigma\in (0,1)$. We let $\delta:=\frac{\sigma}{4}\in \left(0,\frac{1}{4}\right)$ and pick $\alpha:=\delta^{-2}$, $\beta:=\delta$ and $m:=\lceil \delta^{-3}\rceil$. Moreover, we choose 
\begin{equation}
\tilde{\tau}:=\frac{(1-\beta)\cdot \sigma}{4\cdot m\cdot\alpha^4}=\dfrac{\left(1-\frac{\sigma}{4}\right)\cdot \sigma^9}{4^9\cdot\left\lceil \frac{4^3}{\sigma^3}\right\rceil}\in (0,1). \label{EqChoiceTau}
\end{equation}
Assume that there exist $k_0\in\mathbb{N}_{\geq 3}$ and a polynomial time $\tilde{\tau}\cdot (k+1)$-approximation algorithm \texttt{MIS} for the unweighted $k$-Set Packing problem for $k\geq k_0$.
Pick an instance $(\mathcal{S},w)$ of the weighted $k$-Set Packing problem for $k\geq k_0$, let $G=(V,E)$ be its conflict graph and let $A^*$ be an independent set in $G$ of maximum weight. Moreover, let $A$ be the solution maintained by Algorithm~\ref{LocalImprovementAlgo}.
In each iteration, for each $u\in V$, we sort the vertices in $N(u,A)$ by non-decreasing weight and denote the resulting linear order by $\preceq_u$. Then $\forall v,v'\in N(u,A):v\preceq_u v'\Rightarrow w(v)\geq w(v')$. Now, for $i=1,\dots,m$, define
\begin{align*}
n_i:\{u\in V: |N(u,A)|\geq i\}&\rightarrow A\\
u&\mapsto \text{the $i$-th element of $N(u,A)$ with respect to $\preceq_u$},
\end{align*}
that is, $n_i$ maps $u$ to an $i$-th heaviest neighbor in $A$, if exists. Observe that $n_1$ and $n_2$ coincide with the maps $n$ and $n_2$ from our previous analysis. Now, we can re-define the notion of helpful vertices. In doing so, we consider the maximal initial segment of $N(u,A)$ with respect to $\preceq_u$ such that all vertices $v$ contained in this initial segment satisfy $\alpha^{-1}\cdot w(u)\leq w(v)\leq \alpha\cdot w(u)$. Note that it might happen that the upper bound is violated by $n_1(u)$, causing us to define $\mathrm{help}(u)=\emptyset$, even though $n_2(u)$ then falls within the required weight range. We further truncate our initial segment after at most $m$ vertices, which gives a candidate set $\{n_i(u),1\leq i < i_{end}^u\}$ for $\mathrm{help}(u)$. If this candidate set has the additional property that the sum of the squared weights of the remaining neighbors of $u$ in $A$ is small, that is, bounded by $\beta\cdot w^2(u)$, then we select it as $\mathrm{help}(u)$. Otherwise, we set $\mathrm{help}(u):=\emptyset$.
\begin{definition}[helpful vertex]
	Let $u\in V\setminus A$ and let \[i_{end}^u:=\min\big\{|N(u,A)|+1,m+1,\min\{i\in\{1,\dots,|N(u,A)|\}:w(n_i(u))\not\in[\alpha^{-1}\cdot w(u),\alpha\cdot w(u)]\}\big\},\] where $\min\emptyset:=\infty$. We set
	\[\mathrm{help}(u):=\begin{cases} \{n_i(u),1\leq i < i_{end}^u\} &, w^2(\{n_i(u):i_{end}^u\leq i \leq |N(u,A)|\})\leq \beta\cdot w^2(u)\\
	\emptyset&,\text{otherwise}
	\end{cases}.\] Moreover, for $v\in A$, we define $\mathrm{help}(v):=\{u\in V\setminus A: v\in\mathrm{help}(u)\}$.\label{ReDefHelpful}
\end{definition}
Next, we re-define the support of $u\in V\setminus A$ to contain all neighbors of $u$ in $A$ that it is not helpful for.
\begin{definition}[support]
	For $u\in V\setminus A$, we define $\mathrm{supp}(u):=N(u,A)\setminus\mathrm{help}(u).$	\label{ReDefSupport}
\end{definition}
For the rest of the analysis, unless explicitly stated otherwise, we consider the state of $A$ in the last iteration of Algorithm~\ref{LocalImprovementAlgo}, in which it is not modified anymore. In particular, no claw improves $A$.
In this situation, Lemma~\ref{LemProfitNew} tells us that each vertex $u\in A^*\setminus A$ can reimburse all of the vertices it supports by a $\frac{1}{2}-\delta$ fraction of their weight by distributing the slack in the inequality \[\frac{1}{2}\cdot\left(\sum_{v\in A} \contr{u}{v}-2\cdot \chrg{u}{n_1(u)}\right)\geq 0.\]
\begin{lemma}
	For $u\in A^*\setminus A$, we have $\sum_{v\in A} \contr{u}{v}-2\cdot \chrg{u}{n_1(u)}\geq (1-2\cdot\delta)\cdot w(\mathrm{supp}(u))$.\label{LemProfitNew}
\end{lemma}
We first show the following auxiliary statement:
\begin{lemma}
	Let $u\in A^*\setminus A$ such that $w(N(u,A))\geq \delta^{-1}\cdot w(u)$. \\Then $\sum_{v\in A} \contr{u}{v}-2\cdot \chrg{u}{n_1(u)}\geq (1-2\cdot\delta)\cdot w(\mathrm{supp}(u))$. \label{LemAuxProfit}
\end{lemma}
\begin{proof}
	\begin{align*}
	&\phantom{=}\sum_{v\in N(u,A)}\contr{u}{v}-2\cdot\chrg{u}{n_1(u)}\stackrel{\eqref{EqBound1}}{\geq} w(N(u,A))-2\cdot w(u)\\&=\left(1-2\cdot\frac{w(u)}{w(N(u,A))}\right)\cdot w(N(u,A))\geq \left(1-2\cdot\frac{w(u)}{\delta^{-1}\cdot w(u)}\right)\cdot w(\mathrm{supp}(u))\\
	&= (1-2\cdot\delta)\cdot w(\mathrm{supp}(u)).
	\end{align*}
\end{proof}
\begin{proof}[Proof of Lemma~\ref{LemProfitNew}]
	As there is no claw-shaped improvement when Algorithm~\ref{LocalImprovementAlgo} terminates, we know that $A$ is a maximal independent set and $N(u,A)\neq \emptyset$.\\
	\textbf{Case 1:} $w(n_1(u)) > \alpha\cdot w(u)$. Then $w(N(u,A))\geq w(n_1(u))>\alpha\cdot w(u)=\delta^{-2}\cdot w(u) \geq \delta^{-1}\cdot w(u)$, as $\delta\in (0,1)$, and Lemma~\ref{LemAuxProfit} yields the claim.\\
	\textbf{Case 2:} $w(n_1(u)) < \alpha^{-0.5}\cdot w(u)$.
	The fact that there is no claw-shaped improvement, and in particular, $\{u\}$ cannot constitute one, yields \[w^2(u)\leq w^2(N(u,A))\leq w(n_1(u))\cdot w(N(u,A)) < \alpha^{-0.5}\cdot w(u)\cdot w(N(u,A)).\]
	This implies that $w(N(u,A))>\alpha^{0.5}\cdot w(u) = \delta^{-1}\cdot w(u)$ and we can, again, apply Lemma~\ref{LemAuxProfit} to conclude the desired statement.\\
	\textbf{Case 3:} $\alpha^{-0.5}\cdot w(u)\leq  w(n_1(u))\leq \alpha\cdot w(u)$ and there is $2\leq j\leq \min\{|N(u,A)|,m\}$ such that $w(n_j(u))<\alpha^{-1}\cdot w(u)$. Pick $j$ minimum with this property.\\ If $w^2(\{n_i(u), j\leq i\leq |N(u,A)|\})\leq \beta\cdot w^2(u)$, then \[\mathrm{help}(u)=\{n_i(u),1\leq i < j\},\] and we obtain
	\begingroup\allowdisplaybreaks
	\begin{align*}
	&\phantom{=}\sum_{v\in N(u,A)}\contr{u}{v}\cdot w(n_1(u))-2\cdot\chrg{u}{n_1(u)}\cdot w(n_1(u))\\
	&\geq w^2(u)-w^2(N(u,A)\setminus\{n_1(u)\})-(2\cdot w(u)-w(N(u,A)))\cdot w(n_1(u))\\
	&= (w(u)-w(n_1(u)))^2+w(n_1(u))\cdot w(N(u,A)\setminus\{n_1\})-w^2(N(u,A)\setminus\{n_1(u)\})\\
	&=(w(u)-w(n_1(u)))^2+\sum_{i=2}^{|N(u,A)|} (w(n_1(u))-w(n_i(u)))\cdot w(n_i(u))\\
	&\geq \sum_{i=j}^{|N(u,A)|} (w(n_1(u))-w(n_i(u)))\cdot w(n_i(u))\\
	&\geq \sum_{i=j}^{|N(u,A)|} (w(n_1(u))-\alpha^{-1}\cdot w(u))\cdot w(n_i(u)) \quad \text{ $|$ $\alpha^{-0.5}\cdot w(u)<w(n_1(u))$}\\
	&\geq \sum_{i=j}^{|N(u,A)|} (1-\alpha^{-0.5})\cdot w(n_1(u))\cdot w(n_i(u))=(1-\delta)\cdot w(n_1(u))\cdot w(\{n_i(u), j\leq i\leq |N(u,A)|\})\\
	&=(1-\delta)\cdot w(n_1(u))\cdot w(\mathrm{supp}(u))\geq (1-2\cdot\delta)\cdot w(n_1(u))\cdot w(\mathrm{supp}(u)).
	\end{align*}\endgroup
	Division by $w(n_1(u))>0$ yields the claim.
	If $w^2(\{n_i(u), j\leq i\leq |N(u,A)|\}) > \beta\cdot w^2(u)$, then
	\[\alpha^{-1}\cdot w(u)\cdot  w(\{n_i(u), j\leq i\leq |N(u,A)|\}) > w^2(\{n_i(u), j\leq i\leq |N(u,A)|\})> \beta\cdot w^2(u),\] leading to $w(N(u,A))\geq w(\{n_i(u), j\leq i\leq |N(u,A)|\})\geq \alpha\cdot \beta\cdot w(u)=\delta^{-1}\cdot w(u)$.
	Once more, we can apply Lemma~\ref{LemAuxProfit} to conclude the claim.\\
	\textbf{Case 4:} $\alpha^{-0.5}\cdot w(u)\leq w(n_1(u))\leq \alpha\cdot w(u)$ and \[\forall 1\leq j\leq \min\{|N(u,A)|,m\}:w(n_j(u))\geq\alpha^{-1}\cdot w(u).\]
	If $|N(u,A)|\leq m$, all vertices in $N(u,A)$ are helpful for $u$ because $i_{end}^u=|N(u,A)|+1$  and $w^2(\{n_i(u):i^u_{end}\leq i\leq |N(u,A)|\})=w^2(\emptyset)=0\leq \beta\cdot w^2(u)$. In particular, $\mathrm{supp}(u)=\emptyset$ and Proposition~\ref{PropUpperBoundContr} yields the desired statement.
	
	Hence, we may assume $m < |N(u,A)|$ and obtain \[w(N(u,A))\geq \sum_{i=1}^m w(n_i(u))\geq m\cdot\alpha^{-1}\cdot w(u)\geq \delta^{-1}\cdot w(u).\] Applying Lemma~\ref{LemAuxProfit} concludes the proof.
\end{proof}
\begin{lemma}
	When Algorithm~\ref{LocalImprovementAlgo} terminates, we have
	\[w(A^*)\leq \frac{k+1}{2}\cdot w(A)-\left(\frac{1}{2}-\delta\right)\cdot\sum_{v\in A}(k-1-|\mathrm{help}(v)\cap A^*|)\cdot w(v).\]\label{LemGainFromLowDegrees2}
\end{lemma}
\begin{proof}
	The proof is almost identical to the proof of Lemma~\ref{LemGainFromLowDegrees}.
	By Theorem~\ref{TheoApproxFactor}, we know that 
	\begin{align}w(A^*)\leq\frac{k+1}{2}\cdot w(A)&-\frac{1}{2}\cdot \sum_{v\in A} (k-|N(v,A^*)|)\cdot w(v)\notag\\&- \frac{1}{2}\cdot\sum_{u\in A^*}\left(\sum_{v\in A} \contr{u}{v} - 2\cdot \chrg{u}{n_1(u)}\right).\label{EqLemSquareImp1}\end{align}
	We calculate
	\begin{align*}
	&\phantom{=}\frac{1}{2}\cdot\sum_{u\in A^*}\left(\sum_{v\in A} \contr{u}{v} - 2\cdot \chrg{u}{n_1(u)}\right)\\&\stackrel{Lemma~\ref{LemPropPositiveCharges}}{\geq}\frac{1}{2}\cdot\sum_{u\in A^*\setminus A}\left(\sum_{v\in A} \contr{u}{v} - 2\cdot \chrg{u}{n_1(u)}\right)
	\stackrel{Lem.~\ref{LemProfitNew}}{\geq} \sum_{u\in A^*\setminus A} \left(\frac{1}{2}-\delta\right)\cdot w(\mathrm{supp}(u))\\&=\sum_{u\in A^*\setminus A}\sum_{v\in\mathrm{supp}(u)}\left(\frac{1}{2}-\delta\right)\cdot w(v)=\sum_{v\in A} |\{u\in A^*\setminus A:v\in\mathrm{supp}(u)\}|\cdot \left(\frac{1}{2}-\delta\right)\cdot w(v).
	\end{align*}
	Plugging this into \eqref{EqLemSquareImp1} yields
	\[w(A^*)\leq \frac{k+1}{2}\cdot w(A)-\sum_{v\in A}\left((k-|N(v,A^*)|)\cdot \frac{1}{2}\cdot w(v)+|\{u\in A^*\setminus A:v\in\mathrm{supp}(u)\}|\cdot \left(\frac{1}{2}-\delta\right)\cdot w(v)\right),\]
	and to verify the statement of the lemma, it suffices to show that \begin{align*}\forall v\in A: &(k-|N(v,A^*)|)\cdot \frac{1}{2}\cdot w(v)+|\{u\in A^*\setminus A:v\in\mathrm{supp}(u)\}|\cdot \left(\frac{1}{2}-\delta\right)\cdot w(v)\\&\geq \left(\frac{1}{2}-\delta\right)\cdot(k-1-|\mathrm{help}(v)\cap A^*|)\cdot w(v).\end{align*}
	For $v\in A\cap A^*$, we have $|N(v,A^*)|=1$ and, hence, obtain
	\begin{align*}&\phantom{=}(k-|N(v,A^*)|)\cdot \frac{1}{2}\cdot w(v)+|\{u\in A^*\setminus A:v\in\mathrm{supp}(u)\}|\cdot \left(\frac{1}{2}-\delta\right)\cdot w(v)\geq \frac{k-1}{2}\cdot w(v)\\&\geq \left(\frac{1}{2}-\delta\right)\cdot(k-1-|\mathrm{help}(v)\cap A^*|)\cdot w(v).\end{align*}
	
	Next, we consider the case that $v\in A\setminus A^*$. As $A$ is independent, this implies $N(v,A^*)\subseteq A^*\setminus A$.
	Hence, Definition~\ref{ReDefHelpful} and Definition~\ref{ReDefSupport} imply \[N(v,A^*)=(\mathrm{help}(v)\cap A^*)\dot{\cup}\{u\in A^*\setminus A: v\in\mathrm{supp}(u)\}.\]This tells us that
	\begin{align*}
	&\phantom{=}(k-|N(v,A^*)|)\cdot \frac{1}{2}\cdot w(v)+|\{u\in A^*\setminus A:v\in\mathrm{supp}(u)\}|\cdot\left(\frac{1}{2}-\delta\right) \cdot w(v)\\&\geq (k-|N(v,A^*)|)\cdot \left(\frac{1}{2}-\delta\right)\cdot w(v)+(|N(v,A^*)|-|\mathrm{help}(v)\cap A^*|)\cdot \left(\frac{1}{2}-\delta\right)\cdot w(v)\\
	&= (k-|\mathrm{help}(v)\cap A^*|)\cdot \left(\frac{1}{2}-\delta\right)\cdot w(v)\stackrel{\delta\in \left(0,\frac{1}{2}\right)}{\geq} (k-1-|\mathrm{help}(v)\cap A^*|)\cdot \left(\frac{1}{2}-\delta\right)\cdot w(v),
	\end{align*}
	which concludes the proof.
\end{proof}
\begin{corollary}
	If $\sum_{v\in A}|\mathrm{help}(v)\cap A^*|\cdot w(v)\leq \frac{\sigma}{2}\cdot (k-1)\cdot w(A)$, then $w(A^*)\leq \left(1+\frac{\sigma}{2}\cdot(k-1)\right)\cdot w(A).$\label{CorApproxGuarantee}
\end{corollary}
\begin{proof}
	By Lemma~\ref{LemGainFromLowDegrees2}, we get
	\begin{align*}
	w(A^*)&\leq  \frac{k+1}{2}\cdot w(A)-\left(\frac{1}{2}-\delta\right)\cdot\sum_{v\in A}(k-1-|\mathrm{help}(v)\cap A^*|)\cdot w(v)\\ 
	&=\frac{k+1}{2}\cdot w(A)-\left(\frac{1}{2}-\delta\right)\cdot(k-1)\cdot w(A)+\left(\frac{1}{2}-\delta\right)\cdot\sum_{v\in A}|\mathrm{help}(v)\cap A^*|\cdot w(v)\\
	&\stackrel{\delta\in \left(0,\frac{1}{2}\right)}{\leq} w(A)+\delta\cdot (k-1)\cdot w(A)+\left(\frac{1}{2}-\delta\right)\cdot\frac{\sigma}{2}\cdot (k-1) \cdot w(A)\\
	&\stackrel{0<\delta\leq \frac{\sigma}{4}}{\leq} w(A)+\frac{\sigma}{4}\cdot (k-1)\cdot w(A)+\frac{\sigma}{4}\cdot (k-1)\cdot w(A)\\
	&=\left(1+\frac{\sigma}{2}\cdot(k-1)\right)\cdot w(A).
	\end{align*}
\end{proof}
As a consequence, we are done if we can show that the assumption of Corollary~\ref{CorApproxGuarantee} is always met when Algorithm~\ref{LocalImprovementAlgo} terminates. To show this, we just adapt the proof of Lemma~\ref{LemImprovementNextIter} to fit into our new setting.
\begin{lemma}
	If $\sum_{v\in A}|\mathrm{help}(v)\cap A^*|\cdot w(v)> \frac{\sigma}{2}\cdot (k-1)\cdot w(A)$ at the beginning of an iteration of Algorithm~\ref{LocalImprovementAlgo}, then we can find a local improvement in that iteration.\label{LemImprovementNextIterNew}
\end{lemma}
In particular, this means that at the beginning (and end) of the last iteration, we have \[\sum_{v\in A}|\mathrm{help}(v)\cap A^*|\cdot w(v)\leq\frac{\sigma}{2}\cdot (k-1)\cdot w(A)\] and can apply Corollary~\ref{CorApproxGuarantee}.

Lemma~\ref{LemImprovementNextIterNew} follows by combining Lemma~\ref{LemGoodUpperSweepSet1} and Lemma~\ref{LemGoodLowerSweepSet2}.
\begin{lemma}
	If 	$\sum_{v\in A}|\mathrm{help}(v)\cap A^*|\cdot w(v)> \frac{\sigma}{2}\cdot (k-1)\cdot w(A)$ and Algorithm~\ref{AlgoRunIteration} does not find a claw-shaped improvement, (a local improvement of size at most $3$ or a circular improvement,) then there is $L\in \{w(v),v\in V\}$ such that 
	\[|A^*\cap V_{\geq L}|> \frac{\sigma\cdot (k-1)}{2\cdot m\cdot \alpha^2}\cdot |A_{\geq L}|.\]
	In particular, \texttt{MIS}, when applied to $G[V_{\geq L}]$, returns an independent set $\bar{X}$ of cardinality
	\[|\bar{X}| > \frac{\sigma}{4\cdot\tilde{\tau}\cdot m\cdot \alpha^2}\cdot |A_{\geq L}|.\]
	\label{LemGoodUpperSweepSet1}
\end{lemma}
\begin{lemma}
	Assume that the set $\bar{X}$ in line~\ref{LineBarX} of Algorithm~\ref{AlgoRunIteration} satisfies \[|\bar{X}| >\frac{\sigma}{4\cdot\tilde{\tau}\cdot m\cdot \alpha^2}\cdot |A_{\geq L}|.\] Then there exists $U\in\{w(v),v\in V\}$ such that the set $X^{\leq U}$ defined in line~\ref{LineXleqU} yields a local improvement.
	\label{LemGoodLowerSweepSet2}
\end{lemma}
Compliant with Algorithm~\ref{AlgoRunIteration}, let $V_{help}:=\{u\in V\setminus A: \mathrm{help}(u)\neq \emptyset\}$ and for $x\in\mathbb{R}_{\geq 0}$, let \[A_{\geq x}=\{v\in A: w(v)\geq x\}\text{ and }V_{\geq x}:=A_{\geq x}\cup\{u\in V_{help}: w(u)\geq x \wedge \mathrm{help}(u)\subseteq A_{\geq x} \}.\]
\begin{proof}[Proof of Lemma~\ref{LemGoodUpperSweepSet1}]
	The proof proceeds analogously to the proof of Lemma~\ref{LemGoodUpperSweepSet}.
	\begin{claim}
		For $x > 0$ and $v\in A$ with $\alpha^2\cdot x\leq w(v)$, we have $\mathrm{help}(v)\cap A^*\subseteq V_{\geq x}$.\label{ClaimSameDegree2}
	\end{claim}
	\begin{proof}
		Let $u\in \mathrm{help}(v)\cap A^*$. By Definition~\ref{ReDefHelpful}, we know that $w(u)\geq \alpha^{-1}\cdot w(v)\geq \alpha\cdot x\geq x$ and moreover, any $z\in\mathrm{help}(u)$ satisfies $w(z)\geq \alpha^{-1}\cdot w(u)\geq x$, and, hence, $z\in A_{\geq x}$. This implies that $u\in V_{\geq x}$.\end{proof} 
	Our next goal is to show the following statement: 
	\begin{claim}There is $L\in\{w(v),v\in V\}$ such that \[\sum_{v\in A_{\geq L}} |\mathrm{help}(v)\cap A^*\cap V_{\geq L}|> \frac{\sigma\cdot (k-1)}{2\cdot\alpha^2}\cdot |A_{\geq L}|.\] \end{claim}
	\begin{proof}
		We want to apply Lemma~\ref{LemAverage}. To this end, set $S:=A$, $\mu(v):=|\mathrm{help}(v)\cap A^*|$, $\varphi(v):=w(v)$, $\lambda:=\alpha^{-2}$ and $\eta:= \frac{\sigma}{2}\cdot (k-1)$.
		Then Lemma~\ref{LemAverage} tells us that there is $L\in (0,\infty)$ such that \[\sum_{v\in A_{\geq L}} |\mathrm{help}(v)\cap A^*\cap V_{\geq L}|\stackrel{\small\text{Claim~\ref{ClaimSameDegree2}}}{\geq} \sum_{v\in A_{\geq \alpha^2\cdot L}}|\mathrm{help}(v)\cap A^*|\stackrel{\small\text{Lem.~\ref{LemAverage}}}{>}\frac{\sigma}{2} \cdot(k-1)\cdot \alpha^{-2}\cdot |A_{\geq L}|.\] In particular, the strict inequality implies that $A_{\geq L}\neq \emptyset$, and by increasing $L$ to the minimum weight of a vertex currently contained in $A_{\geq L}$, we may assume that $L\in\{w(v),v\in V\}$.
	\end{proof} Given that $|\mathrm{help}(u)|\leq m$ for every $u\in A^*$ by Definition~\ref{ReDefHelpful}, we obtain
	\begin{align*}
	m\cdot |A^*\cap V_{\geq L}|\geq \sum_{u\in (A^*\setminus A)\cap V_{\geq L}} |\mathrm{help}(u)|\stackrel{(*)}{=}\sum_{v\in A_{\geq L}} |\mathrm{help}(v)\cap A^*\cap V_{\geq L}| > \frac{\sigma}{2} \cdot(k-1)\cdot \alpha^{-2}\cdot |A_{\geq L}|.
	\end{align*}
	The equation marked $(*)$ is implied by the facts that $\mathrm{help}(u)\subseteq A_{\geq L}$ for $u\in V_{\geq L}$ , $\mathrm{help}(v)\subseteq V\setminus A$ for $v\in A$, and $u\in\mathrm{help}(v)\Leftrightarrow v\in\mathrm{help}(u)$ for $u\in V\setminus A$ and $v\in A$. Division by $m$ yields the first part of the desired statement. Moreover, as $A^*\cap V_{\geq L}$ is an independent set in $G[V_{\geq L}]$, we know that \texttt{MIS}, applied to $G[V_{\geq L}]$, will output an independent set $\bar{X}$ of cardinality
	\[|\bar{X}|\geq \frac{|A^*\cap V_{\geq L}|}{\tilde{\tau}\cdot (k+1)} > \frac{k-1}{k+1}\cdot \frac{\sigma}{2\cdot\tilde{\tau}\cdot m\cdot \alpha^2}\cdot |A_{\geq L}|\geq \frac{\sigma}{4\cdot\tilde{\tau}\cdot m\cdot \alpha^2}\cdot |A_{\geq L}|.\]
\end{proof}
\begin{proof}[Proof of Lemma~\ref{LemGoodLowerSweepSet2}] Define $X:=\bar{X}\setminus A$.
	As $\bar{X}\cap A\subseteq V_{\geq L}\cap A=A_{\geq L}$, we get \[|X|> \frac{\sigma}{4\cdot\tilde{\tau}\cdot m\cdot\alpha^2}\cdot |A_{\geq L}|-|\bar{X}\cap A|\geq \frac{\sigma}{4\cdot\tilde{\tau}\cdot m\cdot\alpha^2}\cdot |A_{\geq L}\setminus\bar{X}|,\] where the last inequality follows from the fact that $\frac{\sigma}{4\cdot\tilde{\tau}\cdot m\cdot\alpha^2}>1$ by \eqref{EqChoiceTau} and since $\alpha > 1$ and $\beta\in(0,1)$. Moreover, as $\bar{X}$ is independent in $G[V_{\geq L}]$, and, hence, in $G$, no vertex in $X\subseteq \bar{X}\setminus A$ is adjacent to a vertex in $A_{\geq L}\cap\bar{X}$, implying that $N(X,A_{\geq L})\subseteq A_{\geq L}\setminus\bar{X}$. Hence, we obtain
	\[|X|>\frac{\sigma}{4\cdot\tilde{\tau}\cdot m\cdot\alpha^2}\cdot |N(X,A_{\geq L})|.\]
	\begin{claim}There is $U\in\{w(v),v\in V\}$ such that $w^2(X^{\leq U})> \frac{\sigma}{4\cdot\tilde{\tau}\cdot m\cdot\alpha^4}\cdot w^2(N(X^{\leq U},A_{\geq L}))$.\end{claim}
	\begin{proof} We want to apply Lemma~\ref{LemWeightedAverage}. To this end, let $S_1:=X$, $S_2:=N(X,A_{\geq L})$, $\varphi(s):=w^2(s)>0$ for $s\in S_1\cup S_2$, $\eta:= \frac{\sigma}{4\cdot\tilde{\tau}\cdot m\cdot\alpha^2}$ and $\lambda:=\alpha^{-2}$.
		In this setting, Lemma~\ref{LemWeightedAverage} tells us that there is $U>0$ such that \[w^2(X^{\leq U})>  \frac{\sigma}{4\cdot\tilde{\tau}\cdot m\cdot\alpha^4}\cdot w^2(\{v\in N(X,A_{\geq L}): w^2(v)\leq \alpha^2\cdot U\}).\]
		By definition of $V_{\geq L}$, we know that for $u\in X\subseteq V_{\geq L}\setminus A\subseteq V_{help}$, every $v\in N(u,A)$ satisfies $w^2(v)\leq \alpha^2\cdot w^2(u)$ because this holds if $v\in \mathrm{help}(u)$ and for $v\in N(u,A)\setminus \mathrm{help}(u)$, we have
		$w^2(v)\leq w^2(N(u,A)\setminus\mathrm{help}(u))\leq \beta\cdot w^2(u)\leq \alpha^2\cdot w^2(u)$. Hence, \[N(X^{\leq U},A)\subseteq \{v\in N(X,A_{\geq L}): w^2(v)\leq \alpha^2\cdot U\},\] which yields
		\[w^2(X^{\leq U})> \frac{\sigma}{4\cdot\tilde{\tau}\cdot m\cdot\alpha^4}\cdot w^2(N(X^{\leq U}, A_{\geq L})).\] In particular, the strict inequality tells us that $X^{\leq U}\neq \emptyset$ and by decreasing $U$ to the maximum weight among the vertices in $X^{\leq U}$, we can assume that $U\in\{w(v),v\in V\}$.\end{proof}
	To see that $X^{\leq U}$ as implied by the claim constitutes a local improvement of the squared weight function, it remains to bound $w^2(N(X^{\leq U},A\setminus A_{\geq L}))$. To this end, as $X\subseteq V_{\geq L}\setminus A$, we know that for $u\in X$, we have $\emptyset\neq \mathrm{help}(u)\subseteq A_{\geq L}$ and $w^2(N(u,A)\setminus \mathrm{help}(u))\leq \beta\cdot w^2(u)$.  This yields $w^2(N(X^{\leq U},A\setminus A_{\geq L}))\leq \beta\cdot w^2(X^{\leq U})$ and
	\begin{align*}w^2(X^{\leq U})&\stackrel{\eqref{EqChoiceTau}}{=} \left(\frac{4\cdot\tilde{\tau}\cdot m\cdot\alpha^4}{\sigma}+\beta\right)\cdot w^2(X^{\leq U})>w^2(N(X^{\leq U}, A_{\geq L}))+w^2(N(X^{\leq U}, A\setminus A_{\geq L}))\\&=w^2(N(X^{\leq U},A)).\end{align*} This shows that $X^{\leq U}$ yields a local improvement. 	
\end{proof}
\subsection{Proof of Lemma~\ref{LemTildeTau}}
\begin{proof}[Proof of Lemma~\ref{LemTildeTau}]
As both the weighted and the unweighted $k$-Set Packing problem for $k\geq 3$ generalize the 3D-matching problem, by~\cite{BermanKarpinski}, there exists a constant $C>1$\footnote{The precise value can be read from \cite{BermanKarpinski}.} with the property that unless $P=NP$, there is no $C$-approximation for either of the unweighted or the weighted $k$-Set Packing problem. Let $\tilde{\tau}\in (0,1)$ and define $\tau:=\tilde{\tau}\cdot (1+\frac{1}{C-1})^{-1}$. Then $0<\tau<\tilde{\tau}<1$. Let $k_0\in\mathbb{N}_{\geq 3}$ and assume that there exists a polynomial time $1+\tau\cdot (k-1)$-approximation for the unweighted (weighted) $k$-Set Packing problem for $k\geq k_0$. If $1+\tau\cdot (k_0-1)\leq C$, then $P=NP$ and we are done. Hence, assume $1+\tau\cdot (k_0-1) > C$.  Then for every $k\geq k_0$, we have
\begin{align*}\tilde{\tau}\cdot (k+1)&\geq \tilde{\tau}\cdot (k-1)=\tau\cdot \left(1+\frac{1}{C-1}\right)\cdot (k-1)=\tau\cdot (k-1)+\frac{\tau\cdot (k-1)}{C-1}\\&\geq \tau\cdot (k-1)+\frac{\tau\cdot (k_0-1)}{C-1}>\tau\cdot (k-1)+1,\end{align*} which concludes the proof.
\end{proof}
\bibliography{set_packing}
\end{document}